\documentclass[twocolumn,showpacs,nofootinbib,superscriptaddress,preprintnumbers,amsmath,amssymb,epsfig,color]{aa}

\usepackage{natbib}
\bibpunct{(}{)}{;}{a}{}{,} % to follow the A&A style...

\usepackage[usenames]{color}
\usepackage{color,graphicx,epsfig,amsmath,amssymb,fancyheadings}

\newcommand{\beq}{\begin{equation}}
\newcommand{\eeq}{\end{equation}}
\newcommand{\ben}{\begin{eqnarray}}
\newcommand{\een}{\end{eqnarray}}
\newcommand{\bi}{\begin{itemize}}
\newcommand{\ei}{\end{itemize}}

\newcommand{\cf}{\textit{cf.}}
\newcommand{\ie}{\textit{i.e.}}

\newcommand{\pbar}{{\ifmmode \overline{p} \else $\overline{p}$\fi}}
\newcommand{\dbar}{{\ifmmode \overline{d} \else $\overline{d}$\fi}}
\newcommand{\gtilde}{{\ifmmode \tilde{\cal G} \else $\tilde{{\cal G}}$\fi}}
\newcommand{\gtildepos}{{\ifmmode \gtilde^{\; e^+} \else 
    $\gtilde^{\; e^+}$\fi}}
\newcommand{\gtildepbar}{{\ifmmode \gtilde^{\; \pbar} \else 
    $\gtilde^{\; \pbar}$\fi}}

\newcommand{\Msol}{{\ifmmode M_{\odot} \else $M_{\odot}$\fi}}
\newcommand{\Rsol}{{\ifmmode R_{\odot} \else $R_{\odot}$\fi}}
\newcommand{\rhosol}{{\ifmmode \rho_{\odot} \else $\rho_{\odot}$\fi}}
\newcommand{\fsol}{{\ifmmode f_{\odot} \else $f_{\odot}$\fi}}
\newcommand{\gev}{{\ifmmode {\rm GeV} \else ${\rm GeV}$\fi}}

\newcommand{\lcdm}{{\ifmmode \Lambda{\rm CDM} \else $\Lambda{\rm CDM}$\fi}}

\newcommand{\Dvir}{{\ifmmode \Delta_{\rm vir} \else $\Delta_{\rm vir}$\fi}}
\newcommand{\cvir}{{\ifmmode c_{\rm vir} \else $c_{\rm vir}$\fi}}
\newcommand{\Rvir}{{\ifmmode R_{\rm vir} \else $R_{\rm vir}$\fi}}
\newcommand{\Rvirh}{{\ifmmode R_{\rm vir}^{\rm h} \else 
    $R_{\rm vir}^{\rm h}$\fi}}
\newcommand{\Mmin}{{\ifmmode M_{\rm min} \else $M_{\rm min}$\fi}}
\newcommand{\Mmax}{{\ifmmode M_{\rm max} \else $M_{\rm max}$\fi}}

\newcommand{\Mvir}{{\ifmmode M_{\rm vir} \else $M_{\rm vir}$\fi}}
\newcommand{\Mvirh}{{\ifmmode M_{\rm vir}^{\rm h} \else 
    $M_{\rm vir}^{\rm h}$\fi}}
\newcommand{\alpham}{{\ifmmode \alpha_{\rm m} \else $\alpha_{\rm m}$\fi}}

\newcommand{\rhocl}{{\ifmmode \rho_{\rm cl} \else $\rho_{\rm cl}$\fi}}
\newcommand{\Mcl}{{\ifmmode M_{\rm cl} \else $M_{\rm cl}$\fi}}
\newcommand{\Mcltot}{{\ifmmode M_{\rm cl}^{\rm tot} \else 
    $M_{\rm cl}^{\rm tot}$\fi}}
\newcommand{\Ncl}{{\ifmmode N_{\rm cl} \else $N_{\rm cl}$\fi}}
\newcommand{\ncl}{{\ifmmode n_{\rm cl} \else $n_{\rm cl}$\fi}}
\newcommand{\phicl}{{\ifmmode \phi_{\rm cl} \else $\phi_{\rm cl}$\fi}}
\newcommand{\phicltot}{{\ifmmode \phi_{\rm cl}^{\rm tot} \else 
    $\phi_{\rm cl}^{\rm tot}$\fi}}
\newcommand{\Beff}{{\ifmmode B_{\rm eff} \else $B_{\rm eff}$\fi}}

\begin{document}
\input epsf

\title{Full calculation of clumpiness boost factors for antimatter cosmic rays 
  in the light of $\Lambda$CDM N-body simulation results}
\subtitle{Abandoning hope in clumpiness enhancement?}
\author{
  J. Lavalle\inst{1}
  \and Q. Yuan\inst{2}
  \and D. Maurin\inst{3}
  \and X.-J. Bi\inst{2}
} 

\offprints{\tt \\
  lavalle@in2p3.fr\\
  yuanq@mail.ihep.ac.cn\\
  dmaurin@lpnhe.in2p3.fr\\
  bixj@mail.ihep.ac.cn}

\institute{Centre de Physique des Particules ({\sc cppm}), 
  CNRS-IN2P3 / Universit\'e de la M\'editerran\'ee, 163 avenue de Luminy, 
  case 902, 13288 Marseille Cedex 09 | France
  \and Key Laboratory of Particle Astrophysics, 
  Institute of High Energy Physics, Chinese Academy of Sciences,
  P.O.Box 918-3, Beijing 100049 | P.~R.~China
  \and Laboratoire de Physique Nucl\'eaire et Hautes \'Energies 
  ({\sc lpnhe}), 
  CNRS-IN2P3 / Universit\'es Paris VI et Paris VII,
  4 place Jussieu, Tour 33
  75252 Paris Cedex 05 | France
}

\date{Received / Accepted}% It is always \today, today,

\abstract
%
%Text of Context (optional)
%---------------
{Anti-proton and positron Galactic cosmic ray spectra are among the key 
  targets for indirect detection of dark matter. The boost factors,
  corresponding to an enhancement of the signal, and linked to the clumpiness
  properties of the dark matter distribution, have been taken as high as
  thousands in the past. The dramatic impact of these boost factors
  for indirect detection of antiparticles, for instance with the PAMELA 
  satellite or the coming AMS-02 experiment, asks for their detailed 
  calculation.
}
%
%Text of aims
%------------
{We take into account the state-of-the-art results of high resolution
  N-body dark matter simulations to calculate the most likely energy 
  dependent boost factors, which are linked to the cosmic ray propagation 
  properties, for anti-protons and positrons. The results from extreme, but 
  still possible, configurations of the clumpy dark matter component are also 
  discussed.
}
%
%Text of methods
%---------------
{Starting from the mass and space distributions of sub-halos, the anti-proton
  and positron propagators are used to calculate the mean value and the 
  variance of the boost factor for the primary fluxes. We take advantage of the
  statistical method introduced in Lavalle et al. (2007) and cross-check
  the results with Monte Carlo computations.
}
%
%Text of results
%---------------
{By spanning some extreme configurations of sub-halo and propagation
  properties, we find that the average contribution of the clumps is 
  negligible compared to that of the smooth dark matter component. Dark matter
  clumps do not lead to enhancement of the signals, unless they are taken with 
  some extreme (unexpected) properties. This result
  is independent of the nature of the self-annihilating dark matter candidate 
  considered, and provides precise estimates of the theoretical and 
  the statistical uncertainties of the antimatter flux from sub-halos.
}
%
%Text of Conclusions (optional)
%------------------- 
{Spectral distortions can still be expected in antimatter flux measurements,
  but scenarios invoking large and even mild clumpiness boost factors are 
  strongly disfavoured by our analysis. Some very extreme configurations could 
  still lead to large enhancements, e.g. (i) very small clumps with masses 
  $\lesssim 10^{-6}\Msol$ following a $M^{-\alpha}$ mass
  distribution with $\alpha\gtrsim 2$, highly concentrated with internal 
  $r^{-\beta}$ profiles with $\beta\gtrsim 1.5$, 
  and spatially distributed according to the smooth component; or (ii) a big 
  sub-halo of mass $\gtrsim 10^{7}\Msol$ within a distance of $\lesssim$ 1 
  kpc from the Earth. However, they are very unlikely from either theoretical 
  or statistical arguments.
}

\keywords{Dark Matter}

\titlerunning{Antimatter GCRs from DM annihilation: Abandoning hope in 
  clumpiness enhancement?}
\authorrunning{Lavalle et al.}  

\maketitle

\begin{flushleft}
  preprint CPPM-P-2007-02
\end{flushleft}

%/////////////////////////////////////////////////////////////////
\section{Introduction}

The existence of dark matter (DM) has been established by various astronomical 
observations, from galactic to cosmological scales. The evidence come 
from gravitational effects, such as the observation of the rotation curves in 
spiral galaxies and velocity dispersion in elliptical galaxies, the X-ray 
emission and peculiar velocity dispersion of galaxies in the clusters of 
galaxies and the weak lensing effects, all indicating much steeper 
gravitational potentials than those inferred from the luminous matter. 
Recently, there have been two strong smoking guns from the 
\emph{Bullet} cluster system 1E0561 \citep{2004ApJ...604..596C,
  2006ApJ...648L.109C,2006ApJ...652..937B} and a DM ring discovered 
around the cluster CL0024+17 \citep{2007arXiv0705.2171J}, which may 
indicate the existence of DM in the sense that it has first provided means to 
study the dynamics of DM itself. Note, however, that modified gravity 
models might still offer a viable alternative
\citep{2006MNRAS.371..138A,2007ApJ...654L..13A,2007arXiv0704.0381A,
  2007arXiv0706.1279F}.

The nature of DM is still unknown, remains one of the most outstanding puzzles 
in astrophysics and cosmology, and is challenging from the particle physics 
view point. Nevertheless, the unprecedented precision reached in observational 
cosmology in the last decade, thanks to the combined use of different probes 
(CMB, type 1A supernovae, large scale structures, deep surveys, primordial 
abundances, etc.), yields a rather precise estimate of the total amount of 
non-relativistic matter in the Universe, encompassing the standard baryonic 
matter, of which density can be predicted and measured independently 
(for reviews, see e.g. \citealt{2000PhST...85...12T,
2004astro.ph..9426L,2006JPhG...33....1Y}). The overall contribution 
of matter to the (critical) energy density of the Universe is $\sim$ 30\%, 
while the baryonic component accounts for 4\% only. Hence most of the 
matter should be dark and of non-baryonic origin, requiring 
physics beyond the standard model of particle physics. The most attractive 
scenario involves weakly interacting massive particles (WIMPs). An appealing 
idea is that WIMPs could be thermal relics of the early Universe, which 
naturally give rise to a cosmological abundance in the range of the observed 
value if both the interaction strength and the masses are taken at the weak 
scale. Indeed, because of their thermal origin, WIMPs should still (weakly) 
interact with ordinary matter, and even annihilate if they are preserved from 
matter-antimatter asymmetry. Such particles can originate naturally in the 
context of supersymmetric (SUSY) or extra-dimensional (ED) extensions 
of the standard model, independently developed to tackle the issues of 
the unification of interactions and energy scale hierarchical problems. 
Indeed, in such theories, the stability of the proton is very often ensured 
by the conservation of some new discrete symmetry that guaranties the 
lightest exotic particle to be stable. Such 
paradigms provide very good candidates for DM 
(for reviews, see e.g. \citealt{1996PhR...267..195J,2000RPPh...63..793B,
  2005PhR...405..279B}). In particular, the minimal 
SUSY extension of the standard model (MSSM) can yield DM particles, 
the most famous being the neutralino, a Majorana fermion. The 
cosmological constraints on the SUSY parameter space have 
been extensively studied in the literature \citep{2006PhRvD..73k5007B,
2006JHEP...33..0603,2005JHEP...10..020B,2003PhLB..565..176E}:
WIMPs could be detected on the present running or proposed experiments, 
either directly by measuring the recoil energy when they scatter off a 
detector nuclei \citep{2004IJMPA..19.3093M}, or indirectly by observing their 
annihilation products, such as anti-protons, positrons, $\gamma$-rays or 
neutrinos \citep{2005PhR...405..279B,2006RPPh...69.2475C}. 
They may also be generated in the next generation colliders, which is the 
most direct way to probe the existence of new particles. The direct and 
indirect detection methods are viable and complementary to collider 
studies in order to further constrain the nature of DM.

For indirect detection in the Milky Way, since the annihilation rate
is proportional to the square of the DM density, 
the Galactic Centre is believed to be a promising source of DM 
annihilation \citep{1998APh.....9..137B}. However, the existence of the 
central super-massive black hole and the supernova remnant Sgr A$^*$ are 
likely to heavily contaminate the DM signals with high-energy 
standard astrophysical processes \citep{2006PhRvL..97v1102A}. Alternative 
sites, such as the DM dominated dwarf spheroidal galaxies (dSph) 
orbiting close around the Milky Way, or even DM substructures inside 
the Milky Way, could be more favourable.
Indeed, the existence of a myriad of sub-halos throughout galactic-scale host 
halos is a generic prediction of the cold dark matter (CDM) paradigm of 
structure formation in the Universe. High resolution simulations 
\citep[e.g.][]{2006ApJ...649....1D,2007ApJ...657..262D,2007astro.ph..3337D} 
show that for the \lcdm~scenario, the large scale structures form 
hierarchically by continuous merging of smaller halos. As remnants of the 
merging process, about 10\% to 50\% of the total mass of the halo could be in 
the form of sub-halos. Moreover, the centres of sub-halos, like their hosts, 
are found to have high mass densities and therefore, could be ideal targets 
for $\gamma$-rays searches of WIMP annihilation products~\citep[e.g.][and 
references therein]{2004PhRvD..69d3501K,2006NuPhB.741...83B,Bi:2007}. A 
long-standing issue is the possible overall enhancement---\emph{boost 
  factor}---of the signals from the smooth component, due to the presence of 
such inhomogeneities \citep{1993ApJ...411..439S}. The first studies dedicated 
to indirect detection of DM focused essentially on $\gamma$-rays, and more 
marginally on anti-protons, but suffered from the lack of information on DM 
substructures (see e.g. \citealt{1999PhRvD..59d3506B}).
More recently, \citet{2003PhRvD..68j3003B} discussed in more details 
the $\gamma$-rays case, finding boost factors no larger than a few. 
Furthermore, a recent study by \citet{2005Natur.433..389D} reheated the debate 
on clumpiness, because the authors, by means of a very high resolution 
N-body experiment (but stopping at $z=26$), found that the Galaxy could be 
populated by a huge number density of sub-halos as light as the Earth. While 
the survival of such light clumps against tidal effects is still questionable, 
they could yield a significant contribution to the Galactic diffuse 
$\gamma$-ray flux by assuming a very cuspy sub-halo profile \citep{Bi:2006vk}. 
Nevertheless, some recent works also indicate that the current parameter range 
for clumpiness may provide only marginal global effects 
\citep{2007ApJ...657..262D,2007arXiv0706.2101P}.
The aim of the present paper is to provide a detailed study of the impact of 
cosmological sub-halos on the primary antimatter Galactic cosmic ray (GCR) 
flux, as elaborate as that already performed for $\gamma$-rays.

In \citet{2003A&A...404..949M}, the authors noted that the 
difference in propagation properties for \pbar\ and $e^+$ was likely to 
translate into different boost factors for these species. More recently, 
\citet{2007A&A...462..827L} provided a detailed formalism to
tackle the calculation of antimatter CR fluxes, when boosted by DM clumpiness. 
They showed how the uncertainty on the spatial distribution of clumps 
transfers to an uncertainty to the predicted boosted cosmic ray positron 
flux, an effect that depends on energy. More generally, this effect depends 
on the clump number density in a volume bounded by the characteristic 
diffusion length of the involved species. For the sake of clarity,
these authors have used a very simple model, in which all clumps have the 
same internal properties (masses and intrinsic luminosities), and mainly 
stressed the effects coming from their space distribution.
Using this method, \citet{2007PhRvD..76h3506B} fully treated a particular 
class of DM inhomogeneities---the intermediate mass black 
holes~\citep{2005PhRvD..72j3517B}---finding large boosts with huge
variances for the signals: such large variances tag unpredictive scenarios.
This means that in the case of a positive detection, such scenarios can
certainly be tuned to reproduce the data, but generally at the cost
of a vanishingly small associated likelihood for this configuration.

In this paper, we study a more natural DM scenario
\citep[e.g.][]{2005Natur.433..389D}, in which substructures fill
the whole Galaxy down to a minimal mass $M_{\rm min}\gtrsim 10^{-6} \Msol$,
with a mass distribution $dN/d\log M\propto M^{-\alpha_m}$ ($\alpha_m\approx 
0.9$), and a cored spatial distribution. We survey different DM 
configurations in great details by using different sub-halo inner profiles, 
different mass distributions or different concentration models (this has 
already been well studied in the context of gamma-rays, see e.g.
\citealt{2002PhRvD..66l3502U}). It is important to better quantify the boost 
and variance of antimatter signals since the satellite {\sc PAMELA} 
\citep{2006astro.ph..8697P,2007arXiv0708.1808C}, successfully launched in June 
2006, will soon provide new results on antimatter fluxes. The DM 
description suffers uncertainties, and its impact on the calculated fluxes 
adds up to the existing uncertainties from the propagation parameters 
\citep{2001ApJ...563..172D, 2004PhRvD..69f3501D,2005JCAP...09..010L}. 
Regarding this latter issue, {\sc PAMELA} should also update our current 
knowledge of the particles transport in the Galaxy, thanks to 
secondary-to-primary ratio measurements (e.g. B/C). This is crucial for 
the background calculation (standard antimatter production) in order to 
confirm/support any claim of an {\em excess}. Besides, AMS-02 should be 
launched in the coming years, and provide additional crucial information 
on GCR propagation by measuring the radioactive species 
\citep{2007NuPhS.166...19B}.

Below, we take advantage of simplified 
formulations for the \pbar\ \citep[e.g.][]{2006astro.ph..9522M} and $e^+$
\citep[e.g.][]{2007A&A...462..827L} propagators. Using the information
of the mass and space distributions of sub-halos from N-body numerical
simulations (see e.g. the recent Via Lactea simulation,
\citealt{2006ApJ...649....1D,2007ApJ...657..262D,2007astro.ph..3337D}),
we calculate the boost and the variance of the fluxes.
We find that for all plausible choices of the clump properties and
propagation parameters, boost factors for anti-protons and positrons
are close to unity, with small systematic and statistical uncertainties.

The paper is organised as follows. All relevant aspects (for this study)
of the DM distributions in the Galaxy, including
N-body simulation results are discussed in Sect.~\ref{sec:DM}.
The configurations retained are given in Sect.~\ref{subsec:dm_summary},
where the key parameters entering the calculation of
the {\em clumpy flux} (and its variance) are underlined. 
The propagation aspects are treated in Sect.~\ref{sec:propagation}.
The methodology to calculate the antimatter flux, its variance and the 
corresponding boost factors is given in Sect.~\ref{sec:method}, either by 
means of a semi-analytical approach (Sect.~\ref{subsec:methodJulien}) or by 
Monte Carlo (MC) simulations~(Sect.~\ref{subsec:methodMC}). The reader not 
interested in the technical details can directly jump to 
Sect.~\ref{sec:results}, where the results for positrons 
(Sect.~\ref{subsec:positrons}) and anti-protons (Sect.~\ref{subsec:pbar}) are 
presented, highlighting the physical effects coming from clump properties, 
space distribution, mass distribution and GCR propagation. Because of the 
complex origins and the mixing of the relevant physical quantities, such 
details really help to fully understand what kind of information \emph{boost 
  factors} actually encodes. We summarise and conclude in 
Sect.~\ref{sec:conclusions}.

%/////////////////////////////////////////////////////////////////
\section{DM distribution}
\label{sec:DM}

In the last few years, the advent of high resolution N-body simulations have 
increased the number of studies in this field, allowing for 
a better understanding and description of the DM dynamics. 
Even if many issues remain unclear, when comparing simulation results to 
the current observations, collisionless codes now agree at the 10\% level over 
wide dynamic ranges, providing a robust framework for DM studies 
\citep{2007arXiv0706.1270H}.

Throughout this paper, we will separate the WIMP annihilation contribution 
associated with \emph{sub-halos} from that associated with a \emph{smooth} 
component. The former will be related to \emph{any DM inhomogeneity} 
in the Galactic halo, independently of its physical scale | resolved or not 
in N-body simulations | while the latter will refer to \emph{the Galactic DM 
host halo itself}, which will be considered as a continuous fluid (again 
independently of the current resolution of N-body simulations). Although the 
Vlasov (or fluid) limit is likely to be reached when the number of 
particles involved in N-body experiments is huge, one should still be aware 
that such a statement is not trivial at all when dealing with the cosmological 
evolution of structures, and that discreteness might induce important biases 
\citep{2007PhRvD..75f3516J}. Furthermore, one should also keep in mind that 
our DM modelling will rest on (or be extrapolated from) 
N-body experiment results, in the most precise of which the test particle mass 
is not lighter than $\sim 10^4\Msol$ 
\citep[\cf~the \emph{Via Lactea} simulation,][]{2006ApJ...649....1D}, 
and for which the Vlasov limit is not reached at small scales. Nevertheless, 
we will assume throughout this study that the host halo profiles of 
Milky-Way-like galaxies provided by N-body simulations describe a smooth fluid 
(WIMP gas), on top of which some sub-halos may be wandering.

In the following subsections, we summarise the recent results concerning 
(i) generalities about DM distribution in halos of galaxies 
(Sect.~\ref{subsec:smooth_gen}) and (ii) some specific considerations about 
sub-halo description (Sect.~\ref{subsec:clump_cv} and 
\ref{subsec:subhalo_distrib}). 

Given the scope of this work, we will merely consider spherical profiles. For 
sub-halos, several cases will be chosen to encompass some extreme (but still 
plausible) scenarios. This aims at providing realistic estimates of the boost 
factor uncertainties related to the clumpy DM component.

%---------------%
\subsection{Shape and profiles}
\label{subsec:smooth_gen}

%---------------%
\subsubsection{Spherical profiles}

A scale-invariant DM distribution based on N-body numerical simulation
results can be written in a general form as \citep{1996MNRAS.278..488Z}
\begin{equation}
\rho=\frac{\rho_s}{(r/r_s)^{\gamma}[1+(r/r_s)^{\alpha}]^{(\beta-\gamma)
/\alpha}},
\label{eq:profile}
\end{equation}
where $\rho_s$ and $r_s$ are respectively a scale density and a scale radius, 
which can be determined by measuring the relation between the mass of the dark 
halo and the concentration parameter from simulations. Such an empiric law can 
be used for galaxy cluster halos, galaxy halos and for sub-halos. In 
the following, we focus on the central logarithmic slope $\alpha$ of the 
smooth halo component. We will discuss the Galactic scale radius $r_s$ and 
density $\rho_s$ in the section dealing with the concept of concentration 
(cf. Sect.~\ref{subsec:local}).

Navarro, Frenk and White \citep{1997ApJ...490..493N} worked out the following 
set of parameters $(\alpha,\beta,\gamma)=(1,3,1)$, which define the 
NFW profile, with a cusp scaling like $r^{-1}$ at radii smaller than $r_s$. 
\citet{1998ApJ...499L...5M} found another set with 
$(\alpha,\beta,\gamma)=(1.5,3,1.5)$ to fit their simulation results,
which is steeper than NFW at small radii, scaling like $r^{-1.5}$. More recent 
high resolution N-body simulations found that a NFW profile seems to 
underestimate the DM density in the central regions, while a Moore 
profile\footnote{Though those authors have since improved their early time 
  results and parametrisations, we will still use these generic 
  profile names to deal with inner shapes of profiles, for the sake of 
  simplicity.} probably overestimates it (\citealt{2004MNRAS.349.1039N,
  2004MNRAS.353..624D,2005MNRAS.364..665D} and references therein). The mean 
slope of the cusp obtained from various codes is well fitted by a 
$(1,3,\gamma)$ profile, with $\gamma= 1.16\pm0.14$ \citep{2004MNRAS.353..624D},
still in agreement with ($\gamma= 1.3$) analytical similarity solutions 
\citep{2005MNRAS.363.1092A}. However, profiles may not have a universal shape 
(e.g. \citealt{2004MNRAS.349.1039N,2006MNRAS.365..147S}). First, 
from the observational point of view, the relative scatter observed for the 
slope for four nearby low-mass spiral galaxies is 
0.44~\citep{2005ApJ...621..757S}, three times larger than in simulations. 
Second, it was also recently stressed that asymptotic slopes may not be
reached at all \citep{2004MNRAS.349.1039N,2006MNRAS.365..147S,
  2006AJ....132.2685M,2006AJ....132.2701G,2007ApJ...663L..53R}: according to 
\citet{2006AJ....132.2701G}, the Einasto function describes a simulated DM 
halo better than a NFW-like model.

Closer to the Galactic centre, the super-massive black hole dominates the 
mass ($r<r_{\rm BH}\approx 2$~pc$)$. The adiabatic growth of the black hole, 
if taking place in the centre of the DM gravitational potential and 
without any merger, could lead to an enhanced DM density in this 
region (slope as steep as $\sim 2.3-2.4$, dubbed {\em spike}). Nonetheless, 
recent works seem to prefer a final $r^{-1.5}$ behaviour for the DM 
density in the inner regions (see \citealt{2004PhRvL..92t1304M,
  2004PhRvL..93f1302G} and references therein).

Finally, the luminosity of cuspy or spiky halos is singular at the centre of 
the halo. However, a cut-off radius $r_{\rm{cut}}$ naturally appears, within 
which the DM density saturates due to the balance between the annihilation 
rate  $[\langle \sigma v\rangle \rho(r_{\rm cut})/m_\chi]^{-1}$ and the 
gravitational infalling rate of DM particles $(G\bar{\rho})^{-1/2}$
\citep{1992PhLB..294..221B}. Taking $\bar{\rho}$ about 200 times
the critical density, we get 
\begin{equation}
\rho_{\rm sat}= 3.10^{18}
     \left(\frac{m_\chi}{100~\rm GeV}\right) \times
     \left( \frac{10^{-26} {\rm cm}^3~{\rm s}^{-1}}
	  {\langle \sigma v\rangle}\right)
   \Msol~{\rm kpc}^{-3}.
  \label{eq:rho_sat} 
\end{equation}

%---------------%
\subsubsection{Other open questions}

During their history, structures undergo several mergers. The survival of the 
inner cusp of DM in these events has been investigated. The inner 
profile was found to be exceptionally robust, despite the relaxation that 
follows merging processes \citep{2004MNRAS.349.1117B,2006RMxAA..42...41A,
  2006ApJ...641..647K,2007MNRAS.376.1261M,2007ApJ...658..731V}. The 
implications are deep: the characteristic \emph{universal} shape of the DM 
density profile may be set early in the evolution of halos 
\citep{2006ApJ...641..647K}. However, it is still not clear whether the 
central cusp is steepened or flattened when the baryonic distribution is
taken into account. Using N-body hydrodynamical simulations,
\citet{2004ApJ...616...16G}, \citet{2006PhRvD..74l3522G} and 
\citet{2006MNRAS.366.1529M} find that the effect of gas cooling steepens the 
inner density profile to $1.9\pm0.2$, 
while \citet{2006Natur.442..539M} claim that the random bulk motion of gas in 
small primordial galaxies (driven by supernovae explosions) removes the cusp, 
leaving only cored profiles for both small and large galaxies in the present 
Universe.

Several other controversial issues remain and we only briefly quote
them. The first one is the question of the halo evolution in the presence of a 
rotating stellar bar, leading to either a destruction of the cusp
(see \citealt{2006astro.ph.10468S}, \citealt{2005MNRAS.363.1205M}, and 
references therein) or a steepening of the cusp \citep{2004ApJ...616...16G,
  2006ApJ...644..687C}. Some recent simulations including a stellar bar also 
hint at the emergence of a bar-like structure for the DM (DM bar) in the 
central region in the case of a strong stellar bar 
\citep{2006ApJ...644..687C,2007MNRAS.tmp..339A}: this is the second issue,
namely departure from sphericity. Direct observations either favour 
prolatness \citep{2004MNRAS.351..643H,2004ApJ...610L..97H,2005MNRAS.363..146L}
or oblateness  \citep{2005ApJ...619..800J,2007MNRAS.374.1125M,
2007A&A...461..155R}, whereas for pure collisionless simulations, prolatness 
is generally preferred (see also 
\citealt{2007MNRAS.376..215B,2007ApJ...657...56R,
  2007arXiv0705.2037K})\footnote{For further developments on the topic of 
  triaxiality, asymmetries, as well as on the spin of halos, see e.g.
\citet{2004MNRAS.354..522M,2005ApJ...629..219Z,2006ApJ...637..561L,
  2006MNRAS.373...65G,2006astro.ph.11205C,2006PhRvD..74l3522G,
  2007MNRAS.376..215B}. For the dependence of halo parameters on
the environment, see \cite{2007ApJ...654...53M,2007arXiv0704.2595H,
  2007MNRAS.375..489H,2007MNRAS.377.1785R}.}. Prolatness for sub-halos is 
likely to depend on the position in the galaxy, halos being 
more spherical in the outer regions \citep{2007MNRAS.377...50H}. 
Then, more generally, there is some evidence that halos become more spherical 
when the baryonic cooling is taken into account \citep{2004ApJ...611L..73K,
  2006ApJ...646L...9N,2006ApJ...648..807B,2007arXiv0704.3078M,
  2007arXiv0707.0737D}, or when a stellar bar is taken 
into account \citep{2006ApJ...637..582B}, or even during mergers 
\citep{2006ApJ...646L...9N}.

%---------------%
\subsubsection{Simplifying assumptions}
\label{sec:simpl}
It was shown that the choice of one or another DM profile for the 
smooth component \citep{2003A&A...404..949M,2004PhRvD..69f3501D} is not crucial
for the calculated flux of anti-protons and positrons. Indeed, charged 
particles diffuse on magnetic inhomogeneities and fluxes are heavily suppressed
(escape from the Galaxy) when originating far away from us, i.e. those from 
the Galactic centre. For cuspy profiles, the maximal difference is obtained 
between cored isothermal and Moore profiles (a factor $\lesssim 2$, see e.g. 
Fig.~2 of \citealt{2005PhRvD..72f3507B}), the difference between 
isothermal and NFW profiles being even smaller ($\lesssim 20\%$, see 
Table~II in \citealt{2004PhRvD..69f3501D}). Due to the lack of a definitive 
answer for the DM profile in the Galaxy (see the above-discussion), we will 
restrict ourselves in this paper to a spherically symmetric NFW profile 
$(\alpha,\beta,\gamma)=(1,3,1)$ for the galactic smooth distributions. Using 
triaxial halos or different profiles (e.g. $\gamma=1.2$ or any other profile) 
is expected to leave the main conclusions of the paper concerning the effects 
of clump granularity in the Galactic halo unchanged. Sphericity is also 
assumed for the substructures\footnote{Small structures, which formed earlier, 
are expected to be more spherical \citep{2004MNRAS.354..522M,
2006MNRAS.367.1781A}.}. Departure from spherical symmetry is left to a 
forthcoming study.

Finally, we stress that although the existence of a DM spike in the Galactic 
centre is crucial in the context of $\gamma$-ray/neutrino indirect detection
\citep{2005MPLA...20.1021B}, its effect is merely not relevant in this 
study. This is due to the depletion of the signal through the diffusive 
transport of antiparticles, and also to the fact that GCRs originating from 
annihilations in the very tiny extent of this region are only a small 
fraction of the total yield that can reach the Earth (the DM annihilation 
contribution to the GCR flux is integrated over a diffusion volume instead of 
a line of sight for $\gamma$-rays).

%---------------%
\subsection{Concentration parameter and sub-halo description}
\label{subsec:clump_cv}
The concentration parameter is a crucial quantity for computing the 
annihilation rates in (sub)structures. In this section, we actually 
present all relevant parameters that define a sub-halo. We will come back 
to the concentration (and the scale radius) associated with the host smooth 
halo just at the end (Sect.~\ref{subsec:local}).

In the \lcdm\ cosmology, the structures form hierarchically bottom-up via 
gravitational amplification of initial density fluctuations. The properties of 
the emerging structures and their subsequent evolution
may be described by using the virial quantities. Following the approach
and definitions of \citet{2001MNRAS.321..559B}, the two parameters $\rho_s$ 
and $r_s$, defined in Eq.~(\ref{eq:profile}), of a structure of mass 
\Mvir~are expressed in terms of the concentration \cvir~and 
the virial radius \Rvir. This outer radius is defined as the radius within 
which the mean density is $\Dvir(z)$ times the matter density $\Omega_m\rho_c$ 
at redshift $z$. At $z=0$,
\begin{equation}
  \Rvir = \left(\frac{\Mvir}
	{(4\pi/3)\Dvir(0) \Omega_m\rho_c}\right)^{1/3}\;.
	\label{eq:Rvir_def}
\end{equation}
In the following, we use the standard \lcdm\ values ($\Omega_m=0.24$,
$\rho_c=148 \Msol$~kpc$^{-3}$) and $\Dvir(0)\approx 340$ 
\citep{2001MNRAS.321..559B,2006A&A...455...21C}.

The concentration parameter is defined as \citep{2001MNRAS.321..559B}
\begin{equation}
  \cvir \equiv \frac{\Rvir }{r_{-2}}\;,
\end{equation}
where ${r_{-2}}$ is the radius at which 
%$d\ln\rho/d\rho=-2$.
$d/dr \left( r^2 \rho(r) \right)|_{r=r_{-2}} = 0$.
It was found \citep{1997ApJ...490..493N,2001MNRAS.321..559B,
  2001ApJ...554..114E} that $c_{\rm vir}$ strongly correlates with 
$M_{\rm vir}$, with larger concentrations observed in the first structures, 
i.e. in the lighter halos, which have formed in a denser Universe. This 
relation allows to express $\rho_s$ and 
$r_s$ in terms of the sole quantity $M_{\rm vir}$. Note that the 
$c_{\rm vir}-M_{\rm vir}$ is generally given for a NFW profile, for which 
$r^{\rm nfw}_s=r_{-2}$, so that:
\begin{equation}
  r^{\rm nfw}_s(\Mvir)=  \frac{\Rvir (\Mvir)}{\cvir(\Mvir)}\;.
  \label{eq:rs_mvir}
\end{equation}
This is easily transposed to other profiles. For example, for a Moore
profile, $r^{\rm moore}_s=4^{1/3}\/r_{-2}$, and the corresponding scale 
radius for the sub-halo is obtained from rescaling the NFW one, i.e. 
$r^{\rm moore}_s=r^{\rm nfw}_s/0.63$.

The last relation links $\rho_s$ to $\Mvir$. Rewriting
the profile $\rho(r)=\rho_s\times f(r)$, we get 
\begin{equation}
  \rho_s = \frac{\Mvir}{4\pi \int_0^{\rm \Rvir } r^2 f(r) dr}\;.
  \label{eq:rhos_mvir}
\end{equation}
When sub-halos are embedded in a larger host halo, the virial radius does not 
describe the physical radius anymore, and the integration should be performed 
up to the actual gravitational boundary of the object. For the isothermal 
case, for instance, the bound radius $r_b$ would be defined such as 
$\rho(r_b) = 2\rho_{\rm host}(r)$, where $r$ is the sub-halo location. 
Nevertheless, such a change is negligible for small clumps, apart from the 
very central regions of the host halo, of which all details are erased by GCR 
propagation. We will therefore neglect this further, as we have checked that 
it does not affect our results. For a NFW profile, the integration leads to 
(see e.g. Fig.~\ref{fig:clump_parameters1})
\beq
\rho^{\rm nfw}_s =  \Mvir/[4\pi r_s^3 A(\cvir)],
\label{eq:rhos_mvir_nfw}
\eeq
where $A(\cvir)\equiv \ln (1+\cvir)-\cvir/(1+\cvir)$.

Hence, in these models, once the $\cvir-\Mvir$ relation is specified, the 
profile of a clump is fully determined by its virial mass $\Mvir$. The 
behaviour of $r^{\rm nfw}_s(\Mvir)$, $\rho^{\rm nfw}_s$, as well as $\cvir$ 
and other related quantities, are illustrated in 
Fig.~\ref{fig:clump_parameters1}.

%---------------%
\subsubsection{$\cvir-\Mvir$ relation: B01 and ENS01 models}
\label{subsubsec:cvir_mvir}
We will use the two toy models B01 \citep{2001MNRAS.321..559B} and ENS01 
\citep{2001ApJ...554..114E}, which are based on N-body simulations; we refer 
the reader to these two papers for a detailed description.

These models predict that the halo concentration decreases with the halo
mass (see also \citealt{1997ApJ...490..493N}). Note that this behaviour has 
been observationally confirmed recently at the cluster scale
\citep{2007ApJ...664..123B,2007MNRAS.379..190C}, albeit with a slightly
higher $\cvir-\Mvir$ normalisation than predicted~\citep{2007MNRAS.379..190C}.
In subsequent N-body simulations  \citep{2002ApJ...568...52W,
  2005MNRAS.357..387K,2006ApJ...652...71W,2007MNRAS.378...55M}, a good 
agreement was found with the B01 model, adjusted to a slightly lower 
normalisation $\sim 15-20\%$\footnote{The concentration is modified when 
  considering dark energy with various values of $\omega$ 
  \citep{2004A&A...416..853D}, the constant equation of state, but still 
  remains consistent with B01 model \citep{2006ApJ...652...71W}.}. On the 
other hand, the ENS01 model is excluded in \citet{2007MNRAS.378...55M} 
(because of a too shallow slope), but preferred from the analysis of 
analytic lens models in~\citet{2007arXiv0705.3169F}. The state-of-the-art 
results for halo concentrations come from the recent {\em Millennium 
  Simulation}~\citep{2007arXiv0706.2919N}. The analysis at $z=0$ shows a clear 
disagreement with B01 for high halo masses $\Mvir \gtrsim 10^{13}\Msol$
with a better match with ENS01. Still, no conclusion can be drawn to favour
one model or another at the low mass end~\citep{2007arXiv0706.2919N}.

In any case, both toy models are likely to be not realistic enough. For
example, \citet{2007ApJ...657...56R} showed that, to some extent, the 
evolution of $\cvir$ could forget the initial conditions depending on the 
degree of violence in its merger events. The dependence of dark halo 
clustering on the concentration parameter also affects the relation 
\citep{2004MNRAS.350.1385S,
  2006ApJ...652...71W,2007MNRAS.377L...5G,2007ApJ...657..664J,
  2007astro.ph..3337D}, but this is sub-dominant compared to the observed 
dispersion of $\cvir$ (\citealt{2007MNRAS.377.1785R,2007MNRAS.378...55M} and 
references therein).

However, as our goal is to bracket the uncertainties due to the clumpy
contribution, we will stick to the simple descriptions of B01 and ENS01,
which give respectively an {\em upper limit} and a {\em lower limit}
on the concentration for the lower masses (see also Fig.~1 in 
\citealt{2006A&A...455...21C})\footnote{Their figure corresponds to
slightly modified B01 and ENS01, which are not retained here.}.
For our purpose, it is sufficient (and convenient) to use a fitted polynomial 
form at $z=0$, to encompass the two extreme cases (see 
Fig.~\ref{fig:clump_parameters1} for an illustration of the B01 relation):
\beq
  \ln (\cvir ) =  \sum_{i=0}^{4} C_i \times \large[ 
    \ln\left(\frac{\Mcl}{\Msol}\right) \large]^{i}
\label{eq:cvir_fit}
\eeq
with
\ben
C_{i}^{\rm B01} & = & \{ 4.34,\, -0.0384,\, -3.91\times 10^{-4},\, 
- 2.2\times 10^{-6},\nonumber\\
& & - 5.5\times 10^{-7}\}
\label{eq:cvirB01}
\een
and
\beq
C_{i}^{\rm ENS01} = \{3.14,\;-0.018,\; -4.06\times 10^{-4},\;0,\;0\}\,.
\label{eq:cvirENS01} 
\eeq
%Note that, because of the definition of \Rvir\ which tracks the critical 
%energy density, the concentration parameter evolves linearly with redshift, 
%so that any given concentration at redshift $z_1$ can be extrapolated at 
%redshift $z_2 < z_1$ with the relation $\cvir(\Mcl,z_2) = 
%(z_1-z_2 +1)\times\cvir(\Mcl,z_1)$, when neglecting merging 
%and tidal effects. For instance, \citet{2005Natur.433..389D}, found 
%concentrations of $\sim 3$ in a simulation in which they resolved clumps 
%down to $10^{-6}\Msol$ and down to $z=26$ ; this provides a reference
%value of $\cvir(z=0) \sim 78$, which is clearly included in the concentration 
%range that we consider here (for B01, 
%$\cvir^{\rm B01}(10^{-6}\Msol) = 119$|see 
%Tab.~\ref{tab:clump_parameters}|while for ENS01 
%$\cvir^{\rm ENS01}(10^{-6}\Msol) = 27$).

%---------------%
\subsubsection{Scale radius and local DM density in the Milky Way}
\label{subsec:local}
The DM smooth halo of the Galaxy also follows the previous relation,
but a more precise description of the halo properties is in principle
possible from in-situ observations. However, the determination of the mass 
distributions in the Galaxy remains challenging, especially if no assumption 
is made on the DM profile.

In \citet{1998MNRAS.294..429D}, the authors fitted a
multi-parameter mass model to the available kinematic data for the Galaxy.
They found a wide variety of models surviving the fitting process,
showing that the mass distribution within the Milky Way is still ill-defined.
The case of the NFW profile, which performs as good as any other profile
(see Model 2c in their Table~4), was best fitted with a scale radius
$r_s=21.8$~kpc and $\rhosol\approx 0.27$~GeV~cm$^{-3}$.
A more recent analysis \citep{2005MNRAS.364..433B}
using a new sample of 240 halo objects (including globular clusters,
satellite galaxies and stars) found that the isothermal profile
was ruled out for a constant velocity anisotropy of DM.
This conclusion is disputed by \citet{2006MNRAS.369.1688D}.
Nevertheless, both analyses come to similar conclusions concerning
the best fit profiles. For example, for the NFW profile,
\citet{2006MNRAS.370.1055B} find $r_s=17.3$~kpc, $\rhosol\sim 0.4$ 
~GeV~cm$^{-3}$ (corresponding to $M_{\rm vir}=1.5 \times 10^{12}M_\odot$ for
$\Rvir =312$~kpc). These values are in agreement with those found for 
simulated halos with similar mass ranges (see e.g.
\citealt{2004MNRAS.349.1039N,2007ApJ...657..262D}):
in the {\em Via Lactea} run, $r_s\approx 25$~kpc, $\rhosol\sim 
0.2$~GeV~cm$^{-3}$ (corresponding to $M_{\rm vir}=1.77 \times 10^{12}M_\odot$ 
for $\Rvir =389$~kpc). Finally note that a recent analysis 
\citep{2007NewA...12..507C} based on the dynamics of dwarf-spheroidals derives 
$\rhosol\approx [0.25-0.4]$~GeV~cm$^{-3}$.

We shall fix the parameters of the Milky Way smooth DM distribution.
In agreement with the previous values, we define our reference model
with $r_s=20$~kpc and normalise it to the local density
$\rhosol= 0.3$~GeV~cm$^{-3}$ at $\Rsol=8.0$~kpc. 
This allows the calculation of the mass within any radius $r$,
which gives the virial halo radius $\Rvirh $ when combined with 
Eq.~(\ref{eq:Rvir_def}): $\Rvirh =280$~kpc (so that $c_{\rm vir}=14$),
$M_{\rm vir}^{\rm h}=1.1 \times 10^{12}M_\odot$.
Varying $\rhosol$ only changes the overall normalisation of all fluxes,
whereas modifying $r_s$ would slightly change the spatial distribution,
which does not affect the conclusions of this paper.

%---------------%
\subsection{Number density of clumps $\ncl(\Mcl ,r)$}
\label{subsec:subhalo_distrib}

High resolution simulations have revealed that a large number of 
self-bound substructures survived in the Galactic halo 
\citep{1998MNRAS.299..728T,1999ApJ...522...82K,1999ApJ...524L..19M,
2001MNRAS.328..726S,2003ApJ...598...49Z,2004MNRAS.348..333D,
2004ApJ...609..482K,2006astro.ph..4393W,
2007MNRAS.378...55M,2007ApJ...657..262D}.
   
The mass and spatial distribution of sub-halos shown by these simulations
can be approximated as (e.g. \citealt{2004MNRAS.352..535D})
\begin{equation}
  \frac{d\Ncl(r,\Mcl)}{dVd\Mcl} \equiv  \frac{d\ncl(r,\Mcl)}{d\Mcl} =
  \Ncl \times \frac{d{\cal P}_M(\Mcl )}{d\Mcl }
  \times \frac{d{\cal P}_V(r)}{dV},
  \label{eq:distri}
\end{equation}
where the last two quantities are probability functions:
\beq
\int_{M_{\rm min}}^{M_{\rm max}} \frac{d{\cal P}_M(\Mcl )}{d\Mcl }  d\Mcl 
\equiv 1
\eeq
\beq
\int_{0}^{\Rvirh} \frac{d{\cal P}_V(r)}{dV}{dV} \equiv 1.
\eeq
The parameter $\Ncl$ is the total number of clumps within the virial
radius $\Rvirh$ of the Galaxy (see Sect.~\ref{subsbubsec:norm}). This means 
that the mass distribution of sub-halos does not depend of their locations in 
the host halo (tidal effects modify this picture, but only in the very central 
regions of the host halo).

%---------------%
\subsubsection{Mass distribution and cut-off}
\label{subsubsec:mass}

For the mass distribution, the following power-law dependence is observed:
\begin{equation}
\frac{d{\cal P}_M(\Mcl )}{d\Mcl }  = K_M \times
\left(\frac{\Mcl }{\Msol} \right)^{-\alpham}\;.
\label{eq:mass-distrib}
\end{equation}
The factor $K_M$ is such that the previous distribution is normalised 
to 1 for $\Mcl\in[M_{\rm min},M_{\rm max}]$:
\beq
K_M = \frac{1}{\Msol} \times \frac{(\alpham - 1)}
{\displaystyle \left(\frac{M_{\rm min}}{\Msol}\right)^{1-\alpham} -
  \left(\frac{M_{\rm max}}{\Msol}\right)^{1-\alpham}}\;.
\label{eq:norm_dPdM}
\eeq
In the limit $\Mmax \gg \Mmin$, we have 
$K_M \simeq (\alpham -1)\Mmin^{\alpham-1}$.
The logarithmic slope $\alpham\approx 2.0$ (e.g.
\citealt{1999ApJ...524L..19M,2004MNRAS.355..819G,2005MNRAS.359.1537R,
  2006ApJ...649....1D}), but the range of \alpham~values obtained in published 
studies spreads between $1.7$ and $2.1$ (\citealt{2007ApJ...659.1082S}, and 
references therein). For their Milky Way simulation, 
\citet{2007ApJ...657..262D} find $\sim 1.9$. Note that this is in agreement 
with the value $\alpham=1.91\pm 0.03$ found in \citet{2007ApJ...659.1082S}. 
However, when using an improved identification method of sub-halos on the same 
simulations, the latter authors conclude to a shallower dependence
$\alpham=1.79\pm0.04$.

The mass distribution covers a wide range, from the heaviest sub-halo mass
in the Galaxy, $M_{\rm max} \sim 10^{10}M_{\odot}$ 
(e.g. \citealt{1999ApJ...524L..19M,2005Natur.433..389D}),
down to a mass $M_{\rm min}$, of which the value is still debated.
At an early stage of structure formation, a cut-offs on the lower masses
appears due to (i) the diffusion of the DM particles
(collisional damping) out of a fluctuation and (ii) free streaming
(\citealt{2001PhRvD..64h3507H,2003PhRvD..68j3003B,2007JCAP...04...16B},
and references therein). The first process occurs after freeze-out of the 
DM particles, when it is still in kinetic equilibrium for some time
with the thermal bath (leptons, quarks, gauge bosons). Elastic and inelastic 
scattering on fast particles results in momentum exchange such that DM 
particles diffuse in space, leading to a cut-off mass $M_{\rm D}$ for the 
structures. After kinetic decoupling, the particles move freely
in the expanding Universe background and the temperature
of this decoupling sets the free streaming
cut-off $M_{\rm fs}$ of the mass spectrum. Both cut-off depend
on the DM candidate properties. For neutralinos,
\citet{2003PhRvD..68j3003B,2004MNRAS.353L..23G,2005JCAP...08..003G}
find $M_{\rm D}\approx 10^{-12}-10^{-10} M_\odot$ and
$M_{\rm fs}\approx 10^{-8}-10^{-6}M_\odot$.
This lower mass is slightly increased when taking into account
acoustic oscillations owing to the initial coupling between
the CDM and the radiation field 
\citep{2005PhRvD..71j3520L,2006PhRvD..74f3509B}.
A more careful analysis of the temperature of kinetic decoupling taking into 
account a more realistic range of variations of the particle-physics models
consistent with cosmological data was recently done in 
\citet{2006PhRvL..97c1301P}. Considering SUSY models (MSSM and 
mSUGRA) as well as models with universal extra dimensions (UED), these authors 
found the range $M_{\rm fs}\in [10^{-12}-10^{-4}]\; M_\odot$. 

To follow the history of these tiny substructures,
\citet{2005Natur.433..389D} performed a high resolution N-body simulation.
The authors were able, for the first time, to
resolve a Milky-Way size dark halo down to the free-streaming
stage. They report survival from the smallest structures 
(injected down to $M_{\rm min} \sim 10^{-6}M_{\odot}$, size $\sim 0.01$~pc)
at $z=26$.
However, tidal destruction of the lightest clumps and encounters with 
stars are still possible at late stages. In an analytical model, 
\citet{2006PhRvD..73f3504B} compared the strength of tidal stripping (i) 
during the hierarchical clustering, (ii) by stars from the stellar bulge, 
(iii) by stars from the halo and (iv) by the Galactic disk. They found that 
the last of these processes was the most effective, predicting that only 17\% 
of the Earth-mass clumps survived the tidal destruction. Note that the 
efficiency of tidal disruption depends on the mass of the clump but also on 
its environment (position in the Galaxy) so that, in principle, 
Eq.~(\ref{eq:distri}) cannot be used. Indeed, tidal stripping is more 
efficient towards the Galactic centre: for example, 
\citet{2006PhRvD..73f3504B} predict no {\em light} clumps at the radial 
distance $r\lesssim 3$~kpc. However, the fraction of surviving clumps is still 
controversial. Several recent studies have focused on the fate of these 
Earth-mass clumps. Although some of them conclude to near-complete destruction 
\citep{2007ApJ...654..697Z,2007MNRAS.375.1146A}, some others
underlined their resilience \citep{2003ApJ...584..541H,2007MNRAS.375.1111G,
  2007MNRAS.375..191G} in the Galactic potential. In the latter case, it is 
likely that the inner density slope of cuspy satellite halos remains unchanged,
even if the halo loses a lot of its mass~\citep{2004ApJ...608..663K}.

In any case, as we have already emphasised, the contribution of
the central regions of the Galaxy is suppressed by the diffusive
transport (for charged particles), therefore it is expected to
be unimportant. This assumption is reinforced by the fact that,
compared to the smooth distribution that is cuspy, the clump distribution
might be cored (see Sect.~\ref{subsbubsec:spat} below). We checked that
taking or not taking into account a significant destruction of low mass
clumps---as modelled and described, in e.g.
\citealt{2006NuPhB.741...83B}---left the results unchanged.
Thus, for our purpose, Eq.~(\ref{eq:dPdV})
is a good enough description of the clump distribution.
The mass distribution is then fully characterised by its slope
$\alpham$ and its minimal mass cut-off $M_{\rm min}$.

                    %######%
\subsubsection{Spatial distribution of clumps}
\label{subsbubsec:spat}
In most N-body experiments, the spatial distribution of clumps is found 
to be anti-biased with respect to the DM density, at least down 
to the smallest clumps resolved ($\sim 10^6\Msol$) at the moment
(\citealt{2000ApJ...544..616G,2004MNRAS.348..333D,2004MNRAS.351..399G,
  2004MNRAS.355..819G,2004MNRAS.352..535D}, and references therein). It is 
parametrised as (spherical symmetry is assumed)
\begin{equation}
\frac{d{\cal P}_V(r)}{4 \pi r^2 dr} = K_V \times \left[ 1+
  \left(\frac{r}{r_H}\right)^2\right]^{-1}
\label{eq:dPdV}
\end{equation}
where $r_H$, the core radius, is a fraction of the virial halo radius
$\Rvirh $. The constant $K_V$ is chosen here to ensure 
normalisation to unity when integrating over $\Rvirh $:
\[
 K_V \equiv 
  \left\{
    4\pi r_H^3 \times \left[ \frac{\Rvirh }{r_H}-\tan^{-1}\left(\frac{\Rvirh }
      {r_H}\right) \right]
    \right\}^{-1}\;.
\]
\citet{2004MNRAS.352..535D} found
$r_H \approx 0.14 \; \Rvirh $ for galactic-like sub-halos.
This bias could be due to the fact that, on average,
tidal mass loss experienced by sub-halos is larger in the inner regions
than near and beyond the virial
radius. This result seems to be largely unaffected
by the baryon dissipation (\citealt{2005ApJ...618..557N}, but see
\citealt{2006astro.ph..4393W} for a slightly different conclusion from
a SPH simulation).

However, some recent studies
argue that this cored distribution could be a selection bias 
\citep{2007arXiv0705.2037K} or a limitation of collisionless simulations 
\citep{2006MNRAS.366.1529M,2007ApJ...659.1082S}. For example, 
\citet{2007arXiv0705.2037K} find in their Via Lactea run a spatial 
distribution that matches the prolate shape of the host halo. The same trend 
is observed in \citet{2006MNRAS.366.1529M}, where the dissipation of the 
baryons greatly enhances the survival of the sub-halos. These authors
(see also \citealt{2005ApJ...618..557N})
find that the clumps profile is well fitted by a NFW, even if the latter is
still less concentrated ($\cvir\approx6.5$) than their simulated
overall mass distribution ($\cvir=9.6$).
Indeed, the smallest clumps are likely to follow the smooth DM spatial 
distribution, and such an assumption has very often been used in analytical 
studies of DM clumpiness effects on gamma-ray production (e.g. 
\citealt{2003PhRvD..68j3003B}). For the sake of completeness,
such a configuration will also be used later for the calculations, and
to be conservative, the space distribution of clumps will be taken to be 
exactly that of the smooth component (same global concentration relation).

%---------------%
\subsubsection{Clump number normalisation $\Ncl$}
\label{subsbubsec:norm}

The parameter $\Ncl$ is often determined by adopting the number of sub-halos
within a mass range. For example, \citet{1999ApJ...524L..19M}
found 500 sub-halos with bound masses $\gtrsim 10^8 \Msol $.
The recent {\em Via Lactea} simulation of \citet{2006ApJ...649....1D} gives
$\Ncl(> M_{\rm ref}) =6.4\times 10^{-3}\; (M_{\rm ref} /1.8 \times 10^{12} 
M_\odot)^{-1}$, which corresponds to $\Ncl(>10^8 \Msol )\approx 115$.

In a more general context of various masses of host halos, several
simulations (\citealt{2005MNRAS.359.1029V}, and references therein) are
compatible with the value $\Ncl(> M_{\rm ref}) = 0.017 \times (M_{\rm ref} /
M_{\rm host})^{-0.91}$. Taking a mass $M_{\rm host}=1.1 \times 10^{12}M_\odot$ 
for the Galaxy leads to $\Ncl(>10^8 \Mcl )\approx 81$.

For definitiveness, we choose to set the normalisation $\Ncl$
such as $\Ncl(\Mcl >10^8 \Msol \equiv M_{\rm ref} ) = 100 \equiv N_{\rm ref}$. 
Taking an upper bound of $M_{\rm max} = 10^{10}\Msol$, we get for 
$\alpham\neq 1$:
\ben
\label{eq:Ncl}
\Ncl &=& \frac{N_{\rm ref}}{K_M} \times \frac{(\alpham - 1)}
{\displaystyle M_{\rm ref}^{1-\alpham} - M_{\rm max}^{1-\alpham}}\\
& \simeq & (N_{\rm ref} M_{\rm ref}^{\alpham-1})\times\Mmin^{1-\alpham} 
\;({\rm if}\; \Mmin \ll \Mmax)\label{eq:Ncl_approx}
\een
where $K_M$ is the normalisation given in Eq.~(\ref{eq:norm_dPdM}). 
For instance, taking $\{M_{\rm min},M_{\rm ref},M_{\rm max}\} = 
\{10^{-6},10^{8},10^{10}\}\,\Msol$ and $\alpham = 1.9$,
we find $\Ncl \simeq 4\times 10^{14}$ clumps, consistent with values obtained 
by \citet{2005Natur.433..389D}.

%/////////////////////////////////////////////////////////////////
\section{DM modelling choices and salient features}
\label{subsec:dm_summary}

Having discussed in detail the values, uncertainties and
relevance of various parameters entering the DM distributions (both smooth
and clumpy), we now summarise the reference configurations 
used as inputs of this paper (Sect.~\ref{subsec:ref-conf}). Two main
consequences are observed: 
the index of the mass distribution strongly affects the mass
fraction of DM in clumps (Sect.~\ref{subsec:fracDMcl}), whereas 
the $c_{\rm vir}-M_{\rm vir}$ relation impacts on the luminosity profile 
(Sect.~\ref{subsec:lum-profile}).

%---------------%
\subsection{Reference configurations}
\label{subsec:ref-conf}

The distance of the Sun to the Galactic centre is fixed
to $\Rsol=8.0$~kpc.
Whatever the clump configuration, the virial radius of the dark halo
in the Galaxy is set to $\Rvir^{\rm h}=280$~kpc, and the local
DM density (smooth and clump altogether) to
$\rhosol=0.3$~GeV~cm$^{-3}$ (Sect.~\ref{subsec:local}).

%---------------%
\subsubsection{The smooth component}
\label{sec:ref_smooth}
It is chosen as a NFW (see discussion in Sect.~\ref{sec:simpl})
with an inner radius $r_s= 20$~kpc ($c_{\rm vir}=14$).
In the absence of any clump---we denote $\rho^0_{\rm sm}(r)$
the corresponding smooth distribution---and with the above values for 
$\rhosol$, $\Rvir^{\rm h}$ and $r_s$, we recover 
$\Mvirh\equiv M^0_{\rm sm}=1.1 \times 10^{12}M_\odot$.
The fraction $f$ is usually defined as the fraction
of DM taken from the smooth profile and redistributed into the clumps.
The smooth contribution in this configuration is then
$\rho_{\rm sm}(r)=(1-f)\rho^0_{\rm sm}(r)$, such that 
$M_{\rm sm}=(1-f) \Mvirh$.

%---------------%
\subsubsection{The sub-halo component}
\label{sec:cl_summary}
If the spatial density of clumps is $\propto\rho_{\rm sm}(r)$,
the redistribution of the fraction $f$ of the DM into clumps is
straightforwardly written as $\rhocl = f \rho^0_{\rm sm}(r)$.
Note that in this case, we have a local density of clumps
$f \rhosol$ and $M_{\rm cl}^{\rm tot} = f\Mvirh$
(such that $M_{\rm cl}^{\rm tot}+M_{\rm sm}=\Mvirh$).
We elaborate on the important case when the two distributions
are different in the next Sect.~\ref{sec:f_expanded}. Otherwise,
the clumps parameters are as follows:
\begin{enumerate}
  \item The inner profile of the clumps $\rho_{\rm cl}(r)$ is
     taken as a NFW or a Moore. The saturation density is
     taken from Eq.~(\ref{eq:rho_sat}), with $\rho_{\rm sat}\sim
     10^{19}M_{\odot}$~kpc$^{-3}$ for typical WIMP parameters\footnote{See 
       discussion in Sect.~\ref{subsec:lum-profile} for the consequences of 
       varying $\rho_{\rm sat}$ and considering different inner profiles
       (e.g. a Moore inner profile).}. The scale parameters $r_s$
     and $\rho_s$ depend solely on $c_{\rm vir}$ and $\Mvir$
     through Eqs.~(\ref{eq:rs_mvir}) and~(\ref{eq:rhos_mvir}).
     The concentration $c_{\rm vir}$ depends on the virial mass $\Mvir$,
     as provided by the B01 and ENS01 models (see 
     Sect.~\ref{subsubsec:cvir_mvir},
     Eqs.~\ref{eq:cvirB01}~and~\ref{eq:cvirENS01}).
   \item The clump numerical density $\ncl (\Mcl ,r)$
     is given by Eq.~(\ref{eq:mass-distrib}) with $r_H= 0.1\times
     \Rvir^{\rm h} = 28$~kpc. $\Ncl$ is set from the condition 
     $\Ncl(>10^8 \Msol)=100$, with a clump mass upper boundary of 
     $10^{10}\Msol$. 
     The logarithmic slope of the mass distribution is $\alpham \in [1.8-2.0]$,
     and we will survey minimal clump masses starting from
		 $M_{\rm min}=10^{-6}\Msol$.
     These last two parameters completely set the mass fraction $f_M$ of the
		 virial mass in clumps (see Sect.~\ref{subsec:fracDMcl} below),
     defined as $M_{\rm cl}^{\rm tot} = f_M \/ \Mvirh$ (note that
		 $f_M$ does not necessarily coincide with $f$, see below).
     \end{enumerate}

A synthetic view of the relevant parameters retained in this study are 
proposed in Tables~\ref{table:fixed_values} and \ref{table:models}.
Note that all varying parameters come from the clumpy distribution 
(spatial distribution, $M_{\rm Min}$, $\alpham$, inner profile and the 
$\cvir-\Mvir$ relation).
\begin{table*}
\centering
\begin{tabular}{c c c c}
\hline\hline
    {\bf Cosmology} &  {\bf DM (Milky Way)}  &  {\bf Clumps}  & 
    {\bf Smooth DM halo} \\
$\Omega_m=0.24$ &   $\Rvir=280$~kpc &  Global$^\dagger$: cored, 
    $\alpham\!\!=\!\!1.9$,$M_{\rm min}\!\!=\!\!10^{-6}\Msol$   &   
    NFW (1,3,1)\\
    $\rho_c=148 \;\Msol$~kpc$^{-3}$ & $\Mvir=1.1 \times 10^{12}M_\odot$ &  
    Inner profile: NFW, B01$^\ddagger$ &$r_s=20$~kpc ($\cvir=14$) \\
    $\Dvir(0) = 340$ &    $\rhosol(\Rsol)^\S=0.3$~GeV~cm$^{-3}$&  
    $\Ncl(>10^8 \Msol)=100$ & $\rho_{\rm sat}^{~~~\star}\sim 10^{19}
    \Msol~{\rm kpc}^{-3}$ \\\hline
\end{tabular}
\\$^\dagger$ {\scriptsize Number density of clumps as defined in 
  Eq.~(\ref{eq:distri}).}
\\$^\ddagger$ {\scriptsize $\cvir-\Mvir$ relation, see 
  Eq.~(\ref{eq:cvir_fit}).}
\\$^\S$ {\scriptsize The distance of the sun to the galactic centre is set to 
  $\Rsol=8.0~{\rm kpc}$.}
\\$^\star$ {\scriptsize Full expression for the saturation density is given by 
  Eq.~(\ref{eq:rho_sat}). It also applies to the clumps inner profile.}
\caption{Useful parameters and reference configuration of the DM modelling
(see text for further details).}.
\label{table:fixed_values}
\end{table*}
The configurations, for which we will calculate the boost factors, are listed 
in Table~\ref{table:models}. 
\begin{table}
\centering
\begin{tabular}{c c}
\hline\hline
    Clump description       &        Values                 \\ \hline
 $d{\cal P}_V(r)/dV$        &      Cored$^\ddagger$ or NFW   \\
 Inner profile              &    NFW$^\ddagger$  or Moore    \\
 $\alpham$                  &  $[1.8-1.9^\ddagger-2.0]$        \\
 $M_{\rm min}$              &  $[10^{-6}~^\ddagger-1-10^6]~\Msol$ \\
 $c_{\rm vir}-M_{\rm vir}$  &   B01$^\ddagger$ or ENS01      \\
\hline
\end{tabular}
\\$^\ddagger$~{\scriptsize Reference configuration.}
\caption{Description of the various configurations used in the paper
for the sub-halo parameters.}
\label{table:models}
\end{table}

%---------------%
\subsubsection{Defining the local fraction of DM $\fsol$ in clumps}
\label{sec:f_expanded}
If the smooth (e.g. NFW) and the sub-halos (e.g. cored) spatial distributions 
are different, the mass fraction of DM in sub-halos within $r<\Rvirh$ is not 
constant, but depends on the galactocentric radius $r$. The point is that in 
order to compute boost factors, one would naively want to subtract any 
fraction of DM added in the form of clumps to the smooth component, and 
compare this new setup to the case in which DM is only smooth. A 
clear definition of that fraction is crucial before going further. Indeed, we 
show hereafter that if not treated carefully, there is a source of ambiguity 
in the interpretation of the resulting boost factor.

Let us first introduce the total mass carried by the clumps within $\Rvirh$,
defined as
\beq
M_{\rm cl}^{\rm tot} \equiv \Ncl \int_{M_{\rm min}}^{M_{\rm max}} d\Mcl\, 
\Mcl \, 
\frac{d{\cal P}_M(\Mcl )}{d\Mcl } = \Ncl \langle \Mcl\rangle .
\eeq
The quantity $\langle \Mcl\rangle $ is the mean clump mass associated with the mass range 
$[M_{\rm min}-M_{\rm max}]$, the mass probability distribution
(Eq.~\ref{eq:mass-distrib}), and $\Ncl$ is the total number of clumps 
(Eq.~\ref{eq:Ncl}). Without loss of generality, the total density profile of 
DM may be expressed as
\beq
\rho_{\rm tot}(r) = (1 - f) \rho_{\rm sm}(r) + 
M_{\rm cl}^{\rm tot}\frac{d{\cal P}_V(r)}{dV},
\label{eq:rho_tot_old}
\eeq
where $f$ is a DM fraction subtracted to the smooth component, which 
we discuss later on. The quantity $M_{\rm cl}^{\rm tot}d{\cal P}_V(r)/dV$ is 
merely the averaged mass density profile of the whole sub-halo population, 
obtained from integrating Eq.~(\ref{eq:distri}) over the whole mass range of 
clumps.

Two observational constraints can help to define what kind of fraction $f$ 
is needed for consistency: the total mass of the Galaxy \Mvirh, and the local 
density \rhosol. If one wants to ensure that the total mass is left 
unchanged when adding clumps, then $f$ is the \emph{mass fraction} $f_M$ 
given by:
\begin{equation}
f = f_M \equiv \frac{\Mcltot}{\Mvirh}\;.
\label{eq:frac_M}
\end{equation}
Otherwise, if one prefers to have a constant local matter density, then $f$ 
is defined as a \emph{local density fraction} $\fsol$ as follows:
\begin{equation}
f = \fsol \equiv \frac{\Mcltot}{\rhosol}\times 
\frac{d{\cal P}_V(\Rsol)}{dV}.
\label{eq:fsol_def}
\end{equation}

First, if the spatial distribution of clumps tracks the smooth profile, 
then we have by definition $d{\cal P}_V(r)/dV = \rho_{\rm sm}(r)/\Mvirh$, 
and $f_M = \fsol$. Using either one of these fraction concepts is therefore 
equivalent: in other words, the halo mass and the mass density (at any $r$) 
are conserved when sub-halos are added.

Now, if the two distributions spatially differ, it is no longer possible
to fulfil both constraints. We have no choice but to abandon
either the halo mass to be constant, or the local density  to be constant. Let 
us see what happens when one of the two above conditions,  
Eq.~(\ref{eq:frac_M}) or Eq.~(\ref{eq:fsol_def}), is plugged in 
Eq.~(\ref{eq:rho_tot_old}).

\paragraph{First option | ensuring $\Mvirh$ is constant:}
this choice sets that the fraction to be used is the \emph{mass fraction} 
defined in Eq.~(\ref{eq:frac_M}). An immediate consequence is that the total 
local DM density is now slightly modified. Plugging this condition 
(Eq.~\ref{eq:frac_M}) in Eq.~(\ref{eq:rho_tot_old}) for $r=\Rsol$ leads to
\beq
\rho_{\rm tot}(\Rsol) =  \left[ 1 - f_M \left( 1
-\frac{\Mvirh}{\rhosol}\frac{d{\cal P}_V(\Rsol)}{dV}\right)\right] 
\times \rhosol.
\eeq
Taking our reference configuration, i.e. the NFW profile for the smooth 
component (Sect.~\ref{sec:ref_smooth}) and a cored profile for the clump
component (Sect.~\ref{sec:cl_summary}), gives 
\beq
\rho_{\rm tot}(\Rsol) =  \left[ 1 - f_M \left(  1
-5.5\times 10^{-2}\right)\right] \times\rhosol  \simeq (1 - f_M)\rhosol.
\eeq
Such a modification of the local DM density translates the fact that 
the clumpy contribution is locally almost negligible, in that case. This will 
occur every time the clump distribution is flatter than the parent one. As the 
local density may certainly vary within a factor of two 
(Sect.~\ref{subsec:local}), even putting up to 50\% of the DM mass in clumps 
is acceptable. Such a hypothesis is very often made in the literature dealing 
with $\gamma$-rays, but the previous consequence is almost never mentioned.

However, this choice is not judicious in our study. Doing so would 
even bring additional confusion to the issue of boost factors. Indeed, 
unlike $\gamma$-rays, we remind that for primary cosmic antimatter, the flux 
is very sensitive to the local density (see App.~\ref{app:volumes}). Assuming 
for a while that the smooth component locally dominates the clumpy one (it 
will actually be shown later to be the case, see e.g. 
Fig.~\ref{fig:luminosity_profiles}), the calculated mean boost factor would 
be $B_{\rm eff} \sim (1-f_M)^2$ (see Eqs.~\ref{eq:boost} and~\ref{eq:bsol}), 
which would result in a number significantly less than unity. This would 
consequently lead to a \emph{damping} factor instead of a an enhancement, 
which would bring about misleading interpretations.

\paragraph{Second option | ensuring $\rho(Rsol)$ to be constant:}
to avoid the above situation, we may use the concept of \emph{local density 
faction} instead, defined by Eq.~(\ref{eq:fsol_def}). It comes to demanding 
$\rho_{\rm tot}(\Rsol)=\rhosol$. 
The boost factor now asymptotically goes to $(1-\fsol)^2\sim 1$ if the 
clump contribution is negligible. However, this normalisation has to face 
again an unavoidable issue: the total halo mass will be modified by the 
adjunction of clumps to the smooth component. Plugging back $\fsol$ in 
Eq.~(\ref{eq:rho_tot_old}), integrating over the virial volume and using 
again $f_M$ (see Eq.~\ref{eq:frac_M}), we get
\begin{equation}
  M^{\rm tot} = \left( 1- \fsol + f_M\right) \times \Mvirh 
	\simeq (1 + f_M)	\times  \Mvirh.
  \label{eq:additionalMass}
\end{equation}
With the different clump configurations used throughout 
the paper, the total halo mass within the virial radius can be increased up 
to $f_M\lesssim 50\%$ level (see curves $\alpham\lesssim 2.0$ in
Fig.~\ref{fig:frac_clumps}). Such values remain within current estimates of 
the total mass of the dark halo, as recalled in  Sect.~\ref{subsec:local}.

\paragraph{Closing the case:} we conclude by reminding that neither of these 
choices is better than the other. Both are, somehow, equally artificial. 
Indeed, there is only, if so, one {\em true} distribution of smooth and clump 
DM in the Galaxy. The ambiguity appears because we wish to compare the 
calculated fluxes to a hypothetical configuration with no clumps. In the 
context of antimatter fluxes, as explained, the second option (ensuring the 
same local DM density whatever the configurations) makes more sense,
as it leads to boost factor values {\em asymptotically} reaching 1.
This second option is retained throughout the paper\footnote{Let us say it 
  again: should the first option have been retained, we would have ended up 
  with boost factors smaller than one!}.

Finally, before closing the DM section, let us discuss how the
various configurations gathered in Table~\ref{table:models} impact on 
some generic properties for the clumps (mass fraction $f_M$ and luminosities).

%---------------%
\subsection{Mass fraction $f_M$ in sub-halos}
\label{subsec:fracDMcl}

The minimal mass $M_{\rm min}$ of the clumps able to form---and to survive 
tidal disruption---is a crucial parameter (see also next subsection). Along 
with the slope $\alpham$ appearing in the mass distribution
Eq.~(\ref{eq:mass-distrib}), it sets the 
fraction of DM in clumps, $f_M\equiv  M_{\rm tot}^{\rm cl}/\Mvirh$.

\begin{figure}[t!]
\begin{center}
\includegraphics[width=\columnwidth, clip]{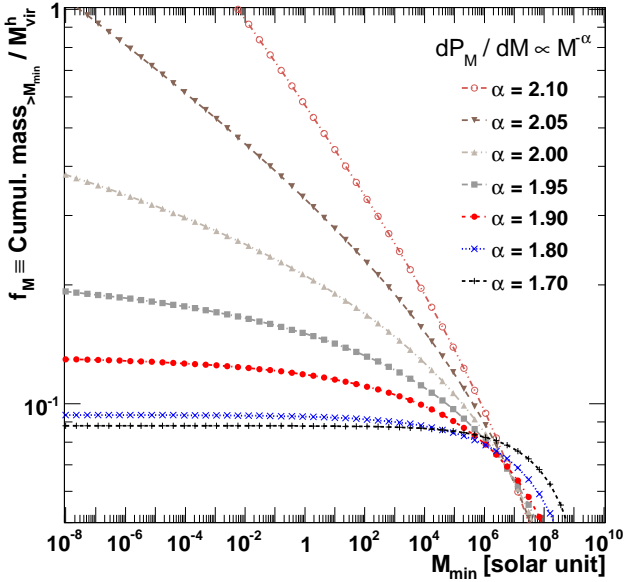}
\caption{\small Mass fraction $f_M$ as a function of \Mmin, for different 
logarithmic slopes \alpham~of the mass distribution (from $2.1$ down to 
$1.7$---top to bottom curves).}
\label{fig:frac_clumps}
\end{center}
\end{figure}

The evolution of  $f_M$ with $\alpham$ and 
$M_{\rm min}$ is shown in Fig.~\ref{fig:frac_clumps}.
The behaviours are in agreement with the figures discussed in several 
simulations. For example, taking a resolved mass $M_{\rm min}\gtrsim 10^{-6} 
\Msol$, \citealt{2007ApJ...659.1082S} (and references therein) found
$f\sim 5-10\%$ for $\alpham=1.8$. As argued in \citet{2007ApJ...657..262D}, 
where a larger value of $\alpham\sim 2.0$ is preferred, 10\% might be only a 
lower limit and this fraction could reach $f_M\sim 50\%$. In the extreme
case of a slope $\alpham=2.1$, all the DM could be distributed in clumps,
even forbidding the existence of clumps of mass smaller than $\lesssim 
10^{-2}\Msol$. This latter configuration is probably not realistic, so we 
choose to limit the study to the range $\alpham \in [1.8-2.0]$. Consequently,
as observed from Fig.~\ref{fig:frac_clumps}, the fraction of mass in clumps 
$M_{\rm tot}^{\rm cl}$, corresponding also to the additional mass added to 
$\Mvirh$ (see Eq.~\ref{eq:additionalMass}), will lie in the range $\sim 10 - 
40$ \%.

%---------------%
\subsection{Luminosity: a closer look on the astrophysical term}
\label{subsec:lum-profile}

Before plugging the propagation, it is interesting to have a look at the 
luminosity of the source terms in the various configurations. The total 
luminosity of DM sources can be separated into a particle physics 
term times an astrophysical term: 
\[
L_{\rm source}(E,r)\equiv (dN/dE)\times {\cal L}_{\rm astro}(r).
\]
The particle physics term is factored out by normalising 
to the local luminosity $L_\odot\equiv L_{\rm source} (\Rsol) \propto 
\rhosol^2$.
The relative astrophysical luminosity is then defined as
\[
{\cal L}(r) \equiv L_{\rm source}(r)/L_\odot={\cal L}_{\rm astro}(r)/
{\cal L}_{\rm astro}(\Rsol).
\]
For short, below, we will continue to call this quantity the luminosity,
and use $M$ instead of $\Mcl$ for the mass clump.

\paragraph{Smooth component:} it is straightforwardly written as
\beq
{\cal L}_{\rm sm}(r) \equiv \frac{\rho^2_{\rm sm}(r)}{\rhosol^2}\;.
\eeq

\paragraph{Substructures:}
it is convenient to separate the total clump contribution 
as the sum of the contributions of each decade of mass (throughout the paper,
the logarithm bins of mass are denoted $M_i\equiv [10^i-10^{i+1}]\Msol$
with $i$ an integer).  Taking into account the space and mass distributions of 
the clumps, the total and per logarithmic mass bins luminosities are defined 
as:
\begin{equation}
{\cal L}_{\rm cl}^{tot} \equiv \sum_i {\cal L}^i_{\rm cl},
\end{equation}
\begin{equation}
{\cal L}^i_{\rm cl}(r) \equiv 
\frac{d{\cal P}_V}{dV}(r) \times \int_{M_i}
\frac{d{\cal L}_{\rm cl}}{d\ln M}(M) \; d(\ln M)\;,
\label{eq:lum_bin}
\end{equation}
with
\begin{equation}
\frac{d{\cal L}_{\rm cl}}{d\ln M}(M) \equiv \Ncl \times M \times 
\frac{d{\cal P}_M}{dM}(M) 
 \times \xi(M) \;,
  \label{eq:lum_cl}
\end{equation}
\begin{equation}
\xi(M) \equiv \int_{V_{\rm cl}} \left[ \frac{\rho_{\rm cl}(r,M)}
  {\rhosol}\right]^2 \; d^3{\bf x}\;.
\label{eq:xi}
\end{equation}
The above Eq.~(\ref{eq:lum_cl}) defines the luminosity mass profile while 
Eq.~(\ref{eq:xi}) defines  an intrinsic effective annihilation 
volume for a clump of mass $M$.
Note that introducing the mean value $\langle \xi^i\rangle _M$ of $\xi$ over 
the $i^{\rm th}$ mass bin, allows the luminosity per logarithmic mass 
Eq.~(\ref{eq:lum_bin}) to be recast under a form where the dimensions appear 
more explicitly:
\beq
{\cal L}^i_{\rm cl}(r) = \Ncl \times \langle \xi^i\rangle _M \times
\frac{d{\cal P}_V}{dV}(r)\;,
\eeq
\beq
\langle \xi^i\rangle _M \equiv \int_{10^i \Msol}^{10^{i+1}\Msol} dM  \xi(M) 
\frac{d{\cal P}_M(M)}{dM}\;.
\label{eq:def_xibar_first}
\eeq

Last, it is useful to introduce a dimensionless intrinsic boost factor 
$B_{\rm cl}(M)$ for a clump (not to be mistaken with the global boost factor 
  defined in Eq.~\ref{eq:boost}). The latter compares the annihilation rate 
of the clump, to the rate that would be obtained for a clump that had the same 
volume, but with a constant DM density $\rhosol$. The {\em local} intrinsic 
boost factor can be expressed as
\beq
B_{\rm cl}(M )  \equiv \xi (M) \times \frac{\rhosol}{M}.
\label{eq:xi_and_boost}
\eeq
We emphasise that such a quantity is meaningful since i) antimatter fluxes
mostly depend on the local DM density and ii) unlike $\gamma$-rays, we do not
look in one specific direction, but rather integrate on the whole clump signal.

Before concluding on resulting luminosities ${\cal L}_{\rm sm}(r)$
and ${\cal L}^i_{\rm cl}$, let us further detail
the various terms appearing in the clump luminosity term.

%---------------%
\subsubsection{Annihilation volume $\xi(M)$}
\label{subsubsec:xi_M}
This quantity is a function of the mass clump $M$, and it depends
on the inner profile (NFW or Moore), the $\cvir-\Mvir$ relation (B01 or ENS01)
and the saturation density $\rho_{\rm sat}$.

\paragraph{Reference configuration (NFW inner profile):} for an NFW profile,
the annihilation volume has a simple analytical expression 
\begin{equation}
\xi^{\rm nfw}(M)= \frac{4\pi}{3} (r_s^{\rm nfw})^3 
\left(\frac{\rho_s^{\rm nfw}}{\rhosol}\right)^2 
\times \left[ \eta^{\rm nfw}(r_{\rm sat}^{\rm nfw}) - \eta^{\rm nfw}(\Rvir ) 
  \right]\;,
\label{eq:xi_nfw}
\end{equation}
where the scale radius $r_s^{\rm nfw}$ and the density at scale radius
$\rho_s^{\rm nfw}$ depend on the clump mass $M$, as given in 
Eqs.~(\ref{eq:rs_mvir}) and  (\ref{eq:rhos_mvir_nfw}). The function 
$\eta^{\rm nfw}(r)$ is defined as
\begin{equation}
 \eta^{\rm nfw}(r)\equiv \left[ 1+\frac{r}{r_s^{\rm nfw}(M)} \right]^{-3},
\end{equation}
where the saturation density for a NFW is given by
\begin{equation}
r^{\rm nfw}_{\rm sat}(M)=r_s^{\rm nfw} (M)\times 
\frac{\rho_s^{\rm nfw}(M)}{\rho_{\rm sat}}.
\end{equation}
It is easily checked that $\xi^{\rm nfw}(M)$ is largely insensitive to
the exact value of $\rho_{\rm sat}$ ($\sim 10^{19}M_\odot$~kpc$^{-3}$). 

The local intrinsic boost factor $B_{\rm cl}^{\rm nfw}(M)$ can also be 
analytically expressed in terms of the virial parameters:
\ben
B_{\rm cl}^{\rm nfw}(M ) &=& \frac{M}{12\pi\rhosol\Rvir^3} \times \nonumber\\
&&
 \frac{\cvir^4 \left(3+\cvir(3+\cvir)\right)}
{(1+\cvir)\lbrack \cvir - (1+\cvir) \ln (1+\cvir)\rbrack^2}\;.
\label{eq:b_boost}
\een
As $\cvir$ is only very slightly mass dependent and 
$\Rvir\propto M ^{1/3}$, the intrinsic boost factor, $B_{\rm cl}$,
is almost constant over a wide range of sub-halo masses. More precisely,
for the NFW case in the B01 model, we find it to scale with the
concentration parameter like
\beq
B_{\rm cl}^{\rm nfw,B01}(M) \simeq 1.29\times 10^{-4} \; 
\left(\cvir^{\rm B01}\right)^{5/2}.
\eeq

\begin{figure}[t!]
\begin{center}
\includegraphics[width=\columnwidth,clip]{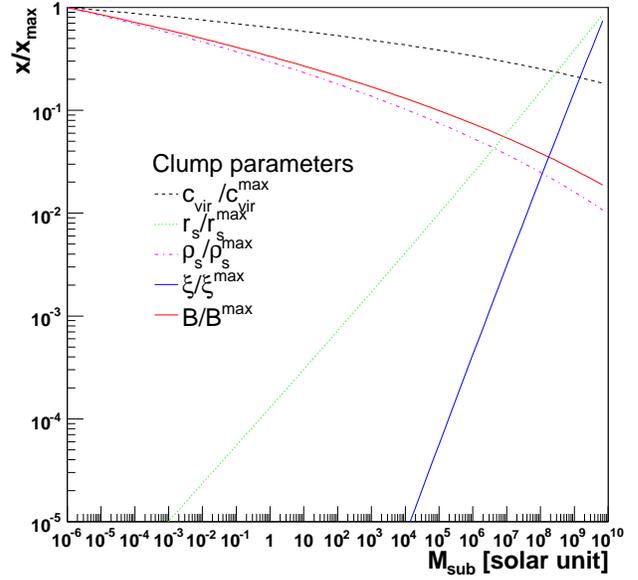}
\caption{\small Same as in Table~\ref{tab:clump_parameters}. Each parameter is
normalised with respect to its maximum value (in the mass range displayed).}
\label{fig:clump_parameters1}
\end{center}
\end{figure}

For illustration purpose, typical values for all above quantities (for various 
masses of clumps) are gathered in Table~\ref{tab:clump_parameters}.
\begin{table}
\centering
\begin{tabular}{c | c c c c | c c}
\hline\hline
$\Mcl$  & $\Rvir $ & $r_s^{\rm nfw}$ & $\rho_s^{\rm nfw}$ & 
$\hspace{-0.1cm}c_{\rm vir}^{\rm B01}\hspace{-0.1cm}$ &
$\xi^{\rm nfw,B01}$ & 
$\hspace{-0.25cm}B_{\rm cl}^{\rm nfw,B01}\hspace{-0.25cm}$ \\
{\scriptsize ($M_\odot$)}  & {\scriptsize(kpc)} & {\scriptsize(kpc)} 
& {\scriptsize\hspace{-0.25cm}($\Msol$~kpc$^{-3}$)\hspace{-0.25cm}} &  & 
{\scriptsize(kpc$^3$)} &  \vspace{0.3mm}\\
 \hline
 $10^{-6}$ & $\hspace{-0.15cm}2.7 \!\cdot\! 10^{-4}\hspace{-0.15cm}$ & 
 $2.3 \!\cdot\!  10^{-6}$ & $1.8 \!\cdot\!  10^{9}$ & 
 $\hspace{-0.1cm}119\hspace{-0.1cm}$ & $\hspace{-0.15cm}2.5 \!\cdot\!  
 10^{-12}\hspace{-0.15cm}$ & 20 \\ 
 $10^{-3}$ & $\hspace{-0.15cm}2.7 \!\cdot\! 10^{-3}\hspace{-0.15cm}$ & 
 $2.8 \!\cdot\!  10^{-5}$ & $1 \!\cdot\!  10^{9}$ & 
 $98$ & $\hspace{-0.15cm}1.6 \!\cdot\!  10^{-9}\hspace{-0.15cm}$ & 12 \\
 $1$     & $\hspace{-0.15cm}2.7 \!\cdot\! 10^{-2}\hspace{-0.15cm}$ & 
 $3.5 \!\cdot\!  10^{-4}$ & $5.4 \!\cdot\!  10^{9}$ & 
 $77$ & $\hspace{-0.15cm}8.6 \!\cdot\!  10^{-7}\hspace{-0.15cm}$ & 6.8 \\
 $10^3$  & $\hspace{-0.15cm}0.27\hspace{-0.15cm}$              & 
 $4.7 \!\cdot\!  10^{-3}$ & $2.5 \!\cdot\!  10^{8}$ & 
 $58$ & $\hspace{-0.15cm}4.3 \!\cdot\!  10^{-4}\hspace{-0.15cm}$ & 3.4 \\
 $10^6$  & $\hspace{-0.15cm}2.7\hspace{-0.15cm}$                & 
 $6.6 \!\cdot\!  10^{-2}$ & $9.9 \!\cdot\!  10^{7}$ & 
 $41$ & $0.19$ & 1.5 \\
 $10^9$  & $\hspace{-0.15cm}27\hspace{-0.15cm}$               & 
 $1$               & $3 \!\cdot\!  10^{7}$ & 
 $26$ & $69$ & 0.5 \\
\hline
\end{tabular}
\caption{\small Sub-halo parameters (reference configuration, i.e. inner 
  NFW and B01 for $\cvir-\Mvir$) for different masses: virial radius  $\Rvir$ 
  (Eq.~\ref{eq:Rvir_def}), scale radius $r^{\rm nfw}_s$ 
  (Eq.~\ref{eq:rs_mvir}), scale density $\rho_s^{\rm nfw}$ 
  (Eq.~\ref{eq:rhos_mvir_nfw}), concentration parameter $\cvir^{B01}$ 
  (Eq.~\ref{eq:cvirB01}), effective volume $\xi^{\rm nfw,B01}$ 
  (Eq.~\ref{eq:xi_nfw}), intrinsic local boost $B_{\rm cl}^{\rm nfw,B01}$ 
  (Eq.~\ref{eq:xi_and_boost}). See text for details.}
\label{tab:clump_parameters}
\end{table}
After applying a suitable renormalisation, the same quantities---some of
which having power law dependencies with $M$---are also displayed in
Fig.~\ref{fig:clump_parameters1}. 
We check that $\xi^{\rm nfw}$ roughly scales like $M$,
a very common feature already emphasised in the literature
that focuses on gamma-rays (see the consequences in the next subsection).
For the intrinsic local boost factor, we read off the last column
in Table~\ref{tab:clump_parameters} that only clumps below
$M\lesssim 10^6\Msol$ may significantly overcome the local annihilation
signal. The first conclusion that can be drawn is that massive clumps,
unless close to the solar neighbourhood (which is very unlikely), will
not be able to {\em boost} the antimatter signals. More importantly, 
the very same parameter allows an upper limit to the total boost 
expected to be set, $B_{\rm cl}^{\rm nfw,B01}<20$.

\begin{figure}[t!]
\begin{center}
\includegraphics[width=\columnwidth]{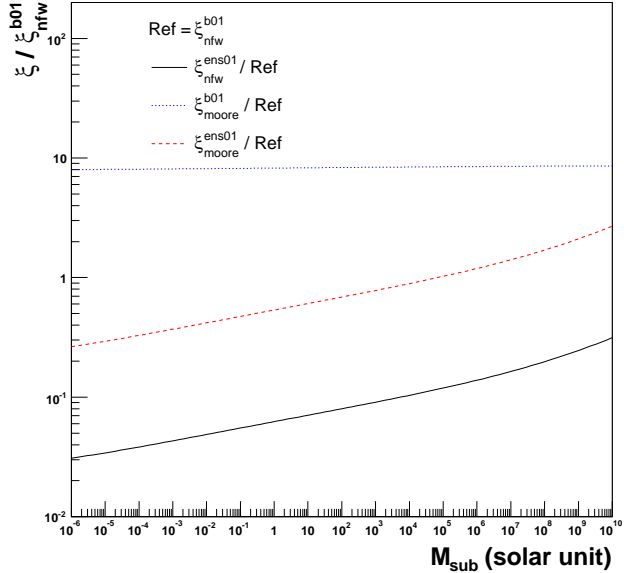}
\caption{\small Ratio of the annihilation volume for three benchmark sub-halo 
  models to the reference one, defined as 
  $\xi_r^X\equiv \xi^{\rm x}/\xi^{\rm ref}=
  B_{\rm}^{\rm x}/B_{\rm cl}^{\rm ref}$. The reference configuration is an 
  inner NFW profile with B01 concentration. The three benchmark configurations 
  are X~$=({\rm Moore,B01})$ (upper curve), X~$=({\rm Moore,ENS01})$ (middle 
  curve) and X~$=({\rm NFW,ENS01})$ (lower curve).
}
\label{fig:clump_parameters2}
\end{center}
\end{figure}

\paragraph{Moore \emph{vs} NFW inner profile | ENS01 \emph{vs} B01:}
these conclusions are very easily extended to other configurations.
We actually find a very simple rescaling factor linking the
annihilation volume $\xi^{\rm moore}(M)$ to the above $\xi^{\rm nfw}(M)$,
when $\rho_{\rm sat}^{\rm ref}=10^{19}~\Msol~{\rm kpc}^{-3}$:
\begin{equation}
\xi^{\rm moore}(M,\rho_{\rm sat})\simeq 8 \times \xi^{\rm nfw}(M)\;.
\label{eq:rescale_xi_moore}
\end{equation}
This does not depend on the $\cvir-\Mvir$ relation, and to a good extent,
neither on the WIMP mass and annihilation cross section, nor on the clump mass,
as illustrated in Fig.~\ref{fig:clump_parameters2} 
Note that, as expected, the ENS01 configuration is lower than B01, implying,
for the intrinsic boost factor, the hierarchy $B_{\rm cl}^{\rm nfw,ENS01} < 
B_{\rm cl}^{\rm nfw,B01} \simeq B_{\rm cl}^{\rm moore,ENS01} < 
B_{\rm cl}^{\rm moore,B01}$ (roughly corresponding to 0.1:1:10). Hence the 
former configuration will provide the minimal boost, the reference 
configuration the median boost and the last configuration the maximal boost.

%---------------%
\subsubsection{$d{\cal L}_{\rm cl}/d\ln M$}

\begin{figure}[t!]
\begin{center}
\includegraphics[width=\columnwidth]{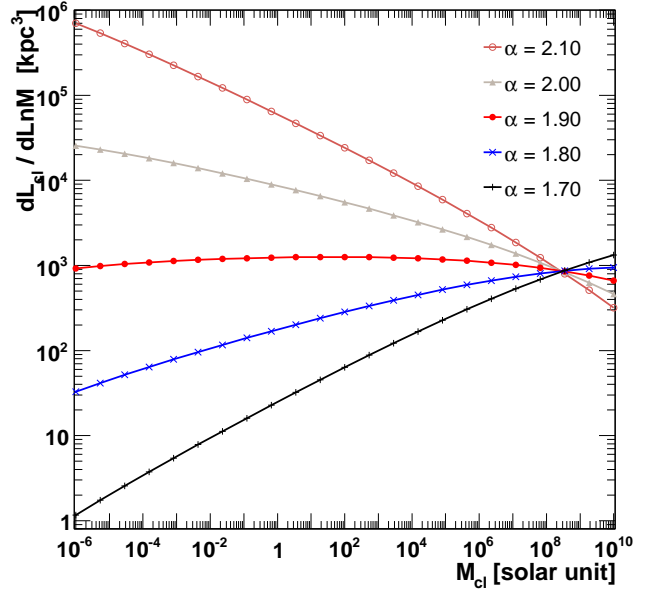}
\caption{\small Differential luminosity (true units in kpc$^3$) of the 
  population of clumps $d{\cal L}_{\rm cl}/d\ln M$ as defined in 
  Eq.~(\ref{eq:lum_cl}). The curves correspond to the NFW-B01 reference
  configuration, for various values of $\alpham$.
  The curves for any other configuration may be obtained
  by multiplying these curves to the ratios shown in 
  Fig.~\ref{fig:clump_parameters2}.}
\label{fig:dLdLnM}
\end{center}
\end{figure}

The differential luminosity Eq.~(\ref{eq:lum_cl}) is evaluated taking into 
account the mass distribution of clumps Eq.~(\ref{eq:mass-distrib}). The result
is shown in Fig.~\ref{fig:dLdLnM} for different values of the slope $\alpham$.
We recover the various trends seen in the literature (see, e.g. Fig.~8 of 
\citealt{2007ApJ...657..262D}, \citealt{2003PhRvD..68j3003B}). In particular, 
the value $\alpham=1.9$, favoured in simulations, shows a roughly constant 
luminosity per decade. For smaller (respectively greater) value of $\alpham$, 
the luminosity will be dominated by the heaviest (lightest) clumps. 
In that case, based upon the understanding gained from the previous 
discussions, the boost is expected to be small (close to unity, from the 
intrinsic boost factor).
At the same time, the variance of the clumpy signal is expected to be large 
(light clumps add no contribution and heavy clumps are scarce).
For larger $\alpham$, the mass $M_{\rm min}$ of the  lightest clump is crucial,
because the latter drives the total luminosity. These large $\alpham$ 
configurations are expected to give the largest boost factors. The last step 
is to put together the smooth and clumpy luminosities.

%

%---------------%
\subsubsection{Luminosity profiles ${\cal L}_{\rm sm}(r)$ and 
  ${\cal L}^i_{\rm cl}(r)$ in the Galaxy}
\begin{figure*}[t]
\begin{center}
\includegraphics[width=\columnwidth, clip]{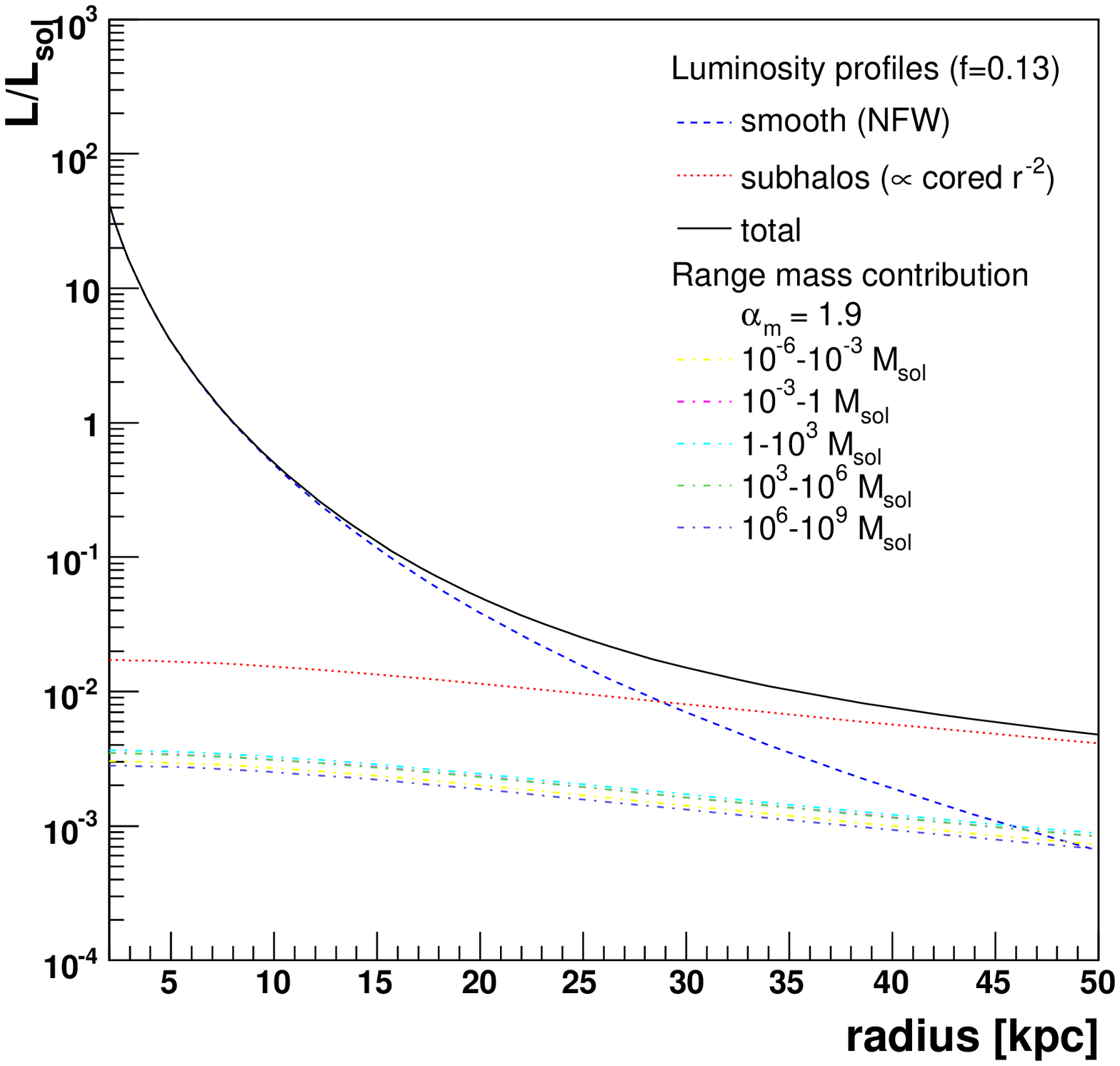}
\includegraphics[width=\columnwidth, clip]{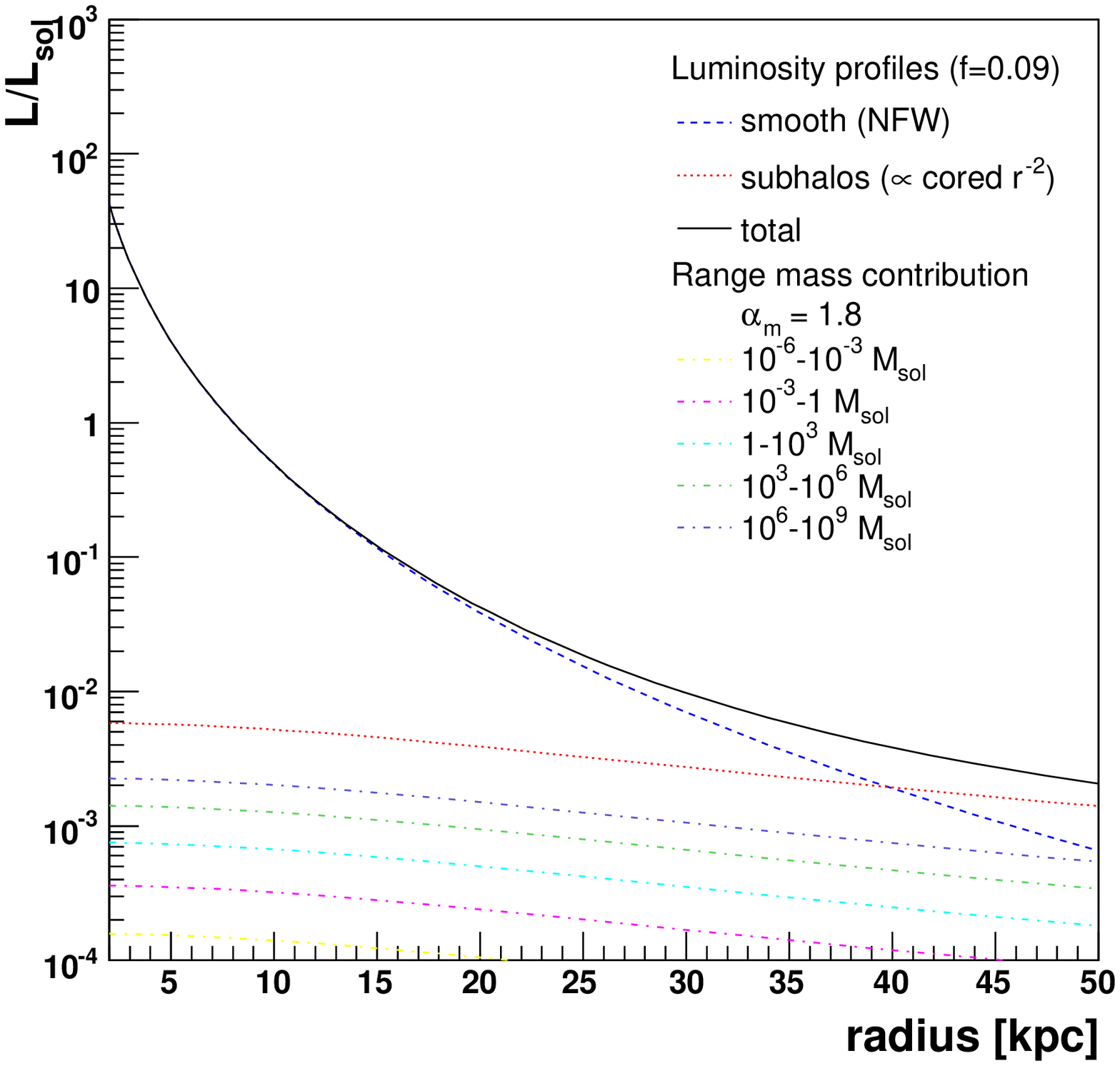}
\includegraphics[width=\columnwidth, clip]{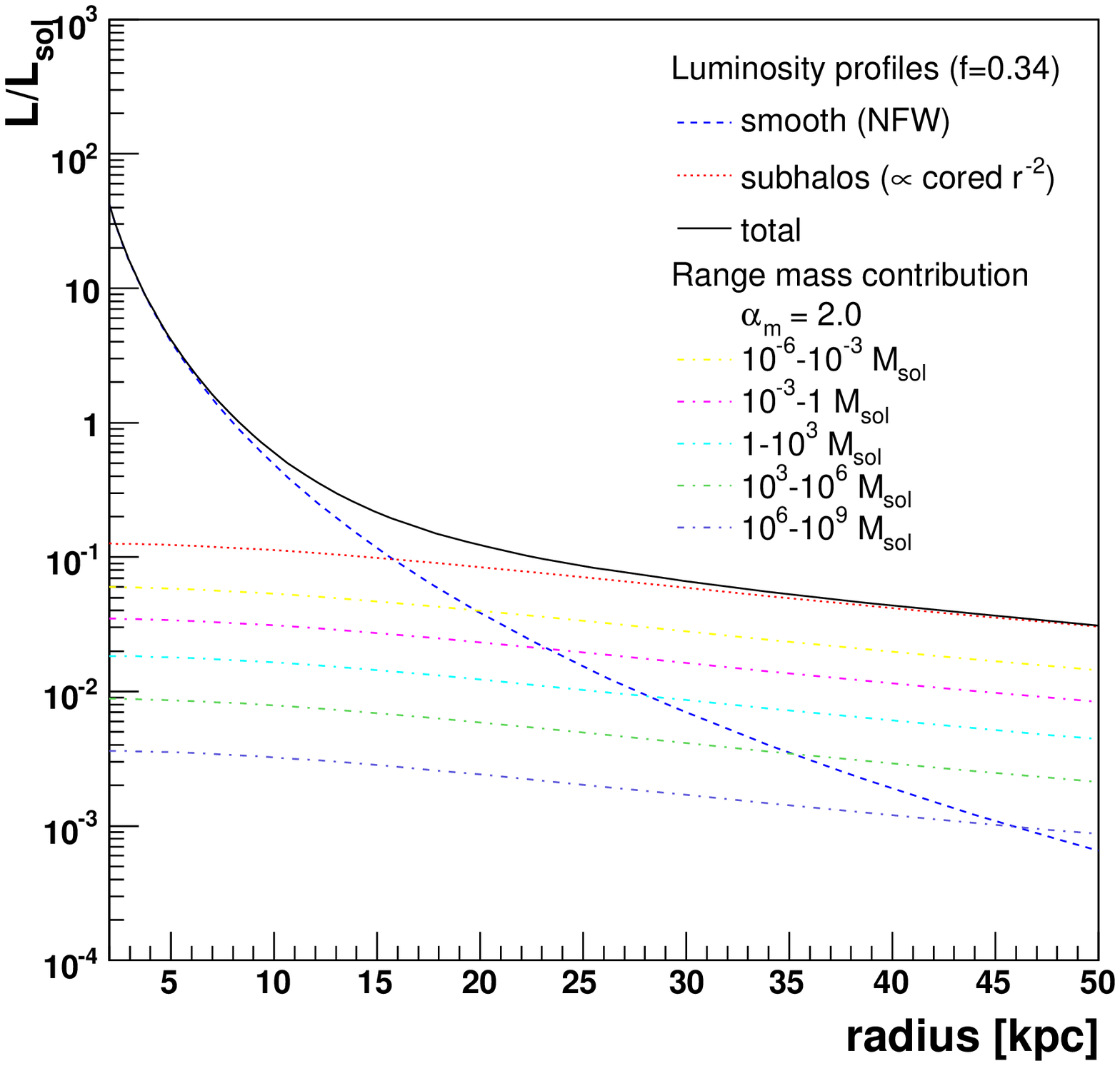}
\includegraphics[width=\columnwidth, clip]{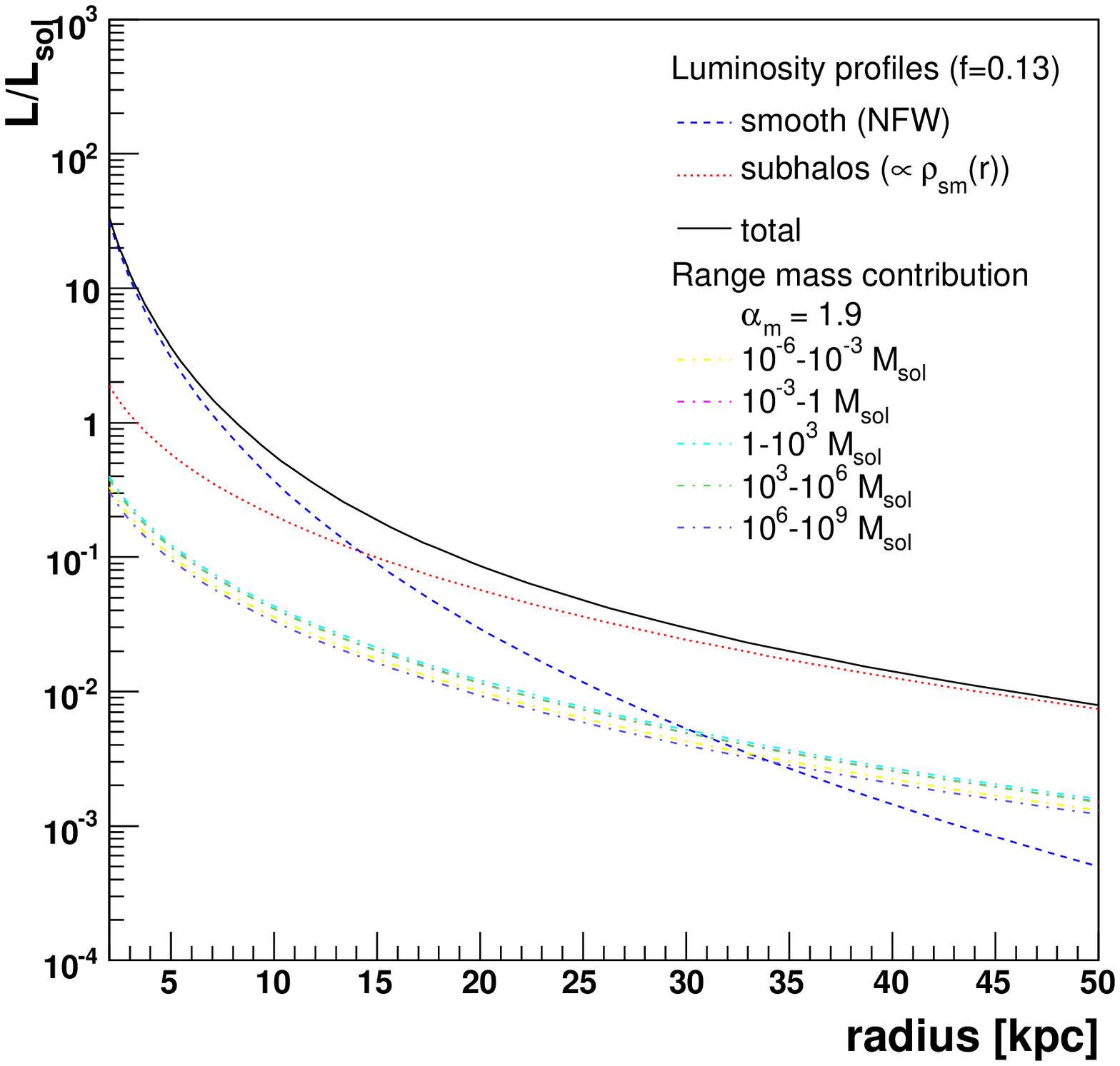}
\caption{\small Relative luminosity profiles as functions of the 
  galactocentric radius $r$, in units of local luminosity 
  $L_{\odot}$. Luminosities are plotted for the smooth DM contribution
	and for clumps in logarithmic mass bins of 3-decade width.
	{\em Top left}: reference configuration (ref).
	{\em Top right}: as ref but $\alpham=1.8$.
	{\em Bottom left}: as ref but $\alpham=2.0$. 
	{\em Bottom right}: as ref but for a spatial distribution of clumps 
	$\propto \rho_{\rm sm}(r)$.}
\label{fig:luminosity_profiles}
\end{center}
\end{figure*}

The last hint at small boost factors for the case of antimatter DM
is given when comparing the smooth and clump luminosities. 
This is first shown for the reference configuration
in Fig.~\ref{fig:luminosity_profiles} (top left panel).
As already emphasised (see also \citealt{2007ApJ...657..262D}), for the 
reference configuration $\alpham=1.9$, the contributions to the average 
annihilation fluxes of any decade mass range will be almost the same on the 
whole range of mass clumps (see also Fig.~1 of \citealt{Yuan:2006ju}). 
Around $r=\Rsol$, the luminosity is completely dominated
by the smooth contribution ($\sim100$ times more than the total clump
luminosity), so that for this configuration, we may predict beforehand (i) 
no boost factor and (ii) a small variance on this boost factor.

The logarithmic slope of the mass distribution $\alpham$ reverses  
hierarchy in the mass contribution: for $\alpham=1.8$ (see
Fig.~\ref{fig:luminosity_profiles}, top right),
the more massive the population of clump the more luminous it is, whereas 
for $\alpham=2$ (see Fig.~\ref{fig:luminosity_profiles}, bottom left), the 
less massive, the more luminous. The trade off is reached close to 
$\alpham = 1.9$. However, the total clump luminosity never reaches the level 
of the smooth one! In the best case ($\alpham=2$), it is 10 times smaller. 
Nevertheless, taking larger \alpham values naturally leads to larger boost, in 
a more general context.

Now, if we now assume that, instead of having a cored profile, the spatial
distribution follows the parent one (NFW), we see in
Fig.~\ref{fig:luminosity_profiles} (bottom right)
that the situation is more favourable for the boost factors.
Keep in mind that this is an upper limit since the the averaged radial mass 
density profile of clumps is believed to be a flatter distribution than the 
smooth one (see Sect.~\ref{subsbubsec:spat}).

Anticipating the results of using ENS01 instead of B01, or using a 
Moore inner profile instead of a NFW, especially from 
Fig.~\ref{fig:clump_parameters2}, we already know that ENS01 will only 
further decrease the total clump luminosity (roughly by a factor of 10 
compared to B01), whereas Moore will increase the total clump luminosity 
(roughly by a factor of 10 compared to NFW).

To summarise, from the general study of the luminosity, we might already 
conclude that no configuration of DM will lead to huge boost factors. 
Pushing all the parameters for the maximum effect, i.e. Moore inner profile, 
NFW spatial distribution, B01 and $\alpham=2.0$, would possibly lead to a 
boost factor of a few, but certainly not a hundred. We can expect all other 
configurations to end up with a boost factor close to unity. The rest of the 
paper is devoted to the full calculations to confirm these expectations.

%/////////////////////////////////////////////////////////////////
\section{Propagation model}
\label{sec:propagation}

In the Galaxy, a charged particle travelling from
its source to the solar neighbourhood is affected by several
processes. The scattering off random magnetic fields leads
to spatial and energy diffusion (reacceleration) and 
particles may also be spatially convected away by the galactic
wind (which induces adiabatic losses).

In this paper, the framework used is the following
(e.g. \citealt{1990acr..book.....B}): for the transport processes
we take a spatial independent diffusion coefficient 
$K(E)=\beta K_0{\cal R}^{\delta}$
(where ${\cal R}=pc/Ze$ is the rigidity) and a constant wind $V_c$ directed
outwards along $z$. Cosmic rays are confined within a diffusive halo $L$,
such as the differential density, $dN/dE\equiv N$, is bound by $N(z=L,r)=0$.
The free parameters of the model are the halo size
$L$ of the Galaxy, the normalisation of the diffusion coefficient $K_0$
and its slope $\delta$, and the constant galactic wind $V_c$ 
(see Sect.~\ref{subsec:propag_param}).
Other processes (such as continuous and catastrophic gain/losses)
are more species-dependent. Hence,
although all charged particles are propagated in the same framework,
due to this dependence, the phenomenology of propagation is completely
different for \pbar\ and positrons.

The reader is referred to \citet{2001ApJ...555..585M} for
a more detailed presentation and motivation of the framework.
Note that this model has been repeatedly and consistently used in several
studies to constrain the propagation parameters
\citep{2001ApJ...555..585M,2002A&A...394.1039M,2002A&A...381..539D}
and examine the consequences \citep{2003A&A...402..971T,2003A&A...404..949M}
for the standard \pbar\ flux \citep{2001ApJ...563..172D},
the exotic \pbar\ and \dbar\ fluxes \citep{2002astro.ph.12111M,
  2006astro.ph..9522M,2004PhRvD..69f3501D,
  2002A&A...388..676B,2005PhRvD..72f3507B,2007PhRvD..75h3006B},
but also for positrons~\citep{2007A&A...462..827L,2007PhRvD..76h3506B}.

%---------------%
\subsection{Propagator and flux for anti-protons}

It was shown in \citet{2006astro.ph..9522M} that neglecting all energy 
redistribution terms (energy losses, reacceleration and tertiary source term)
provides a correct description at sufficiently high energy, while remaining 
good enough down to $\sim$~GeV IS energies (better than 50\% depending on the 
propagation parameters considered). This approximation is retained here.
The only catastrophic losses for anti-protons are spallations---the particle 
does not survive the interaction.

Denoting $\Gamma_{\rm tot}= \sum_{\rm ISM} 
n_{\rm ISM}.v.\sigma^{\bar{p}}_{\rm ISM}$ the destruction rate of \pbar\ in 
the thin gaseous disk ($n_{\rm ISM}=$H, He), the transport equation for a 
point source, defining the propagator, reads \citep{2006astro.ph..9522M}:
\ben
  \left\{ -K\triangle +V_c \frac{\partial }{\partial z}
  +2h \Gamma_{\rm tot} \delta(z)\right\} {\cal G}^{\;\bar{p}} = 
  \delta({\bf r}-{\bf r'}).
\een

For simplicity, we only consider a flux detected at solar position 
${\bf r}_{\odot}=(x_\odot=\Rsol, y_\odot=0, z_\odot=0)$. For a point source 
$S$ at ${\bf r}_S$, the corresponding flux only depends, in cylindrical 
coordinates, on the relative distances $r=|{\bf r}_{\odot}-{\bf r}_S|$ and 
$z=z_S$. The propagator ${\cal G}^{\;\bar{p}}_\odot(r ,z)$ is given by
\ben
\displaystyle {\cal G}^{\;\bar{p}}_\odot(r ,z) &=& 
\frac{\exp^{-k_v z}}{2\pi K L} \times \\\nonumber
\displaystyle &&\sum_{n=0}^\infty c_n^{-1}K_0(r  \sqrt{k_n^2+k_v^2}) 
\sin [k_n L] \, \sin [k_n (L-z)]
\label{ppropa}
\een
where $K_0$ is the modified Bessel function of the second kind. The quantity 
$k_n$ is the solution of
\[ 
2k_n \cos k_n L = -k_d \sin k_n L \;,
\]
and
\begin{equation}
 c_n = 1 - \frac{\sin k_n L \cos k_n L}{k_n L}.
\end{equation}
We also have
\[
 k_v \equiv V_c/(2K) \quad {\rm and }\quad k_d \equiv 2h \;
 \Gamma_{\rm tot}/K + 2 k_v\;.
\]

For a source term $q_\odot(r,z,\theta)Q(E)$ (origin is taken to coincide 
with solar location) the equilibrium spectrum at solar position is finally 
given by 
\beq
\Phi_\odot^{\;\bar{p}}(E)\! =\! \frac{v Q(E)}{4\pi}
\!\times 2\!\int_{0}^L \!\!\!\!\! dz\!\!\int_0^\infty \!\!\!\!\!\! 
rdr \;{\cal G}^{\;\bar{p}}_\odot(r ,z)
 \!\int_0^{2\pi} \!\!\!\!\!\! d\theta \, q_\odot(r,z,\theta)\,.
\eeq

%---------------%
\subsection{Positrons}
Contrarily to nuclear species, there are no catastrophic losses for positrons.
A more crucial point is that propagation of positrons is dominated by energy
losses (e.g. \citealt{1998ApJ...493..694M}). In that case, a monochromatic 
line at the source leads to a spectrum once propagated. This is at variance to 
\pbar\ whose propagator for exotic sources is at constant energy.

The diffusion equation that characterises the evolution of the positron number 
density $N$ per unit energy, with a source term $q({\bf r})Q(E)$, reads
\beq
\label{eq:untilded}
 -K_0 \left(\frac{E}{E_0}\right)^\delta \triangle N  + 
 \frac{\partial}{\partial E} \left\{ \frac{dE}{dt}\;  N \right\} = 
 q({\bf r})Q(E).
\eeq
The first term is simply the diffusion coefficient written as 
$K({\cal R})\approx K_0 (E/E_0)^\delta$. For simplicity, we have also 
neglected the effect of the Galactic wind (see next section for a discussion).

We proceed as in~\citet[][and see references therein]{2007A&A...462..827L}.
The synchrotron and inverse Compton losses can be written as
$dE/dt(E) = - E^{2}/(E_0 \tau_{E})$,  with $E_0=1$~GeV and 
$\tau_{E} \approx 10^{16}\rm{~s}$.
Defining a pseudo-time
\[
\hat{t} \equiv \tau_E \; \frac{{(E/E_0)}^{\delta-1}}{1-\delta}
\]
and applying the following rescaling,
\[
	\hat{N} \equiv {(E/E_0)}^{2} N \quad \text{and}\quad
	\hat{Q}(E) \equiv {(E/E_0)}^{2-\delta} Q(E),
\]
the diffusion equation can be rewritten as
\begin{equation}
\label{eq:tilded}
\frac{\partial}{\partial \hat{t}} \, \hat{N} -K_0\triangle \hat{N} = 
q({\bf r})\hat{Q}(E).
\end{equation}
Thus, instead of finding the solution of Eq.~(\ref{eq:untilded}),
we are left to solve the well-known time-dependant diffusion equation
Eq.~(\ref{eq:tilded}). 

It proves convenient to separate diffusion along the radial and vertical 
direction. Considering a source located at $(x,y,z,\hat{t})$ detected at 
$(\Rsol,0,0,\hat{t}_O)$, the corresponding flux depends only on the radial 
relative distance $r  = |{\bf r}_S-{\bf r}_O|$, the distance of the source 
from the plane $z=z_S$ and the relative pseudo-time $\hat{\tau}=\hat{t}_{E_O}-
\hat{t}_{E_S}$. The Green function ${\cal \hat{G}}_\odot(r ,z,\hat{\tau})$ of 
Eq.~(\ref{eq:tilded}) is then given by:
\beq
\label{eq:pseudo-propag}
{\cal \hat{G}}_\odot(r ,z,\hat{\tau})=
{\displaystyle
\frac{\theta \left( \hat{\tau} \right)}{ 4 \pi K_{0} \hat{\tau}}}
\; \exp \left(-\frac{r^{2}}{4 K_{0} \hat{\tau}} \right) \,\times \,
{\cal G}^{1D}(z,\hat{\tau}).
\eeq

The effect of boundaries along $z=\pm L$ appears in 
${\cal G}^{1D}(z,\hat{\tau})$ only. For convergence properties, two distinct 
regimes are worth considering~\citep{2007A&A...462..827L}:
\begin{enumerate}
\item for sources close to us, it is best to use the so-called electrical 
  image formula (e.g.~\citealt{1999PhRvD..59b3511B}):
  \beq
      {\cal G}^{1D}(z,\hat{\tau})\!\!= \!\!\!
      { \sum_{n = -\infty}^{+\infty}} \!\! \left( -1 \right)^{n} 
      \frac{\theta \left( \hat{\tau} \right)}{\sqrt{ 4 \pi K_{0} 
	  \hat{\tau}}} \;
      \exp \left\{ - \frac{\left( z_n - z \right)^{2}}{4 K_{0} 
	\hat{\tau}}\right\},
      \label{eq:propagator_reduced_1D}
   \eeq
   where $ z_n = 2 L n + \left( -1 \right)^{n} z$;
 \item for far away sources, a more suitable expression is
   \ben
	 {\cal G}^{1D}(z,\hat{\tau})& = &
	 \!\frac{1}{L}\! \sum_{n=1}^{+ \infty} \!
	 e^{- K_0 k_n^2 \hat{\tau}} \phi_{n}(0) \phi_{n}(z)
	 \!+\! \nonumber\\
	 & & e^{-  K_0 {k'}_n^2 \hat{\tau}} {\phi'}_{n}(0) \phi'_{n}(z)\!
	 \label{eq:V_quantum}
   \een
   where 
   \begin{eqnarray}
     \phi_n(z) = \sin \left[ k_n (L-|z|) \right]  \;&\text{;}&\; k_n = 
     \left( n - \frac{1}{2} \right)\frac{\pi}{L}\;\; {\rm (even)} \nonumber\\
	 {\phi'}_n(z) = \sin\left[ {k'}_n (L-z)  \right]  \;&\text{;}&\;  
	 {k'}_n= n \frac{\pi}{L} \;\; {\rm (odd).} \nonumber
   \end{eqnarray}
\end{enumerate}

Coming back to the non-hat quantities, the propagator for a monochromatic 
point source is related to Eq.~(\ref{eq:pseudo-propag}) by
\beq
 {\cal G}^{\; e^+}_\odot(r ,z,E\leftarrow E_S)= \frac{\tau_E E_0}{E^2} \times 
 {\cal \hat{G}}_\odot(r ,z,\hat{\tau}=\hat{t}_E-\hat{t}_{E_S})\;.
\label{epropa}
\eeq
It follows that for a spatial and spectral distribution of sources 
$q_\odot({\bf r})Q(E)$ (origin of coordinates at solar neighbourhood),
the equilibrium spectrum at solar position and energy $E$ is 
given by 
\begin{eqnarray}
\Phi_\odot^{\;e^+}(E)\! &=&\frac{v}{4\pi}
       \times 2\int_{0}^L  dz \int_0^\infty  rdr
			 \\\nonumber
	\int_{E}^{\infty} &dE_S& \left\{ Q(E_S) 
  {\cal G}^{\; e^+}_\odot(r ,z,E\leftarrow E_S)\right\}
	 \int_0^{2\pi}  d\theta  q(r,z,\theta)\;.
\end{eqnarray}

%---------------%
\subsection{Propagation parameters}
\label{subsec:propag_param}
A few important points are reminded concerning the role of the various 
transport parameters on the propagated spectra of the antiparticles created 
in the DM halo. More details can be found in \citet{2004PhRvD..69f3501D,
  2005PhRvD..72f3507B}.

The halo height $L$ determines the total number of sources inside
the diffusive region and the typical distance a GCR can travel before escaping
from the Galaxy (see also App.~\ref{app:volumes}). The galactic wind wipes the 
particles away from the disk, and a similar effect occurs if $V_{c}$ is large 
enough. The parameters $L$, $V_{c}$ and $K_{0}$ are correlated.
In the subset of parameters giving the observed B/C ratio
\citep{2001ApJ...555..585M,2002A&A...394.1039M}, low values of $K_{0}$ 
generally correspond to low $L$ and  $V_{c}$, so that the DM signal is
expected to decrease with decreasing $K_{0}$. On that basis,
extreme and median parameters can be extracted, in the sense
that these parameters lead to the minimal and maximum expected flux,
while the median parameters (best fit to B/C data) provide the most
likely flux. These parameters are recalled in Table~\ref{table:prop}.
\begin{table}[t!]
\begin{center}
{\begin{tabular}{c c c c c}
\hline
\hline
&  $\delta$  & $K_0$ (kpc$^{2}$~Myr$^{-1}$) & $L$ (kpc)   & 
$V_c$ (km~s$^{-1})$  \\
\hline
{\rm max} &  0.46  & 0.0765 & 15 & 5.0    \\
{\rm med} &  0.70  & 0.0112 & 4  & 12.0   \\
{\rm min} &  0.85  & 0.0016 & 1  & 13.5 \\
\hline
\end{tabular}}
\caption{
Propagation parameters giving the maximal, median and minimal antiparticle DM 
fluxes compatible with B/C analysis.
\label{table:prop}}
\end{center}
\end{table}

Having in mind the connection between the propagation parameters and the 
fluxes, we can now justify discarding, for our calculations,
the effects of the wind and reacceleration for the positrons. For example, 
for configurations with small $\delta$, as the effect of the wind is always
negligible for anti-protons, it is also the case  for positrons (their travel
time in the Galaxy is less or at most that of the anti-protons).
For the sets of parameters with larger $\delta$, the effect of the 
wind becomes dominant below $\lesssim 1$~GeV. However, we are mainly 
interested in the high energy regime for positrons. Furthermore, if the 
low-energy behaviour is strongly dominated by convection (as is the case for 
anti-protons when $\delta=0.85$), then it superseeds energy loss effects for 
positrons: in that case, all the conclusion about \pbar\ 
would also hold for $e^+$.

%/////////////////////////////////////////////////////////////////
\section{Methods}
\label{sec:method}

The smooth contribution is straightforwardly calculated,
contrarily to the clumpy contribution that is plagued
by {\em statistical} uncertainties (in the sense that the
position of clumps is a random variable, see Sect.~\ref{subsec:methodJulien}).
The latter issue is the primary concern of this section.

Two complementary approaches are followed to calculate the
{\em Galactic variance} of the clumpy contribution. The first one 
\citep{2007A&A...462..827L} is a semi-analytical calculation of the mean and 
variance from the generic statistical properties of the clumps (spatial
and mass distributions), using the particle propagators that we recalled.
The second one uses the same ingredients, but quantities under scrutiny (mean 
and variance) are obtained by accumulating realisations of a clumpy galactic 
halo. Due to the lack of any clue about the precise location and intrinsic
properties of each individual DM clump, working with statistical tools is well 
motivated. The numerous clumps  can be treated as random objects, which 
average properties are taken here from N-body simulations.

Note that both methods rest on the assumption that clumps
are considered as point-like sources. This is correct while the distance of 
a clump to the Earth is greater than its spatial extension, and if the GCR 
propagation properties do not change within the spatial extension of a clump 
(see Table~\ref{tab:clump_parameters}). As the flux, on average,
is not dominated by nearby substructures, and since for those far away
clumps the spatial dependence of the propagator is smooth enough (diffusive 
process), the point-like source assumption holds. Would
a nearby clump dominate the positron or anti-proton flux---which
is very unlikely according to our calculation---a single source computation 
would be enough to deal with the clumpiness issue. Nevertheless, such a 
case, while easier to calculate, would make the clumpiness itself an 
absolutely unpredictive scenario for the indirect search for DM using 
antimatter GCRs, and is beyond the scope of this paper.

%---------------%
\subsection{Generalities}
\label{subsec:generalities}
Before exposing the methods, it is convenient to define a pseudo-Green 
function, denoted $\tilde{\cal G}$, by absorbing the energy dependence of the 
GCR propagators. To this aim, we define the quantity $dN/dE_S (E_S)$ to be the 
antimatter species spectrum at the source, which is defined here as the number 
of antimatter particles injected per annihilation and per energy unit.

For anti-protons, the pseudo-Green function reads:
\beq
\tilde{\cal G}^{\; \pbar} (E) \equiv  \frac{dN}{dE}(E) \times
  {\cal G}^{\; \pbar}_\odot(r ,z,E)\;.
  \label{eq:gtilde_pbar}
\eeq
The propagation term and the source term can be factorised (no energy mixing 
during propagation, a \pbar\ emitted at $E_S$ is detected at the same energy 
$E_S$). It means that the results for the relative uncertainties on the fluxes 
and for boost factors are independent of the particle physics model. 
Unfortunately, this is not the case for positrons:
\beq
\tilde{\cal G}^{\; e^+}(E) \equiv \int_{E}^{\infty}  dE_S  
\left\{ \frac{dN}{dE_S}(E_S) 
  \times {\cal G}^{\; e^+}_\odot(r ,z,E\leftarrow E_S)\right\}\;.
  \label{eq:gtilde_pos}
\eeq
The integral characterises energy losses and the source spectrum cannot be 
factored out of the integrand. Nevertheless, in the following, in order to 
keep the discussion at the most general possible level, we will mainly focus 
on a monochromatic line of positrons at $E_S$, i.e. 
$dN/dE_S (E_S)=\delta(E-E_S)$. The results for positrons are thus forced 
to be independent of any particle physics model. As we will see in 
Sect.~\ref{sec:results}, as the boost factor is close to unity for all 
energies, convolving the propagator with a realistic DM source spectrum
would yield a similar boost factor as obtained from the monochromatic line.

In the following, we will make use of $\tilde{\cal G}$, where the energy 
dependence is implicit for any species.

The total GCR flux $\phi_{\rm{tot}}$ originating from DM annihilations may be 
separated into two contributions, for the smooth component and for clumps:
\begin{equation}
\phi_{\rm{tot}}=\phi_{\rm{sm}}+\phi_{\rm{cl}}.
\label{eq:flux}
\end{equation}

\paragraph{Smooth contribution:}
The {\em smooth} contribution to the flux $\phi_{\rm{sm}}$ is calculated using 
the smooth density profile $\rho_{\rm{sm}}$:
\begin{equation}
\phi_{\rm{sm}}= \frac{v}{4\pi}\, S 
\int {\rm d}^3{\bf x}\left(\frac{\rho_{\rm{sm}}}{\rhosol}\right)^2({\bf x})\;
\tilde{\cal G}({\bf x_{\odot}\leftarrow x})\;,
\end{equation}
where $v$ is the cosmic ray velocity, $\tilde{\cal G}$ is the pseudo-Green 
function, defined above for \pbar\ and $e^+$, and $S$ is a particle physics 
coefficient\footnote{So that $dN/dE\times S\equiv Q(E)$, as used
  in Sect.~\ref{sec:propagation}.} depending on the WIMP model\footnote{This 
  differs from the convention used in \citet{2007A&A...462..827L} for which 
  $(v/4\pi)$ is included in $S$.}:
\begin{equation}
\label{eq:def_S}
S  \equiv  \delta\frac{\langle \sigma v\rangle }{2}
\left(\frac{\rhosol}{m_{\rm wimp}}\right)^2.
\end{equation}
The normalisation is thus chosen with respect to the local DM 
density $\rhosol =0.3$~GeV~cm$^{-3}$ and $\delta = 1\;(1/2)$ if the WIMP of 
mass $m_{\rm wimp}$ is a Majorana (respectively Dirac) particle. $S$ is 
actually counting the number of annihilations occurring in an infinitely small 
volume, in which the DM density is set to $\rhosol$.

\paragraph{Sub-halo contribution:}
The Galactic halo is populated by a constellation of many clumps, whose 
positions and masses are actually unknown. Nevertheless, if \Ncl~is the number
of clumps in a certain diffusive volume in the Galaxy, then their total 
contribution to the flux reads:
\beq
\phi^{\rm tot}_{\rm cl}= \sum_{i=1}^{\Ncl} \phi_i\;.
\label{eq:phicltot}
\eeq
The spatial dependence of the propagator is smooth enough (diffusive process) 
so that $\tilde{\cal G}$ may be considered constant over the clump scale: 
each clump behaves as a point-like source. The cosmic ray flux measured at 
the Earth from the $i^{th}$ clump is therefore given by:
\begin{equation}
\phi_i({\bf x}_{\odot}) = \frac{v}{4\pi}\, \times S \times \xi_i \times 
\tilde{\cal G}({\bf x}_{\odot}\leftarrow {\bf x}_i)\;,
\label{eq:phi_cl}
\end{equation}
where $\xi_{i}$ is the effective annihilation volume defined by 
Eq.~(\ref{eq:xi}).

%---------------%
\subsection{Semi-analytical calculation of the flux and the boost 
  factor, and associated variances, due to sub-halos}
\label{subsec:methodJulien}

In this section, we apply the 
formalism developed in \citet{2007A&A...462..827L} in order to predict how 
boosted the antimatter cosmic ray fluxes should be when adding sub-halos. 

%---------------%
\subsubsection{Whole sub-halo flux $\langle \phi^{\rm tot}_{\rm cl}\rangle $ }
The propagator describes the probability for a cosmic ray injected at position 
${\bf x}_S$ with energy $E_S$ to be detected at the Earth (${\bf x}_{\odot}$) 
with energy $E$ (recalling that for anti-protons, $E_S=E$, as they do not 
loose energy). 

As the intrinsic luminosity of a clump is entirely set once its mass is known,
the effective volume $\xi_i$ can be expressed as $\xi_i (M_{{\rm cl},i})$. 
Thus, given Eq.~(\ref{eq:phi_cl}), the flux associated with a single 
clump is a stochastic variable that depends on two probability 
distributions: the space and the mass distributions 
(Sect.~\ref{subsec:subhalo_distrib}). This is summarised in the following 
equation:
\beq
\frac{dP_{\phi}}{ d\phi_{\rm cl}} =  \frac{dP({\bf x}_{\rm cl},\Mcl ) }
     {d^3 {\bf x}_{\rm cl} d \Mcl}
 = \frac{1}{4\pi r^2}\frac{d{\cal P}_V(r)}{ dr} \times 
     \frac{d {\cal P}_M(\Mcl )}{ d \Mcl} ,
\label{eq:proba_phi}
\eeq
where both distributions, given by Eqs.~(\ref{eq:dPdV}) and
(\ref{eq:mass-distrib}), respectively, are considered 
uncorrelated\footnote{We remind that tidal disruption of a clump in the 
  Galactic centre depends either on its mass and on its location, which 
  induces a small correlation between the mass and the space distributions. 
  Nevertheless, we have checked that it could be neglected for this purpose.}.

The halo is populated by a constellation of many clumps whose total 
contribution to the GCR flux is given by
\beq
\label{eq:phitotcl}
\phi_{\rm cl}^{\rm tot}({\bf x}_{\odot}) = \sum_{i=1}^{N_c} \phi_i = 
\frac{v}{4\pi}\, S\sum_{i=1}^{N_c} \xi_i \times
\tilde{\cal G}({\bf x}_{\odot}\leftarrow{\bf x}_i)\;.
\eeq
Though the previous expression would be the actual expected flux for our 
Galaxy, we do not neither know the number nor the precise locations and the 
masses of clumps in the halo. Nevertheless, the knowledge of their phase space 
distribution can be used to determine the mean value of that flux:
\ben
\langle \phi_{\rm cl}^{\rm tot}\rangle  = \Ncl \times \frac{v}{4\pi} \times 
S \times \langle \xi\rangle _M \times \langle \tilde{\cal G}\rangle _V \, ,
\label{eq:true_mean_flux}
\een
where $\Ncl$ is the number of sub-halos hovering in the DM volume $V$ of 
interest, and
\beq
\label{eq:def_gmean}
\langle \tilde{\cal G}\rangle _V = \langle \tilde{\cal G}\rangle \equiv  
\int_{V}d^3 {\bf x} \; \tilde{\cal G}({\bf x}_{\odot}\leftarrow {\bf x})\times
 \frac{d{\cal P}_V({\bf x})}{dV}\;;
\eeq
\beq
\label{eq:def_ximean}
\langle \xi\rangle _M = \langle \xi\rangle \equiv  
\int_{M_{\rm min}}^{M_{\rm max}} dM \; \xi(M) \frac{d{\cal P}_M(M)}{dM}\;.
\eeq
Equation (\ref{eq:true_mean_flux}) is the mean value of the flux, in the 
statistical sense, due to all clumps in the Galaxy for a given model (space 
and mass distributions)\footnote{The integration volume $V$ is the DM halo 
volume, but in practical calculations, we reduce it to the diffusion volume.}. 
Anticipating the next section, we stress that any MC approach should converge 
to these values when taking a very large number of halo realisations.

%######%
\subsubsection{Variance $\sigma^{\rm tot}_{\rm cl}$ of the whole sub-halo flux}
The fact that we do not know how clumps are actually distributed, in the phase 
space defined by their locations and masses, can be expressed in terms of a 
variance $\sigma^{\rm tot}_{\rm cl}$ associated with their total mean flux 
$\langle \phi^{\rm tot}_{\rm cl}\rangle $. For a single clump, the relative 
flux variance is given by:
\beq
\frac{\sigma_{\rm cl}^2}{\langle \phi_{\rm cl}\rangle ^2} =  
\frac{\sigma_{\tilde{\cal G}}^2}{\langle \tilde{\cal G}\rangle ^2} + 
\frac{\sigma_{\xi}^2}{\langle \xi\rangle ^2} + 
\frac{\sigma_{\tilde{\cal G}}^2}{\langle \tilde{\cal G}\rangle ^2}\times 
\frac{\sigma_{\xi}^2}{\langle \xi\rangle ^2}\;,
\label{eq:sig_ju}
\eeq
where the individual variances affecting $\tilde{\cal G}$ and $\xi$ are 
respectively
\beq
\label{eq:sig_g}
\sigma_{\tilde{\cal G}}^2  =   \int_{\rm halo}d^3 {\bf x} \;
\tilde{\cal G}^2({\bf x}_{\odot}\leftarrow{\bf x})\times
\frac{d{\cal P}_V}{dV} \; - \; \langle \tilde{\cal G}\rangle ^2\;;
\eeq
and
\beq
\label{eq:sig_xi}
\sigma_{\xi}^2  =  \int_{M_{\rm min}}^{M_{\rm max}} dM \;  
\xi^2(M) \times\frac{d{\cal P}_M}{dM} \; - \; \langle \xi\rangle ^2\;.
\eeq
Quantities related to $\xi$ and \gtilde~will be quoted as mass- and 
space-related, respectively (see Sect.~\ref{sec:results}). We will further 
show  that mass-related effects dominate among the relative variances, in 
spite of sizable space-related ones. The (third) crossing term of the right 
hand side of Eq.~(\ref{eq:sig_ju}) will consequently set the full variance of 
the clump flux.

The resulting relative flux variance for the whole population of sub-halos is 
then merely:
\beq
\frac{\sigma_{\rm cl}^{\rm tot}}{\phi_{\rm cl}^{\rm tot}} = 
\frac{1}{\sqrt{\Ncl}} \frac{\sigma_{\rm cl}}{\phi_{\rm cl}}\;.
\label{eq:sig_phicl_tot}
\eeq

%It is worth quoting that the number \Ncl\ used in this calculation 
%has to be the one associated with the volume $V$ encompassing the diffusive 
%halo, and over which the spatial distribution of sub-halos is normalised to 
%unity. This guarantees that only the clumps contributing to the flux are 
%taken into account in the estimate of the global variance.

%######%
\subsubsection{Boost factor $B_{\rm eff}$ and its variance $\sigma_{B}$ }

Once the contribution of sub-halos to the flux is fully determined,
the \emph{boost factor} is easily computed for any species. As cosmic ray 
propagation has an explicit energy dependence, the boost factor is also 
energy-dependent~\citep{2007A&A...462..827L}, and, of course, also depends on 
the cosmic ray species~\citep{2003A&A...404..949M}.

The energy-dependent mean effective boost factor is given by the sum of the 
clumpy and the smooth contribution divided by the flux that would provide the 
only smooth reference halo $\rho^0_{\rm sm}(r)$ 
(see Sect.~\ref{sec:ref_smooth}):
\beq
\label{eq:boost}
B_{\rm eff} = (1-\fsol)^2 + 
\frac{\phi_{\rm cl}^{\rm tot}}{\phi_{\rm sm}}\;,
\eeq
where the local density fraction $\fsol$ has been defined in 
Eq.~(\ref{eq:fsol_def}), in order to keep the local DM density constant 
(see Sect.~\ref{subsec:ref-conf}).

It may be useful to determine the limit for which only an infinitely small 
volume around the Earth $\delta({\bf x}-{\bf x}_\odot)$ is taken into 
account. Actually, this will give a rough estimate of the asymptotic 
(maximum) value of the boost factor for both positrons (at detected 
energies very close to injected energies) and anti-protons (at low 
energies), because we are blind to contributions from regions close to the 
Galactic centre | where the smooth DM density dominates | in this case. This 
\emph{local} asymptotic value is given by:
\beq
B_{\odot} = (1-\fsol)^2 + 
\Ncl \times \langle \xi\rangle _M \times\frac{d{\cal P}}{dV}(\Rsol)\;.
\label{eq:bsol}
\eeq
This expression neither depends on the WIMP model, nor on the 
species. We see that, as $d{\cal P}/dV(\Rsol) = 3.9 \times 10^{-7} \; 
{\rm kpc}^{-3}$ in our reference model, only configurations with $\Ncl 
\times \langle \xi\rangle _M \gtrsim 2.6\times 10^6 \; {\rm kpc}^{3}$ will 
yield a relevant averaged contribution of clumps compared to the smooth 
component. From Table~\ref{tab:clump_parameters}, one can already see that the 
corresponding probability is likely to be very small. We can therefore provide 
a very simple criterion to check whether a boost is likely to appear in any 
(many-object) configuration:
\ben
\label{eq:boost_criterium}
\ncl(\Rsol)\times \langle \xi\rangle  = \Ncl \frac{d{\cal P}}{dV} (\Rsol) \times \langle \xi\rangle  
&\gtrsim & 1\,, \\
\Rightarrow \Beff &\gtrsim& 1.\nonumber
\een

Should the sub-halos spatially track the smooth component, then one would get:
\beq
B_{\odot}^{\rm sm} = (1-f_M)^2 + f_M \times 
\frac{\rhosol \times \langle \xi\rangle _M }{\langle \Mcl\rangle }
\label{eq:bsolsmooth}
\eeq
where $f_M$ is the mass fraction of DM in clumps, and $\langle \Mcl\rangle $ 
is the mean mass of clumps. The previous criterion to get a relevant boost 
factor then becomes merely 
$\langle \xi\rangle  \gtrsim \langle \Mcl\rangle /(f_M \rhosol)$.

The above value of the boost factor fluctuates up to a variance $\sigma_{B}$, 
which reads
\beq
\label{eq:sigma_b}
\sigma_{B}  =  \frac{\sigma_{\rm cl}^{\rm tot}}
      {\phi_{\rm sm}}\;,
\eeq
leading to
\beq
\label{eq:sigma_b_to_b}
\frac{\sigma_{B}}{B}  =  \frac{\sigma_{\rm cl}^{\rm tot}}
      {(1-\fsol)^2\phi_{\rm sm} + \phi_{\rm cl}^{\rm tot}}\;.
\eeq
If the sub-halo contribution dominates over the smooth component, then the 
relative variance of the effective boost factor is roughly equal to that of 
the sub-halo flux. Nevertheless, as soon as sub-halos become irrelevant in the 
flux estimate, the variance of the boost factor is strongly diluted by the 
smooth term. In this case, we obviously find a very small variance associated 
with the boost factor, even when the relative statistical uncertainty on the 
sub-halo flux itself is large.

%---------------%
\subsection{MC approach}
\label{subsec:methodMC}

A complementary approach is to calculate and add explicitly the contribution 
of each clump by MC drawing. Simulating many realisations of the DM 
sub-halos is another way to extract the mean flux as well as the variance
of the clump contribution. The ensuing calculation of boost factors
is as before, but in addition, MC provides the law of probability for the 
stochastic variable $\phicl$ that describes the single clump flux, which 
is hardly inferred from the clump phase space distribution itself due 
to the needed convolution with propagation.

From a technical point of view, it is very inefficient to calculate 
contributions from so many sub-halos (e.g. $\gtrsim 10^{15}$ for the lightest 
ones) one at a time. Indeed, for the clumps in a given mass range, two types 
of contributions exist. For low mass clumps, which are numerous, the variance 
associated with the flux is expected to be small (i.e. $\sigma_{\rm cl}/
\phi_{\rm cl}\ll 1$). In this case, we can spare the effort of averaging many 
configurations and directly compute the flux from a single realisation. 
Conversely, as the mass of the sub-halos increases, the associated number of 
clumps decreases, so that the variance finally become sizable. A threshold 
mass $M_{\rm th}$ needs to be specified, below which the contribution to the 
total variance $\sigma^{\rm tot}_{\rm cl}$ can be neglected: only sub-halos 
that have masses $\Mcl > M_{\rm th}$  need to be calculated for all samplings. 
The value of $M_{\rm th}$ is discussed in App.~\ref{app:volumes}.

For one sample, the total annihilation flux observed in the solar 
neighbourhood may be rewritten as
\begin{equation}
\phi^{\rm tot}_{\rm{cl}}=\phi_{\rm{low}}+\phi_{\rm{high}},\nonumber
\label{eq:flux_cl_highlow}
\end{equation}
where the quantities 
\begin{eqnarray}
\phi_{\rm{low}}&=& \frac{v}{4\pi}\, S\int {\rm d}^3{\bf x}
\int_{M_{\rm{min}}}^{M_{\rm{th}}}
{\rm d}\Mcl\; \tilde{\cal G}({\bf x\rightarrow x_{\odot}})n(\Mcl ,r)\,
\nonumber \\
& &\times\int_{\rm{sub}}\left(\frac{\rho_{\rm{sub}}}{\rho_0}\right)^2
({\bf x^{\prime}}){\rm d}^3
{\bf x^{\prime}} \label{eq:phi_low}\\
\phi_{\rm{high}}&=&\sum_{M_{\rm{i}}>M_{\rm{th}}}\phi_{\rm{i}}\,,
\label{eq:phi_high}
\end{eqnarray}
are the contributions from the low-mass sub-halos component and the high-mass 
sub-halos component, respectively. The number density of clumps was defined in 
Eq.~(\ref{eq:distri}), and the flux from a single clump $\phi_{\rm{i}}$ is 
given by Eq.~(\ref{eq:phi_cl}).

When taking into account all realisations:
\[
\langle\phi_{\rm{high}}\rangle=\left< \sum_{M_{\rm{i}}>M_{\rm{th}}}
\phi_{\rm{i}} \right>_{n}
\]
where $\langle.\rangle_n$ denotes the average over $n\gg1$ realisations of the
spatial distribution.  This leads to:
\begin{equation} 
\langle\phi_{\rm cl}^{\rm tot}\rangle\equiv  \phi_{\rm low} + 
\langle\phi_{\rm high}\rangle
\end{equation}
and
\begin{equation}
\frac{\sigma_{\rm{cl}}^{\rm tot}}{\langle\phi_{\rm cl}^{\rm tot}\rangle}
\simeq \frac{\sigma_{\rm{high}}}{\langle\phi_{\rm cl}^{\rm tot}\rangle},
\label{eq:var}
\end{equation}
where $\sigma_{\rm{high}}$ is the variance associated to high-mass clumps 
($\sigma_{\rm{low}}$ is neglected as underlined above).

The total flux and variance are now given by
\begin{equation} 
\langle\phi_{\rm tot}\rangle\equiv \phi_{\rm sm} + \phi_{\rm low} + 
\langle\phi_{\rm high}\rangle
\end{equation}
and
\begin{equation} 
\frac{\sigma_{\rm tot}}{\langle\phi_{\rm tot}\rangle}\simeq
\frac{\sigma_{\rm{high}}}{\langle\phi_{\rm tot}\rangle}.
\end{equation}

Thanks to this reasoning, a simple picture emerges, and a qualitative 
behaviour of the expected variance nicely complements the discussion
from the previous method. The highest-mass sub-halos, which are rare, carry 
all the variance of the total flux. In App.~\ref{app:m_th}, they are found to 
be in the mass range $\Mcl \gtrsim 10^7 \Msol$. Thus, as soon as the 
integrated luminosity of lower-mass clumps (that depends on $\alpham$) is much 
larger than that of the high-mass ones, the variance of the total clump 
contribution is expected to vanish. According to Fig.~\ref{fig:dLdLnM}, such a 
situation will occur for $\alpham\gtrsim 1.9$. Actually, even for 
$\alpham<1.9$, the variance will still be significantly decreased, because, as 
we already underlined, the local smooth contribution dominates over the clump 
one (see Fig.~\ref{fig:luminosity_profiles}).

%/////////////////////////////////////////////////////////////////
\section{Results and discussion}
\label{sec:results}

Fluxes, boost factors and associated variances have been calculated for both 
positrons and anti-protons using a semi-analytical approach 
(Sect.~\ref{subsec:methodJulien}) and, for the sake of comparison,
MC simulations (Sect.~\ref{subsec:methodMC}).

The coming results are based on a fiducial model for the injection of 
antimatter in the Galaxy, which allows a WIMP-model-independent analysis. In 
practise, for positrons, a monochromatic line of 200 GeV is injected at a rate 
assumed to be proportional to the squared density of DM in 
sources. In order to recover realistic orders of magnitude, especially for 
fluxes, we will also suppose that those positrons originate from, e.g. 
not-s-wave-suppressed annihilations of WIMPs at rest, with masses of 200 GeV 
and annihilation cross-section $\langle \sigma v\rangle /2 = 3\times 10^{-26}$ 
cm$^{3}.$s$^{-1}$ (for instance, Dirac fermions/anti-fermions with only 
trilinear couplings to $e^-\phi^+/e^+\phi^-$, where $\phi^{+/-}$ would be some 
exotic|conjugate|charged scalar fields). Besides, because anti-protons do not 
loose energy, we took their injection spectrum to be constant 
$dN/dT = 1 {\rm GeV}^{-1}$ between kinetic energies 0.1-200 GeV (any spectrum 
could have been taken, as it can be factorised out). One can easily guess what 
the results for any injection spectrum would be (originating from 
hadronisation or fragmentation processes for instance) by a mere rescaling. In 
this case, the WIMP properties can be almost the same as for positrons: fluxes 
have been computed using a Majorana WIMP with a mass of 200 GeV, and an 
annihilation cross-section of $\langle \sigma v\rangle  = 3 \times 10^{-26} 
{\rm cm}^3 / {\rm s}$.

\begin{figure*}[t]
\begin{center}
\includegraphics[width=\columnwidth, clip]
		{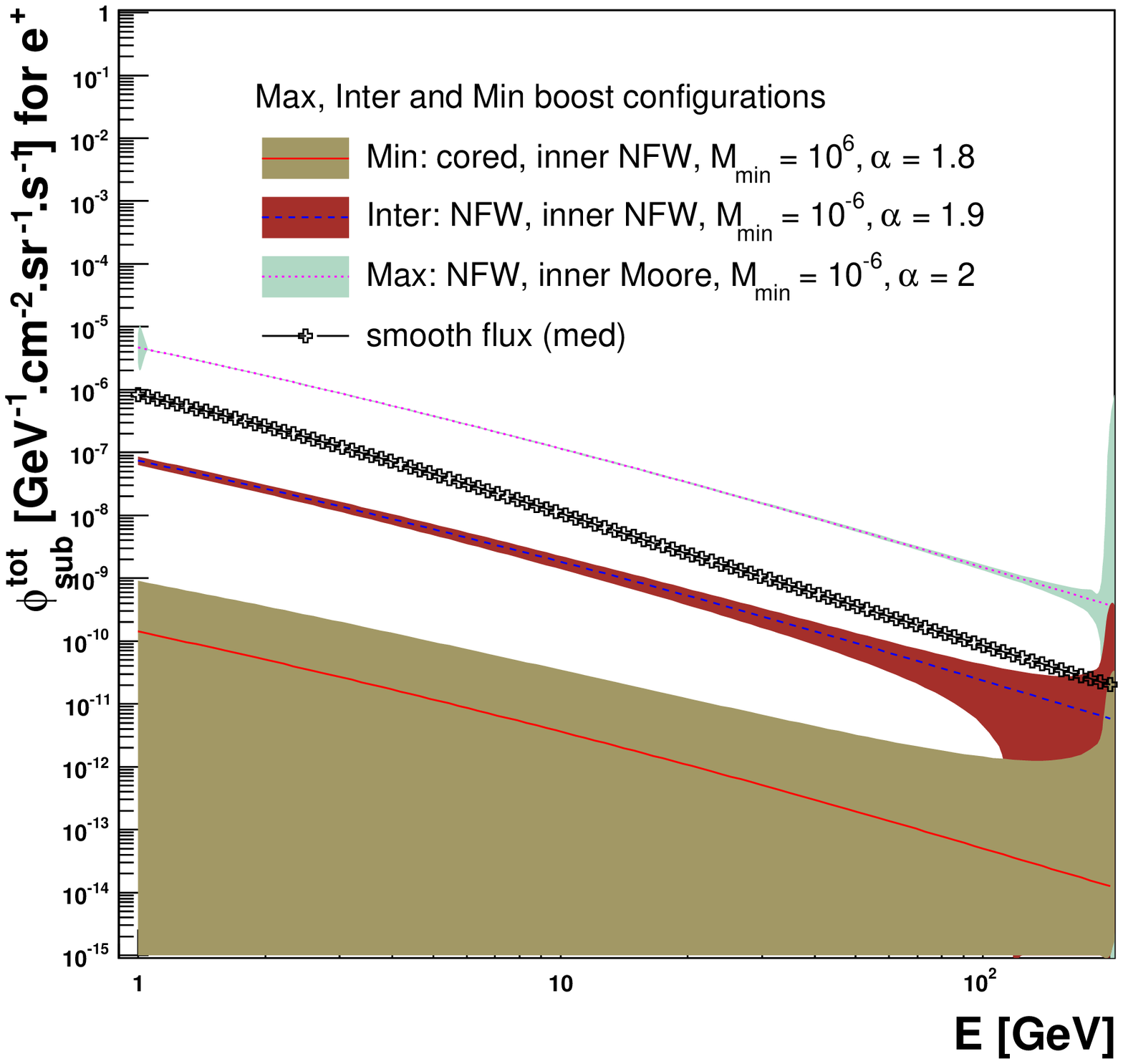}
\includegraphics[width=\columnwidth, clip]
		{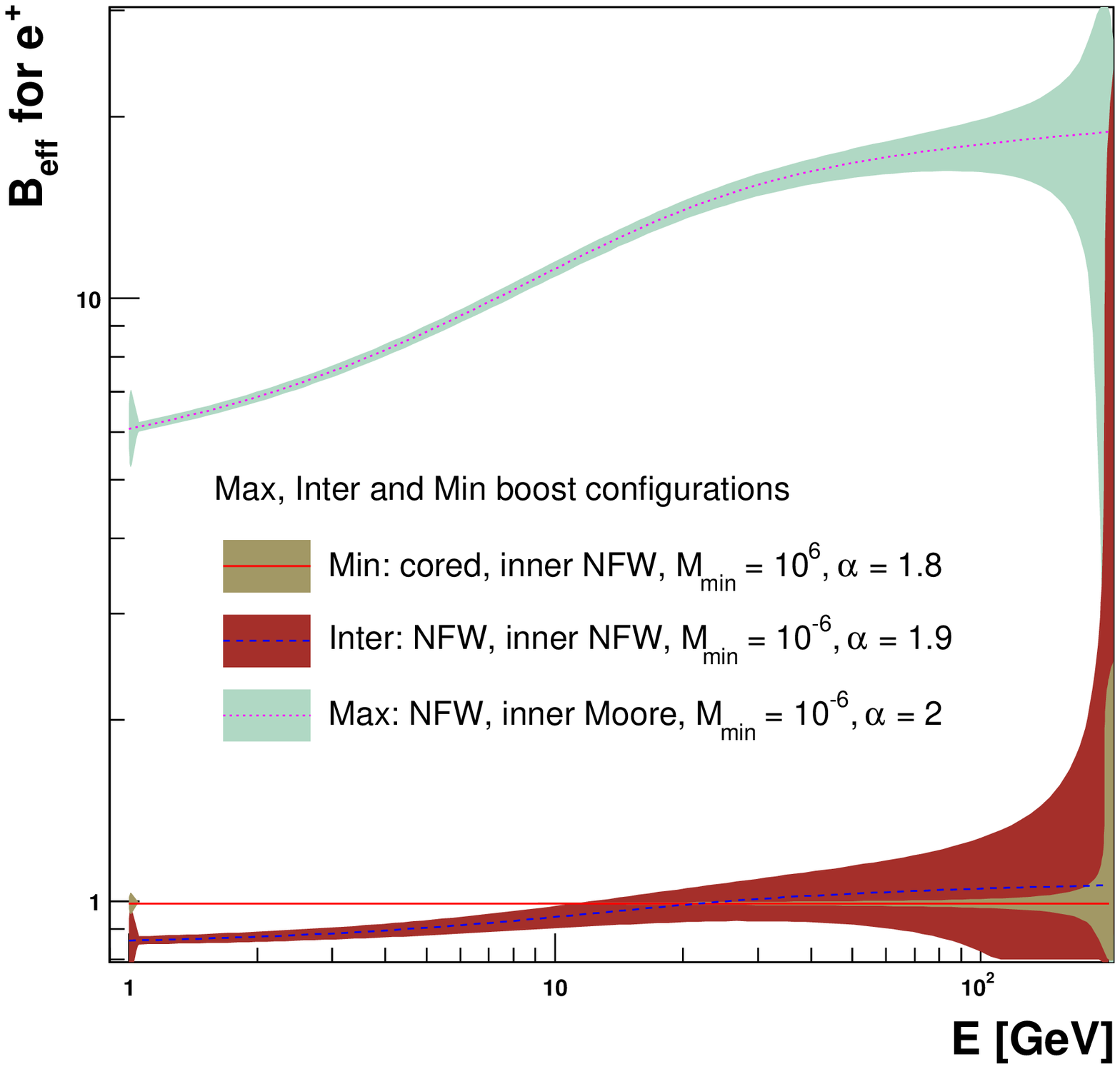}
\includegraphics[width=\columnwidth, clip]
		{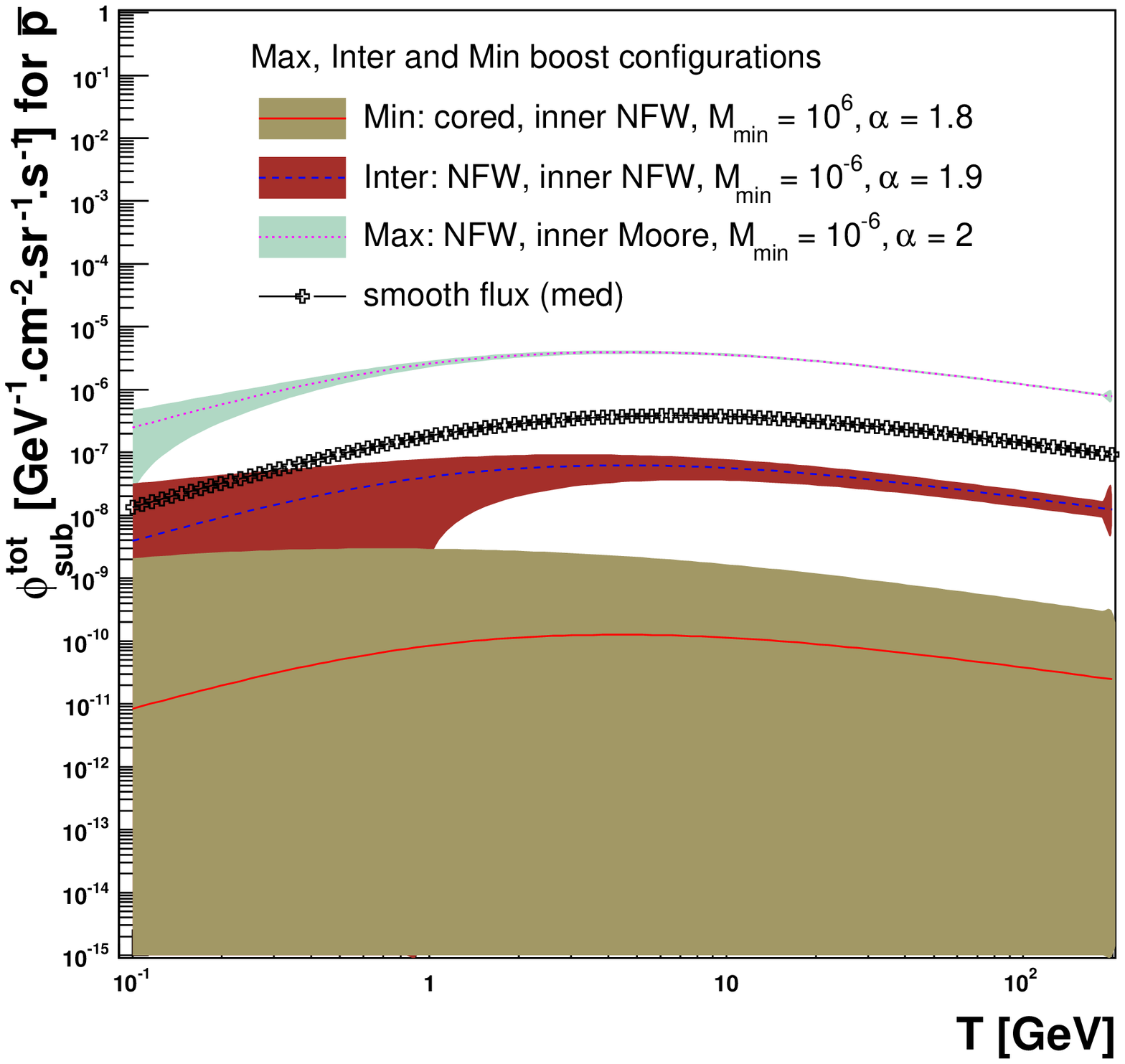}
\includegraphics[width=\columnwidth, clip]
		{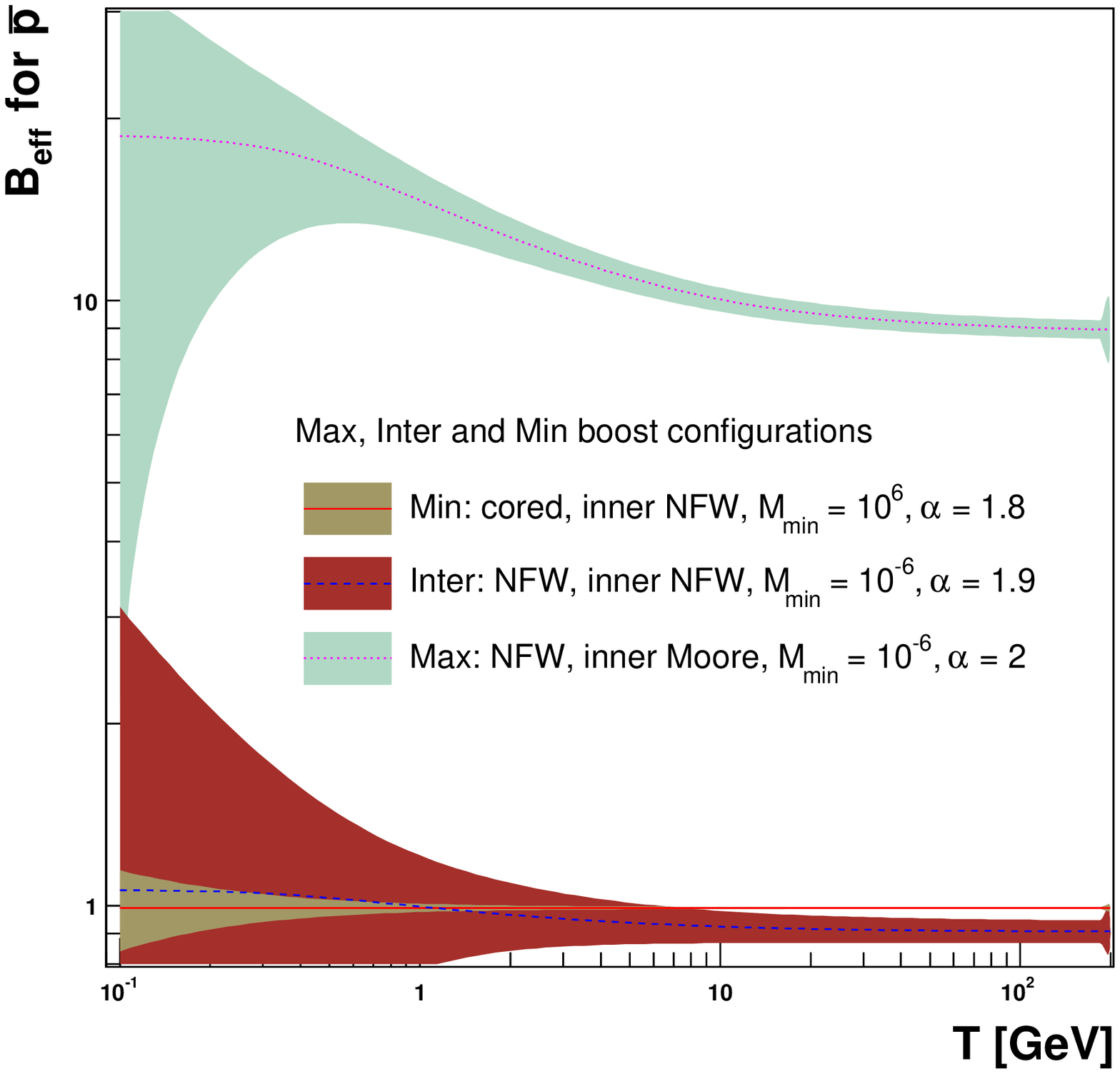}
\caption{\small \emph{Extreme cases for the DM configurations: sub-halo 
    antimatter fluxes associated with the maximal, intermediate and minimal 
    DM configurations (medium set of propagation parameters). Left/right: 
    fluxes/boosts and corresponding $1-\sigma$ contours. Top/bottom: 
    positrons/anti-protons. See the details in the text.}}
\label{fig:extreme_boosts}
\end{center}
\end{figure*}

Before going into the details of the studied configurations, we show in 
Fig.~\ref{fig:extreme_boosts} the extreme cases that we obtained for both 
species (with the medium set of propagation parameters). The first line panels 
are plots of the smooth and sub-halo fluxes and the resulting effective 
boosts, with associated $1-\sigma$  statistical contours. The second line 
panels are the same plots, but for anti-protons. The \emph{maximal} 
configuration is given by: largest \alpham\ (2), cuspiest sub-halo 
inner profile (Moore), smallest \Mmin\ ($10^{-6}\Msol$), spatial distribution 
according to the smooth NFW profile, and the B01 concentration model. The 
\emph{minimal} configuration is the reverse: smallest \alpham\ (1.8), flattest 
inner profile (NFW), greatest \Mmin\ ($10^{6}\Msol$), and smallest local 
number density (cored isothermal profile). The \emph{intermediate} is close to 
the reference configuration, given in Sect.~\ref{sec:cl_summary}, and takes 
the most likely values of parameters according to N-body simulations (except 
for \Mmin, of which the used reference value is $10^{-6}\Msol$; and for the 
spatial distribution of sub-halos which tracks the smooth NFW profile). 

\begin{figure}[t]
\begin{center}
\includegraphics[width=\columnwidth, clip]{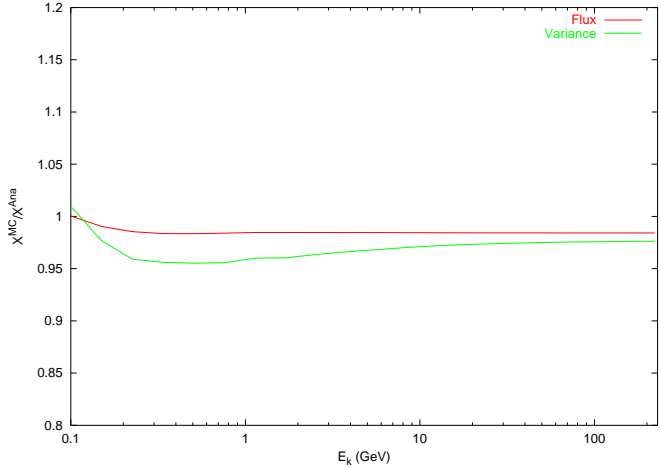}
\caption{\small \emph{Ratio of MC over semi-analytic results for the 
anti-proton flux and associated variance.}}
\label{fig:comp_MC_Ana}
\end{center}
\end{figure}

From this figure, we see that the boost factors obtained are functions of the 
energy and lie between 1 and 20, with small statistical uncertainties. Such 
a range has to be taken as that of theoretical uncertainties affecting 
the DM distribution in the Galaxy. From the approximate 
Eq.~(\ref{eq:bsol}), the asymptotic values obtained are also 1 and 20
(neglecting the density fraction $\fsol$), which are in excellent agreement 
with our full results. Before going into deeper details, it is worth 
emphasising that the maximal value of $\sim 20$ is as large as unlikely, as 
already discussed in Sect.~\ref{sec:DM}. For completeness, we have checked our 
results with MC simulations (see Sect.~\ref{subsec:methodMC}). In 
Fig.~\ref{fig:comp_MC_Ana}, we 
show that the agreement between the MC and the semi-analytic calculation 
is excellent for anti-protons, up to a few percents. It is the same for 
positrons, as already demonstrated in \citet{2007A&A...462..827L,
  2007PhRvD..76h3506B}.

Remembering that the whole sub-halo flux reads 
$\phicltot \propto (\Ncl \langle \xi\rangle _M)\times (\langle \gtilde\rangle _V)$, it makes sense to 
gather 
the impact of the various ingredients into two main physical classes.
\begin{description}
\item[\underline{Mass-related effects} ($\Ncl \langle \xi\rangle _M$):]
  these are encoded in the mean value and the variance of $\xi$, as
  defined in Eqs.~(\ref{eq:def_ximean}) and (\ref{eq:sig_xi}). The relevant 
  parameters to discuss (see Table~\ref{table:models}) are the minimal mass 
  \Mmin~of sub-halos, the logarithmic slope of the mass distribution 
  \alpham, the mass-concentration model and the inner profile.
\item[\underline{Space-related effects} ($\langle \gtilde\rangle _V$):] these 
  are encoded in the mean value and the variance of \gtilde, as defined in 
  Eqs.~(\ref{eq:def_gmean}) and (\ref{eq:sig_g}), which depend on both the 
  propagation model (through the propagator and the propagation parameters) 
  and the spatial distribution of sub-halos, besides of course the antimatter 
  species.
\end{description}
More details on this classification can be found in the appendix (see 
App.~\ref{app:mass_space_complement}). We first discuss the mass-related 
effects. We then focus on each antimatter species, for which we explain 
space-related effects, before describing the whole consequences of clumpiness 
on fluxes and boost factors. When discussing positrons, we will comment on the 
so-called HEAT excess.

%---------------%
\subsection{Mass-related effects}
\label{subsec:mass-like}

Once the space distribution of sub-halos and the propagation model are fixed, 
the propagator mean value $\langle \tilde{\cal G}\rangle _V$ of the sub-halo 
flux Eq.~(\ref{eq:true_mean_flux}) is fully determined, as well as its 
statistical fluctuation $\sigma_G$. Hence, provided the WIMP model 
is also fixed, the only differences from one sub-halo configuration to another 
will be the averaged total amount of antimatter yielded by clumps, given by 
the integrated clump luminosity $L_{\rm cl} \equiv \Ncl \times \langle \xi
\rangle _M $, and its associated fluctuations. Such a quantity depends on two 
parameters only: \Mmin\ and \alpham\ (plus the concentration-mass relation, 
plus the choice of the inner sub-halo profile). A decrease of \Mmin\ enhances 
the total number of clumps in the Galaxy, and an increase of \alpham\ raises 
the relative density of light compared to heavy objects (and the total 
luminosity accordingly because the clump number density is normalised with 
respect to the heaviest clumps, as given in Eq.~\ref{eq:Ncl}). More precisely, 
we find the luminosity to approximately scale with \Mmin only logarithmically 
(see the details in App.~\ref{app:mass-like_effects}) like:
\beq
L_{\rm cl} \propto  M_{\rm ref}^{\alpham-1} \times \ln(\Mmax/\Mmin).
\eeq
Therefore, we do not expect a large variation when spanning (\alpham,\Mmin) 
from the minimal (1.8,$10^6\Msol$) to the maximal (2,$10^{-6}\Msol$) 
parameter sets. Actually, we find the total luminosity to drop by a factor 
of $\sim 40$ only, while running the number of clumps over 13 orders of 
magnitude in the meantime (the above approximate expression sightly 
over-estimates the difference, leading to a factor of $\sim 150$). For 
completeness, an additional factor of 10 appears if one also considers 
inner Moore profiles for clumps (see Sect.~\ref{subsubsec:xi_M}), whereas a 
factor of 1/10 results from using an ENS01 mass-concentration relation.

Regarding the pure mass-related relative fluctuations, given by 
$\Ncl^{-1/2}(\sigma_{\xi}/\langle \xi\rangle )$ (see Eq.~\ref{eq:sig_ju}), we 
would naively expect them to significantly deplete when decreasing \Mmin, 
which enhances the total number of clumps. Furthermore, since the relative 
luminosity of light clumps is raised by increasing \alpham, the relative 
variance should be significantly reduced accordingly. Nevertheless, 
interestingly, we find the mass-related relative fluctuations to roughly scale 
like $(\ln(\Mmax/\Mmin))^{-1}$, and to vary only in the range 0.1-10\% when 
spanning over (\alpham,\Mmin) from minimal to maximal parameter sets (see 
above). The physical interpretation is the following: as \Mmin\ goes up, the 
number of clumps decreases accordingly, but in the meantime, the intrinsic 
clump luminosity ($\propto \xi$), which is fixed by the mass, fluctuates much 
less from clump to clump; there is a trade-off between shrinking the 
statistical sample and reducing the phase space, so that the relative 
mass-related variance remains almost constant. Taking a Moore inner profile 
does not affect the relative mass-related variance, while adopting the ENS01 
mass-concentration relation increases it by a few (see the third line panels 
of Fig.~\ref{fig:mass-like_effects_pos}).

As a consequence, the global flux relative variance given in 
Eqs.(\ref{eq:sig_ju}) and (\ref{eq:sig_phicl_tot}), should vary over two 
orders of magnitude at most, once the GCR propagation is fixed and at a given 
energy. Indeed, as mass-related uncertainties are almost always greater than 
space-related ones (see App.~\ref{app:mass-like_effects}), the dominant 
contribution is the space-mass crossing term $\Ncl^{-1/2} 
(\sigma_\xi/\langle \xi\rangle )(\sigma_\gtilde/\langle \gtilde\rangle )$, so 
that the global relative variance encompasses values in the range 
$(0.1-10\%)\times (\sigma_\gtilde/\langle \gtilde\rangle )$ when varying 
\alpham\ and \Mmin\ from extreme configurations.

From the previous statements together with the luminosity profiles already 
discussed and shown in Fig.~\ref{fig:luminosity_profiles}, scanning over 
the most likely mass-related parameters is unlikely to make the sub-halo 
contribution dominate over the smooth flux, except for extreme configurations 
combining the B01 concentration model, Moore inner profiles, large \alpham\ 
and very small \Mmin.

%---------------%
\subsection{Positrons}
\label{subsec:positrons}

%######%
\subsubsection{Space-related effects for positrons}
\label{subsubsec:space-like_pos}

The space-related effects for positrons come through the averaging of the 
propagator $\langle \gtildepos\rangle _V$ over the sub-halo spatial 
distribution. We summarise here a more detailed discussion that will be found 
in the appendix (see App.~\ref{app:space-like_effects_pos}). The relevant 
scale is the propagation scale $\lambda_D$ that depends on both diffusion and 
energy loss processes for positrons. $\lambda_D$ is obviously larger for 
larger diffusion coefficients, and smaller when the detected energy gets 
closer to the injected energy. Since it is of the order of kpc, we can safely 
focus on local quantities. Actually, $\langle \gtildepos\rangle _V$ encodes an 
effective detection volume bound by $\lambda_D$ and weighted by the clump 
spatial probability function $d{\cal P}/dV(r)$ in the solar neighbourhood. In 
the limit of infinite 3D diffusion, and when the propagation length is small 
enough, we find in App.~\ref{app:space-like_effects_pos} that $\langle 
\gtildepos\rangle \simeq (\tau_E/\epsilon^2) \times d{\cal P} (\Rsol)/dV$. 
Hence, the averaged propagator increases linearly with the local value of the 
clump spatial probability function. As $d{\cal P}/dV(\Rsol) \langle  
\rhosol/\Mvirh \rangle $ for the reference case (clumps are spatially 
distributed according to a cored isothermal profile), we see that given 
mass-related parameters, a configuration in which the clumps track the smooth 
profile will give a higher flux.

Regarding the pure space-related relative variance for a single object 
$\sigma_\gtilde/\langle \gtilde\rangle  $, we find it to scale like 
$(\lambda_D^3 \times d{\cal P}(\Rsol)/dV)^{-1/2}$, thus, decreasing when the 
effective detection volume or the clump local spatial probability increase 
(detected energies much  lower than injected ones). When 
taking the whole contribution, an additional factor of $\Ncl^{-1/2}$ 
reduces the global variance, and the picture becomes very simple: the 
relative space-related variance scales like $N_{\rm obs}^{-1/2}$, one over 
the square root of the number of clumps contributing to the signal at the 
Earth. It is maximal at high energy for positrons.

To summarise, the space-related contribution for positrons increases with 
the diffusion coefficient, and with the clump local space probability 
function. The relative space-related variance decreases when the propagation 
length raises (at low energy for positrons), because a larger number of 
sub-halos can contribute to the signal at the Earth.

%######%
\subsubsection{Overall effect on the positron flux: boost factor estimate}
\label{subsubsec:pos_tot}

Taking this fiducial injection model, we assess the different effects 
and draw four typical plots, which will compose four specific panels in the 
next figures, from left to right: positron flux, relative flux variance, 
boost factor, relative boost variance (as functions of the positron detected 
energy).

\begin{figure*}[t!]
\begin{center}
\includegraphics[width=0.5\columnwidth, clip]
		{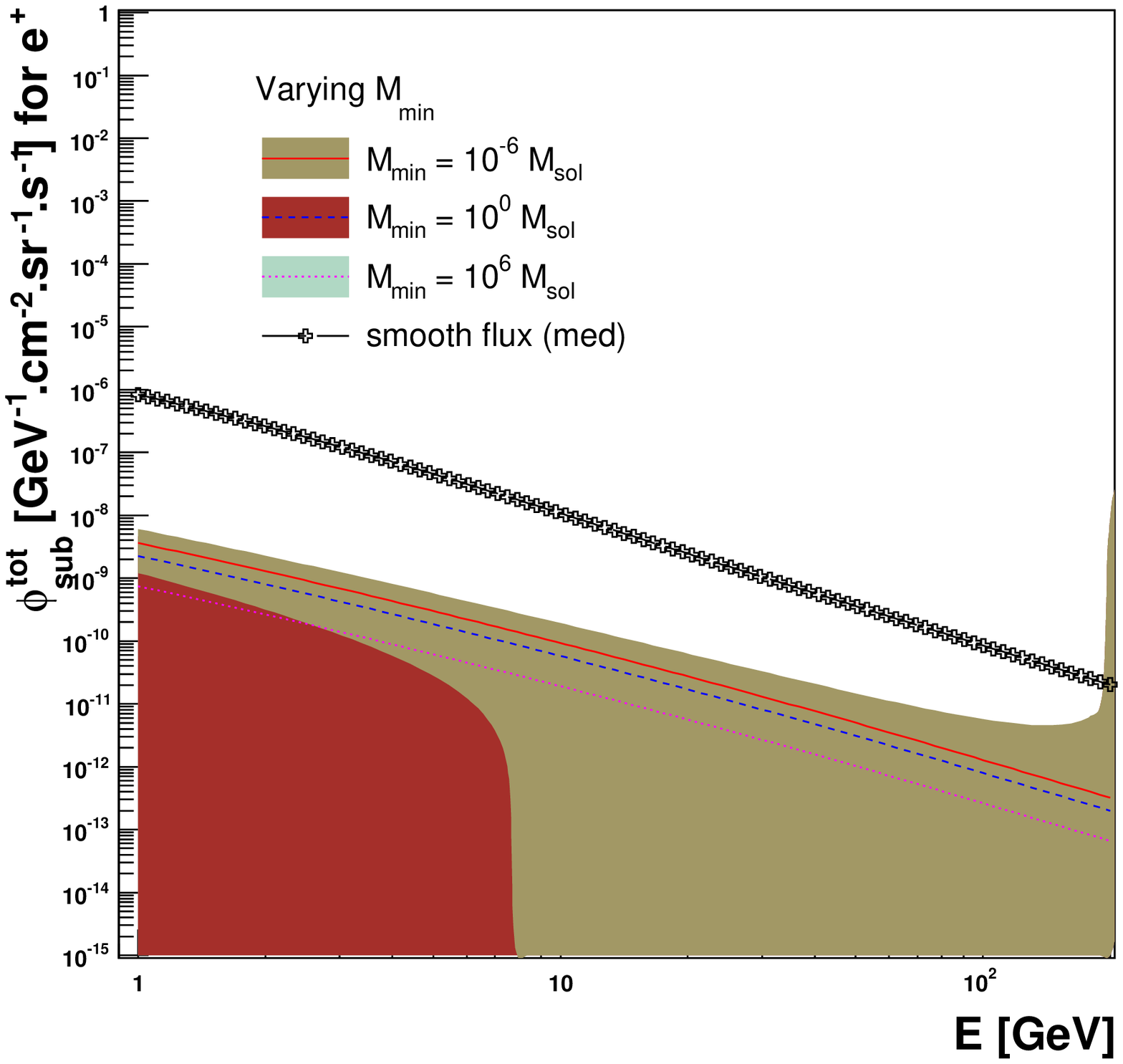}
\includegraphics[width=0.5\columnwidth, clip]
		{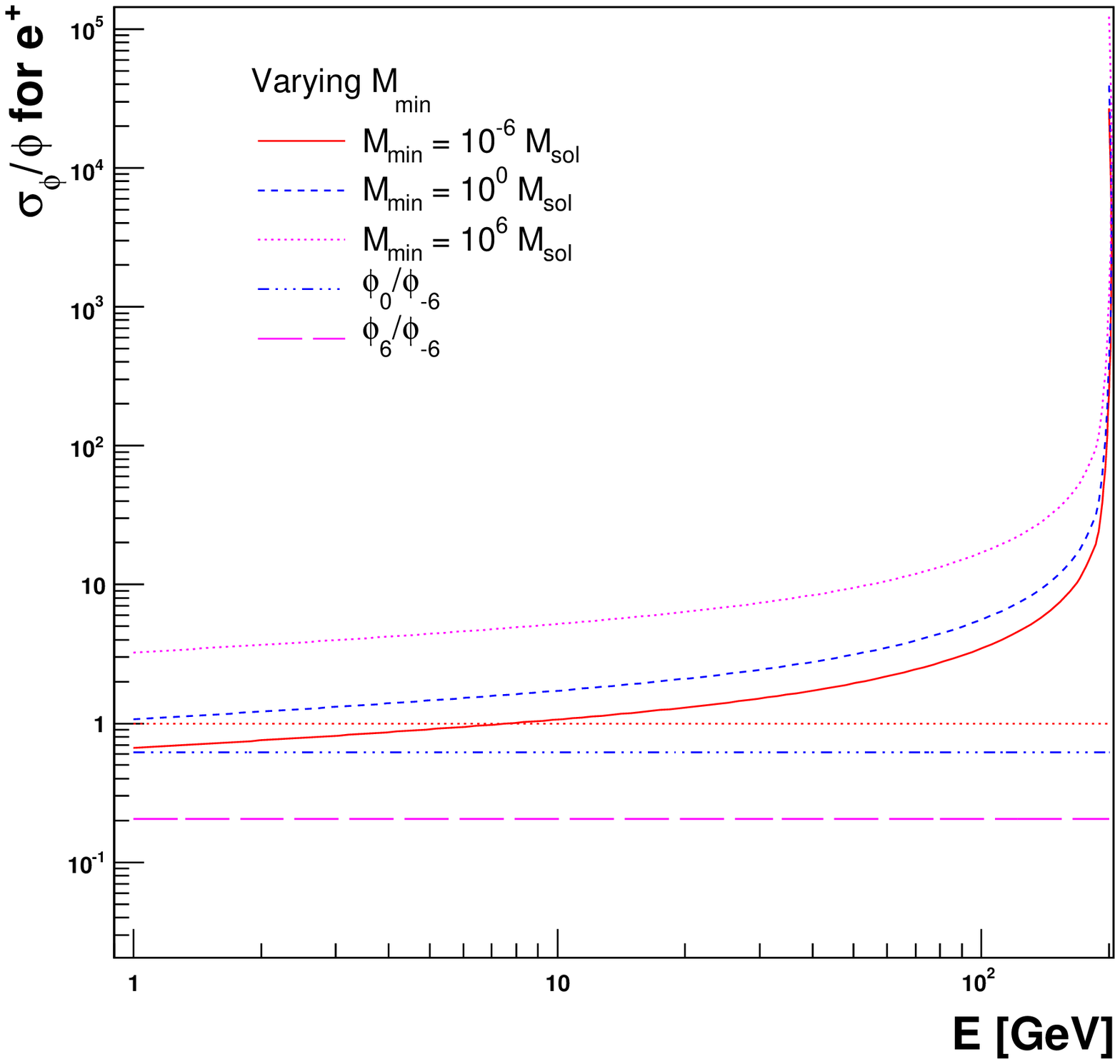}
\includegraphics[width=0.5\columnwidth, clip]
		{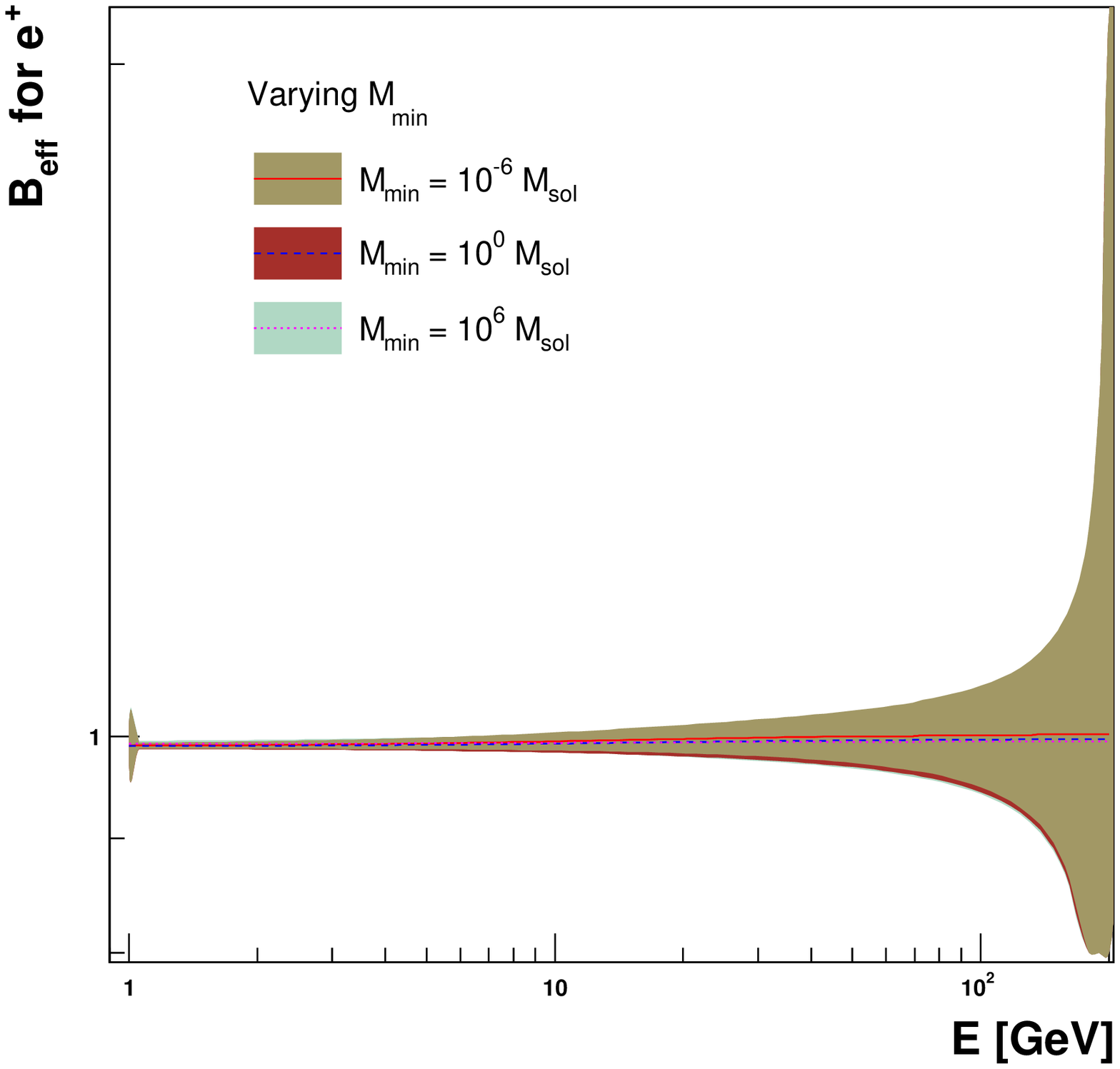}
\includegraphics[width=0.5\columnwidth, clip]
		{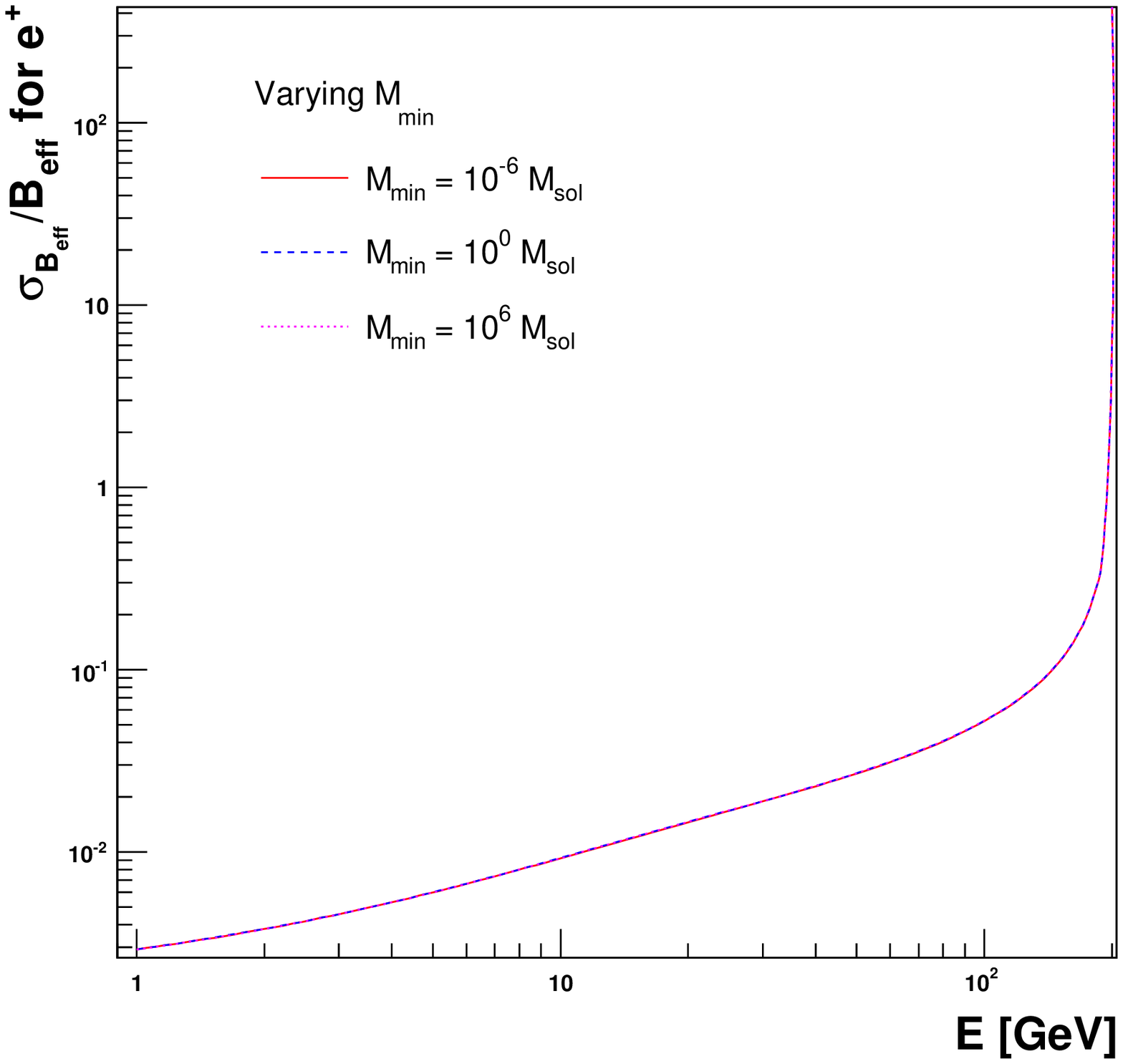}
\includegraphics[width=0.5\columnwidth, clip]
		{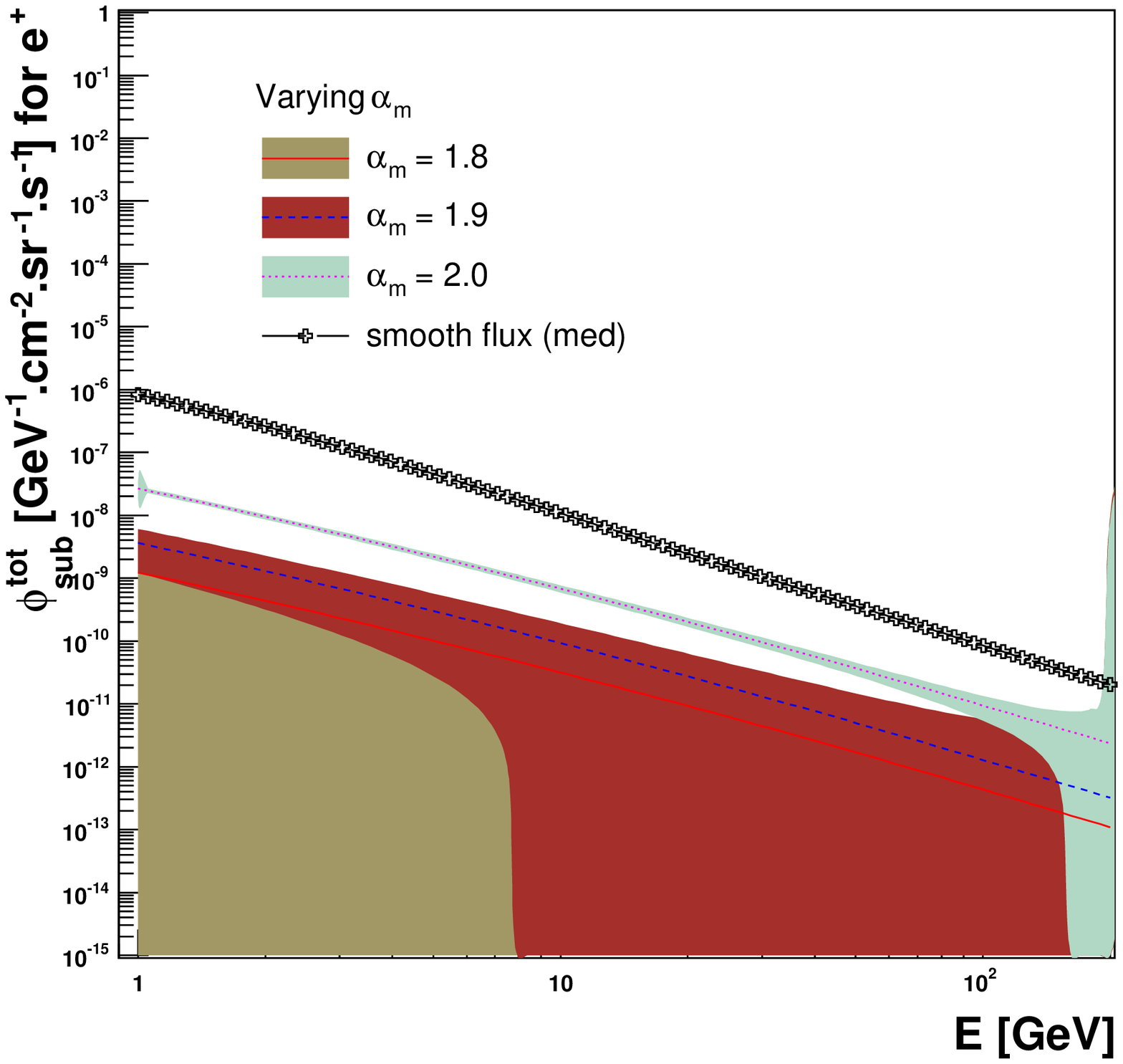}
\includegraphics[width=0.5\columnwidth, clip]
		{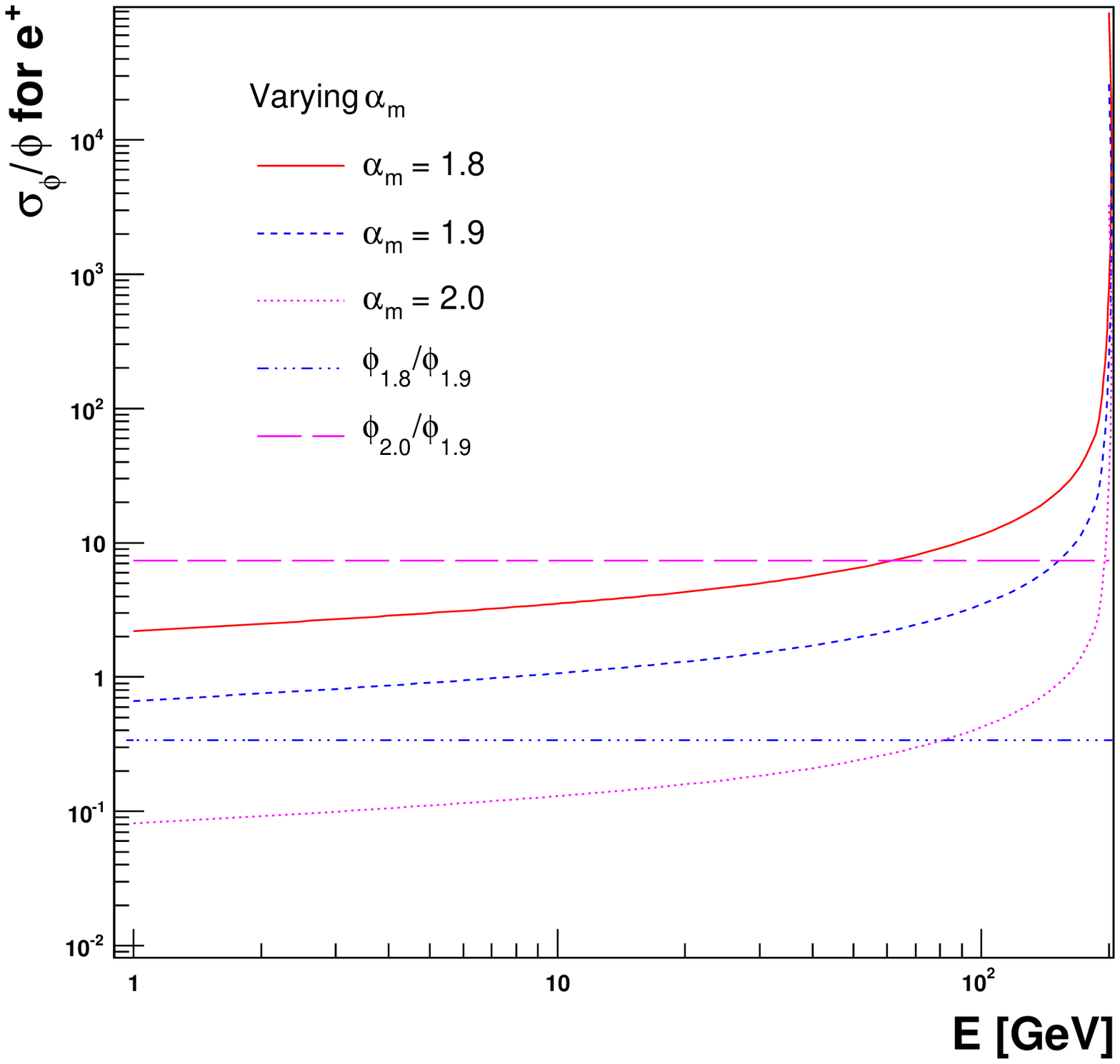}
\includegraphics[width=0.5\columnwidth, clip]
		{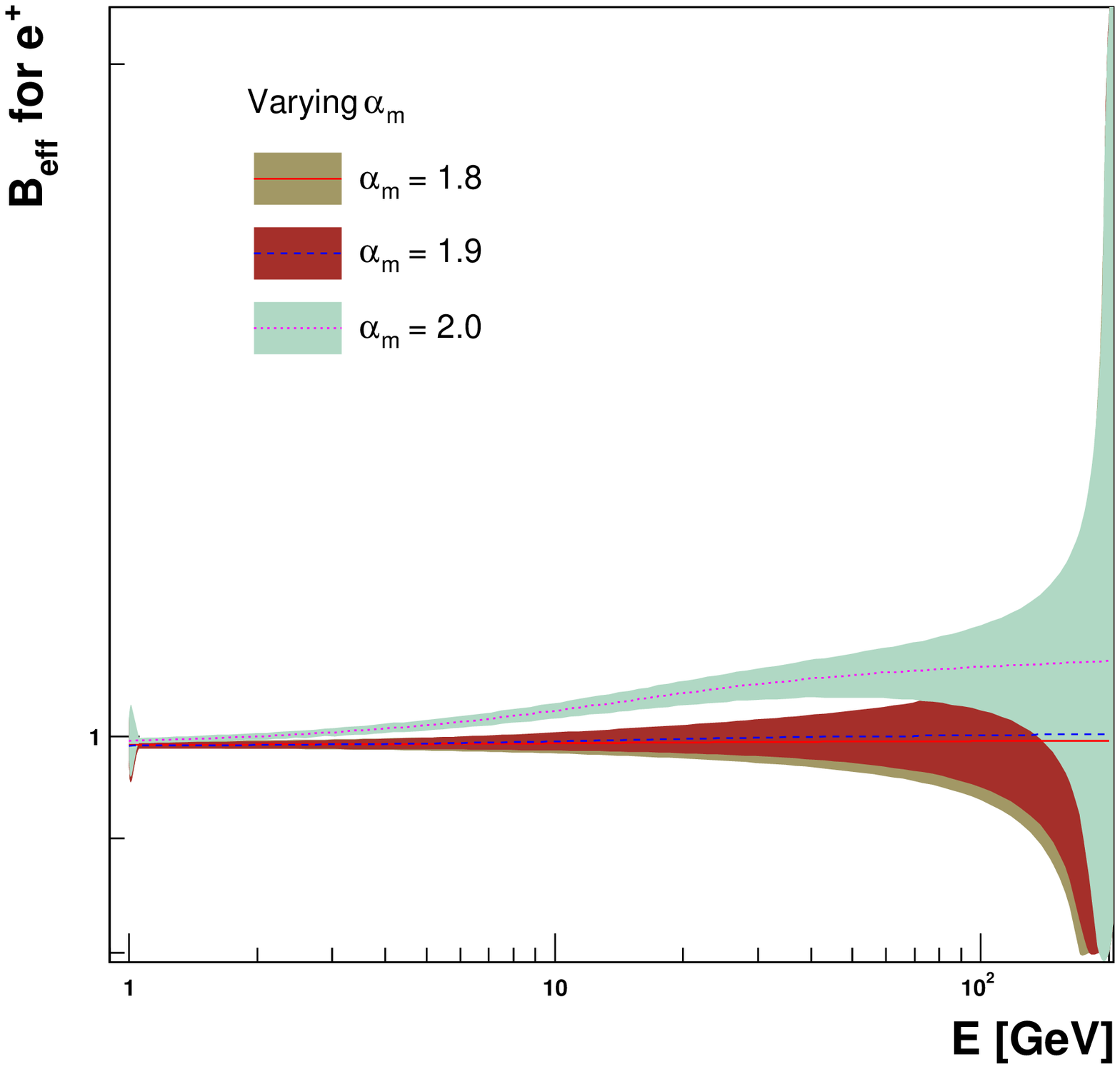}
\includegraphics[width=0.5\columnwidth, clip]
		{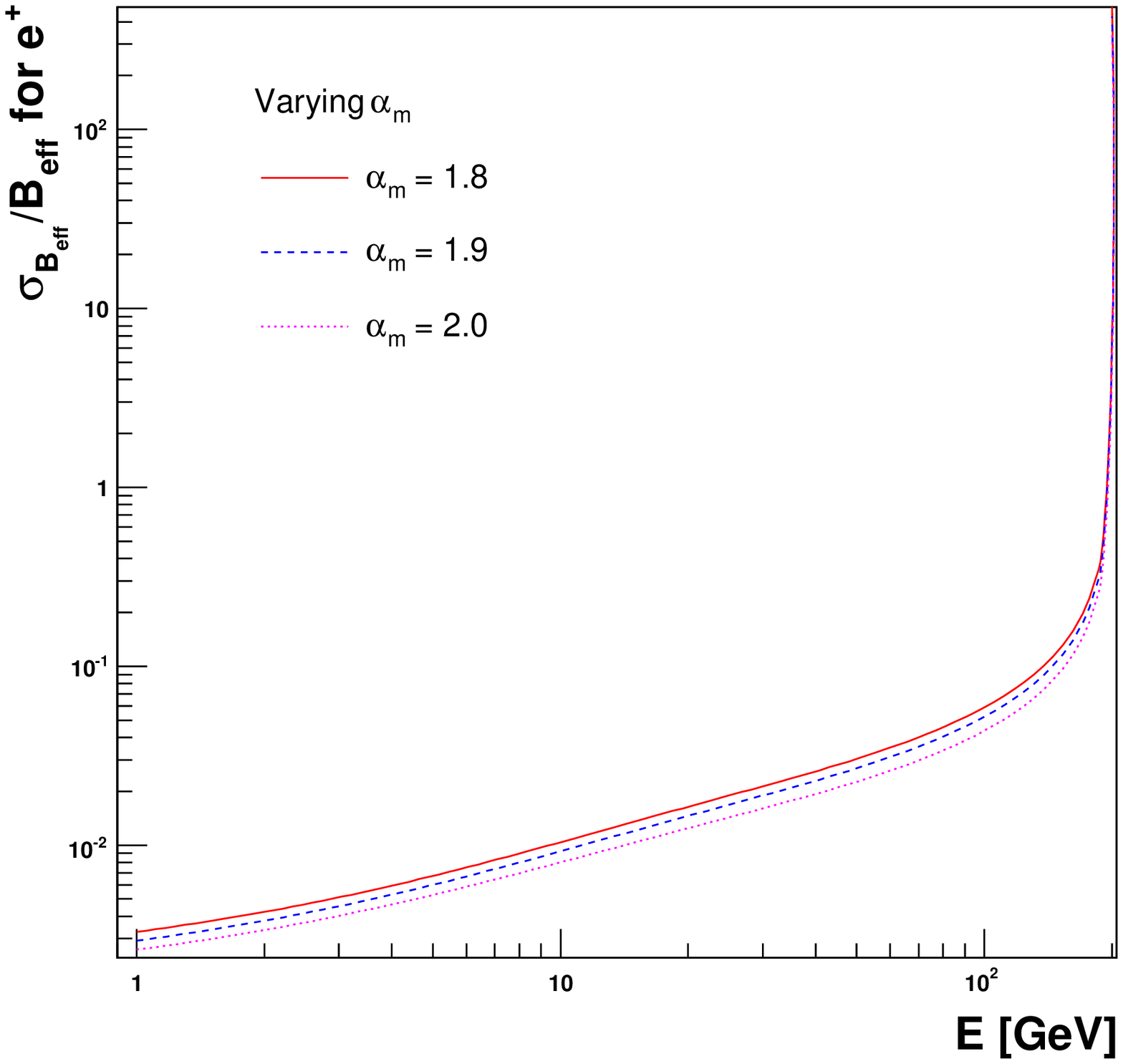}
\includegraphics[width=0.5\columnwidth, clip]
		{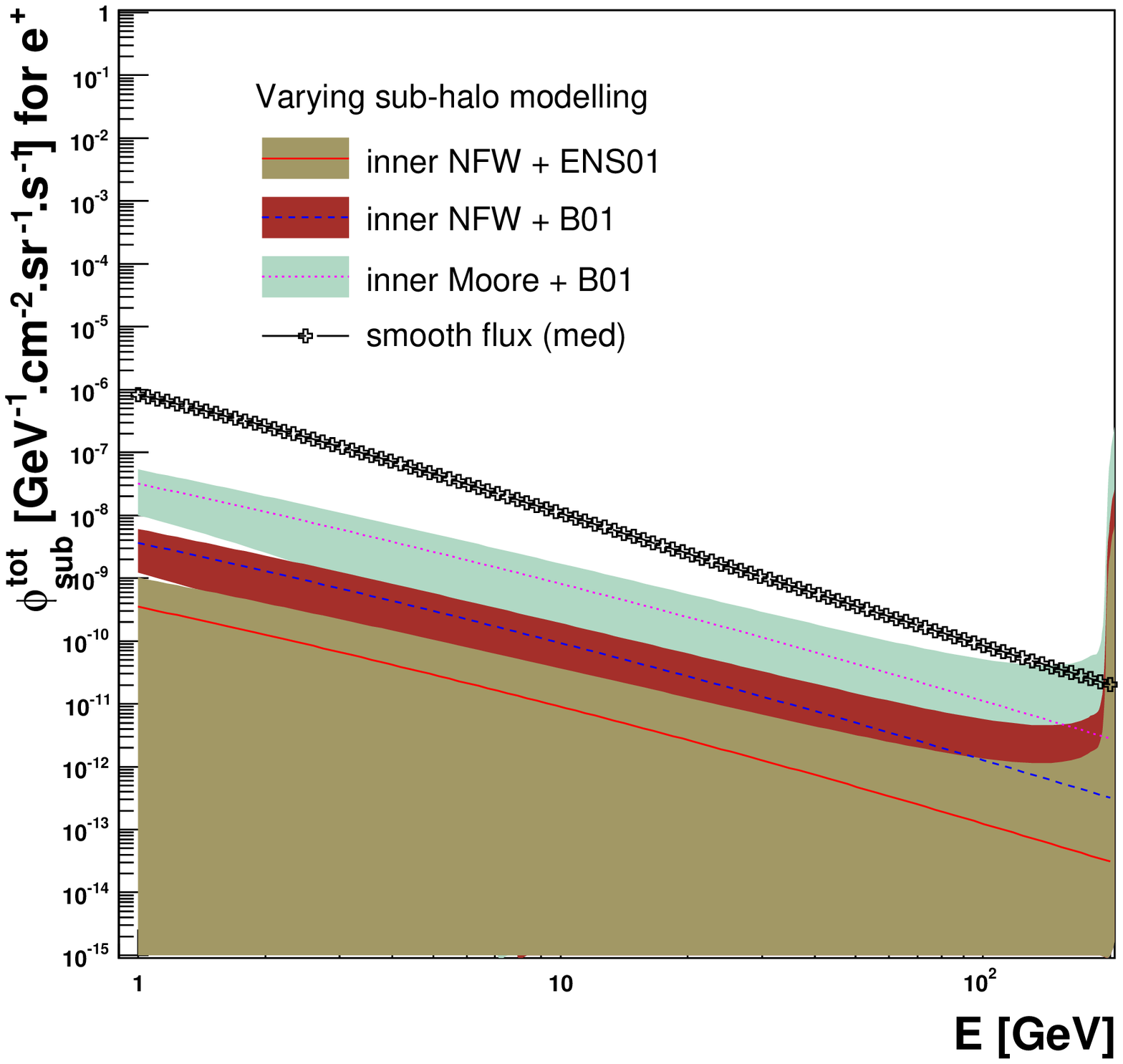}
\includegraphics[width=0.5\columnwidth, clip]
		{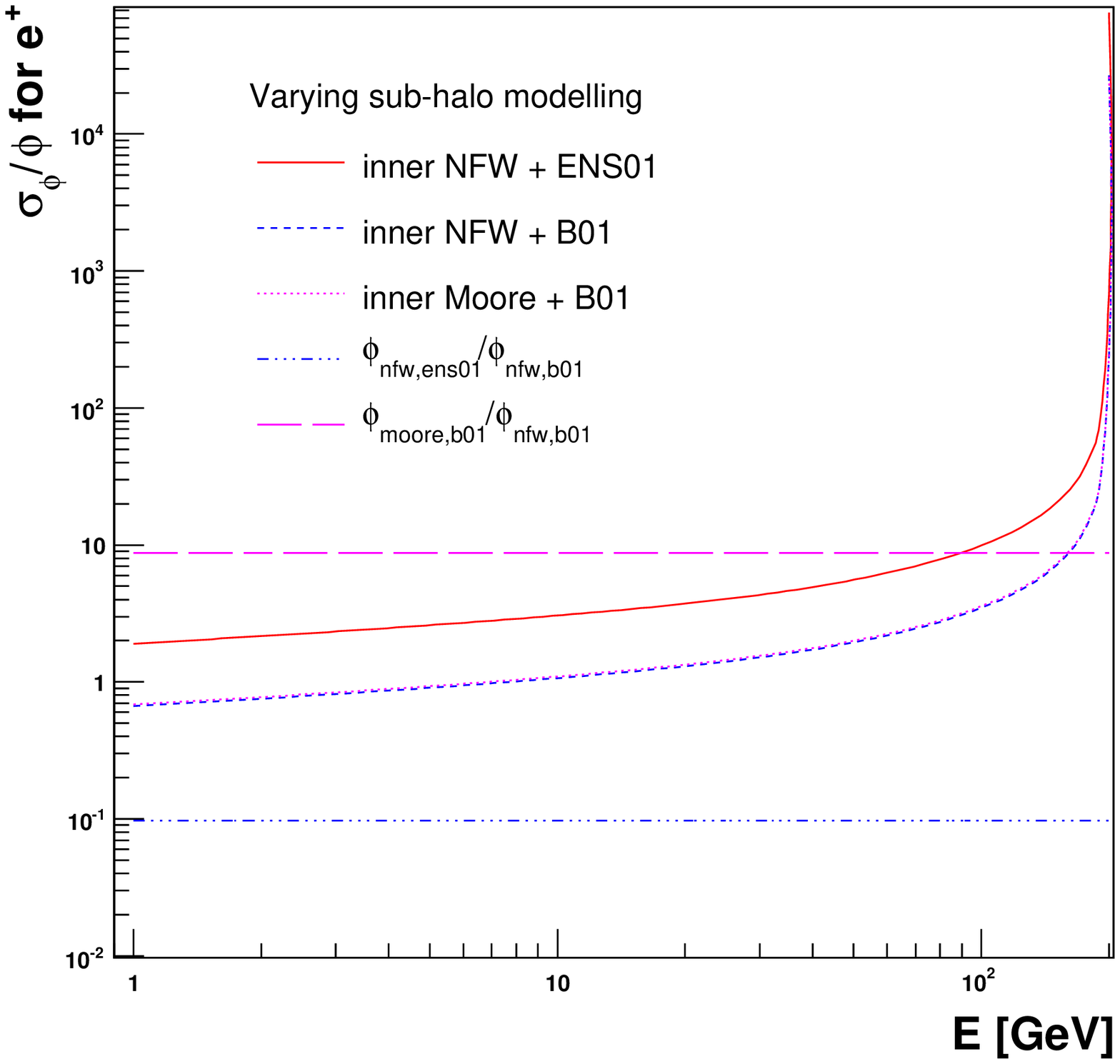}
\includegraphics[width=0.5\columnwidth, clip]
		{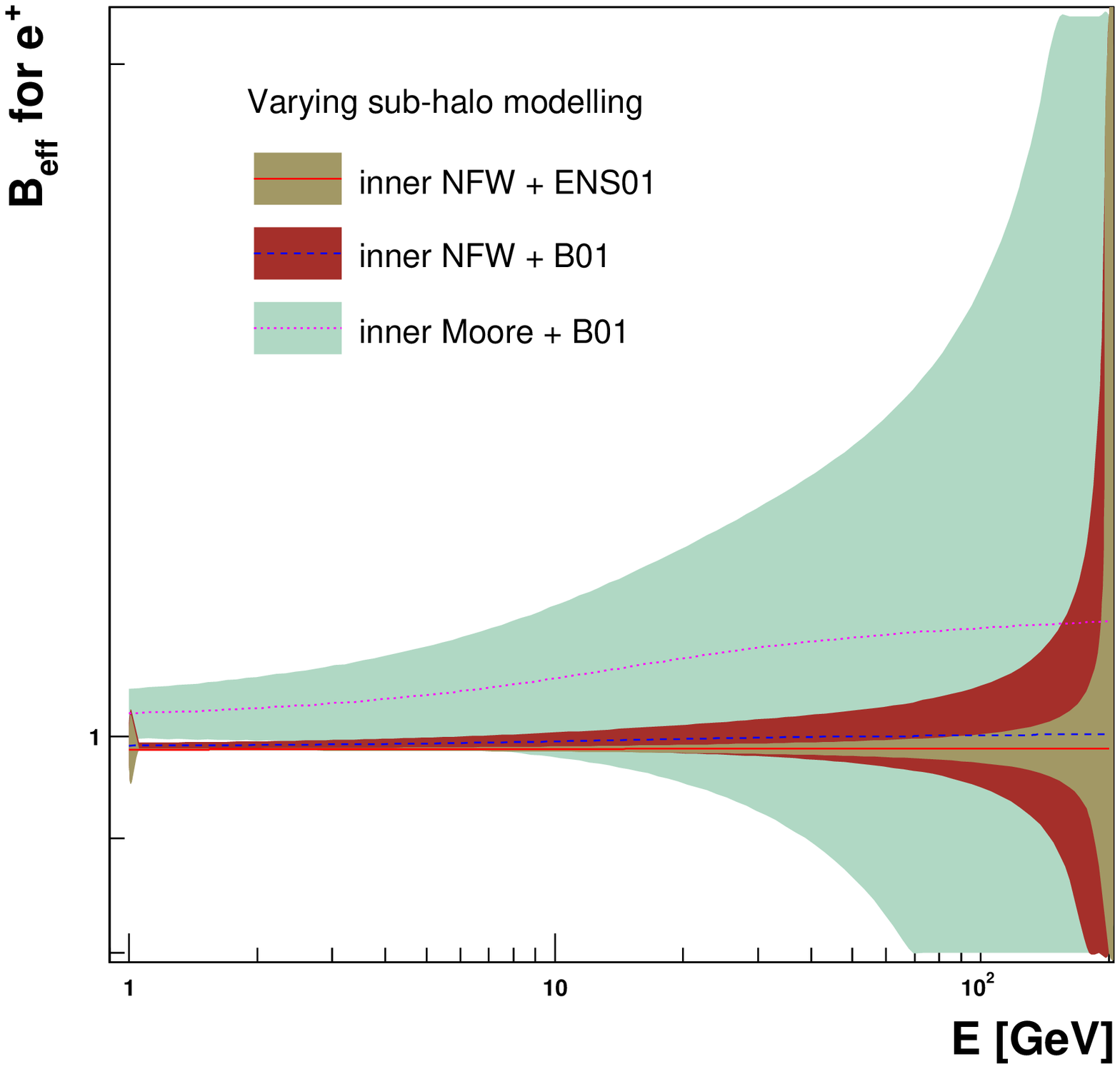}
\includegraphics[width=0.5\columnwidth, clip]
		{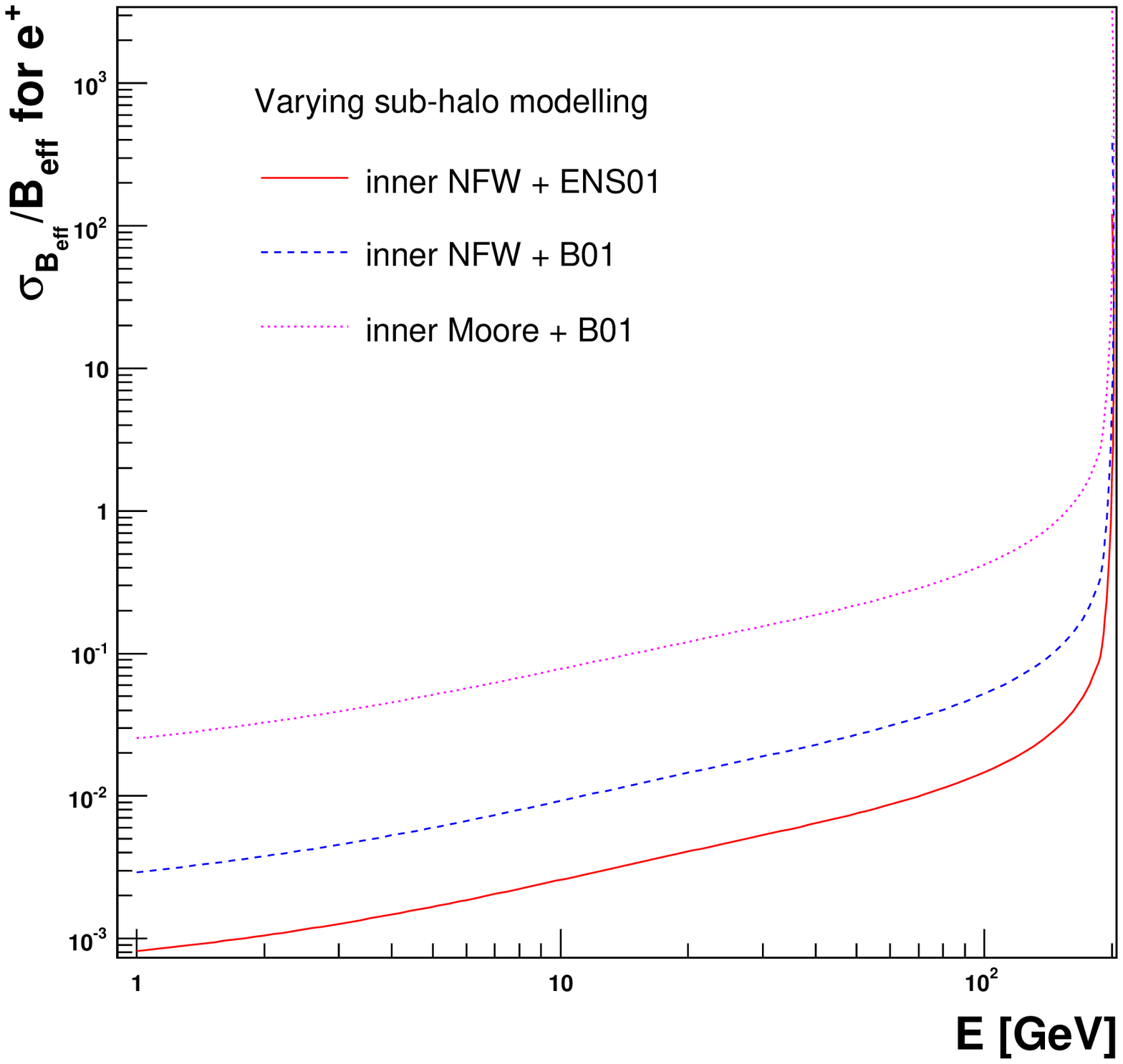}
\caption{\small \emph{Mass-related effects on the positron flux (first column) 
    with a focus on the corresponding relative uncertainty (second column), 
    for positron lines of 200 GeV injected at sources, with a rate 
    corresponding to standard values of WIMP annihilation, and the associated 
    boost factor with its relative variance (third and fourth columns). The 
    plain contours account for one standard deviation.
    First row: 
    Effect of changing the minimal mass of sub-halos, $\Mmin= 10^{-6}$, $1$ 
    and $10^{-6}$ \Msol . Second row: Effect of changing the logarithmic 
    slope \alpham\ of the sub-halo mass function, with $\alpham= 1.8$, 
    1.9 and 2.0. Third row: Effect of changing the sub-halo inner 
    properties ; the inner profile is taken to be either NFW or Moore, and 
    the concentration model varies from B01 to ENS01.}}
\label{fig:mass-like_effects_pos}
\end{center}
\end{figure*}

\begin{figure*}[t!]
\begin{center}
\includegraphics[width=0.5\columnwidth, clip]
		{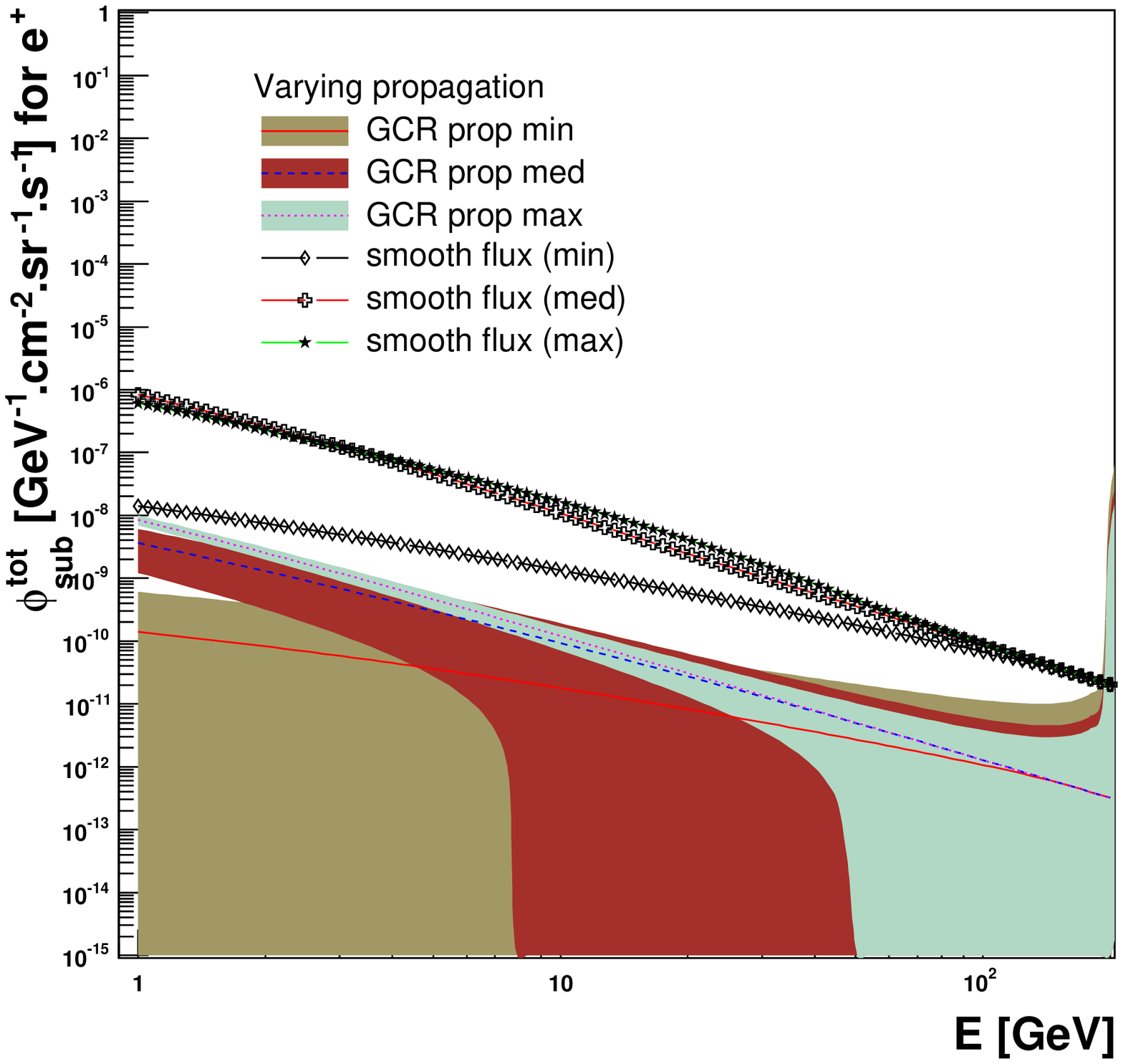}
\includegraphics[width=0.5\columnwidth, clip]
		{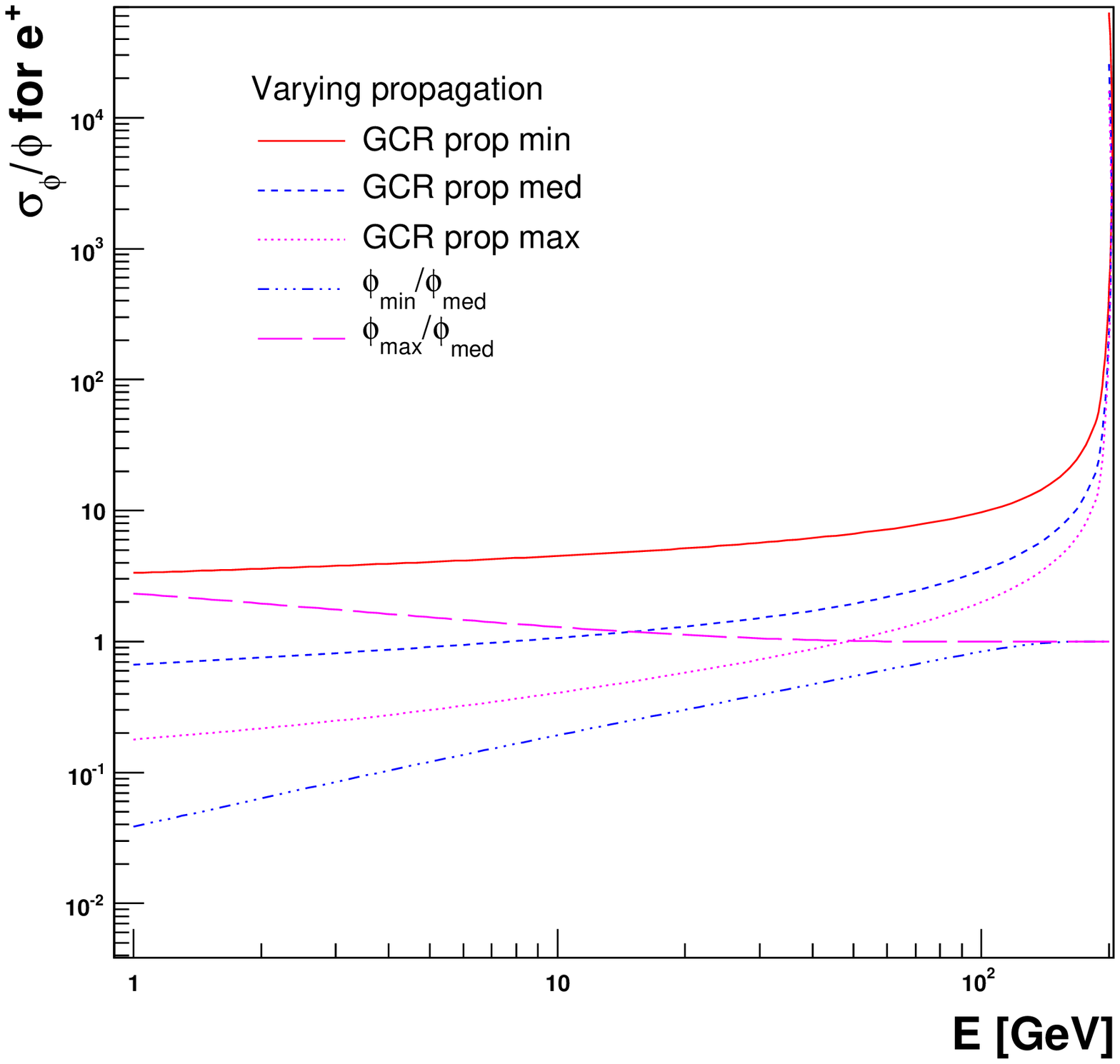}
\includegraphics[width=0.5\columnwidth, clip]
		{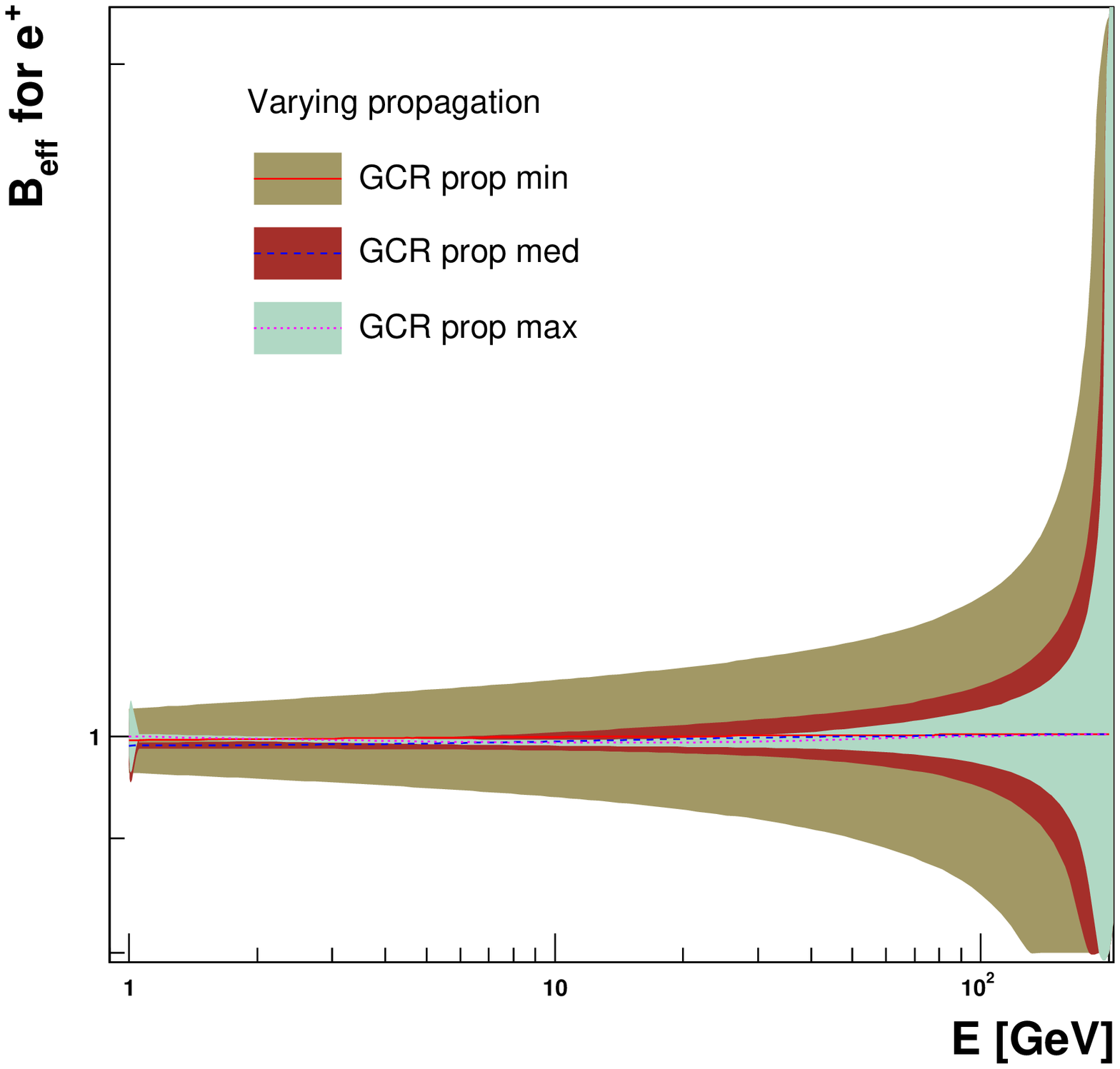}
\includegraphics[width=0.5\columnwidth, clip]
		{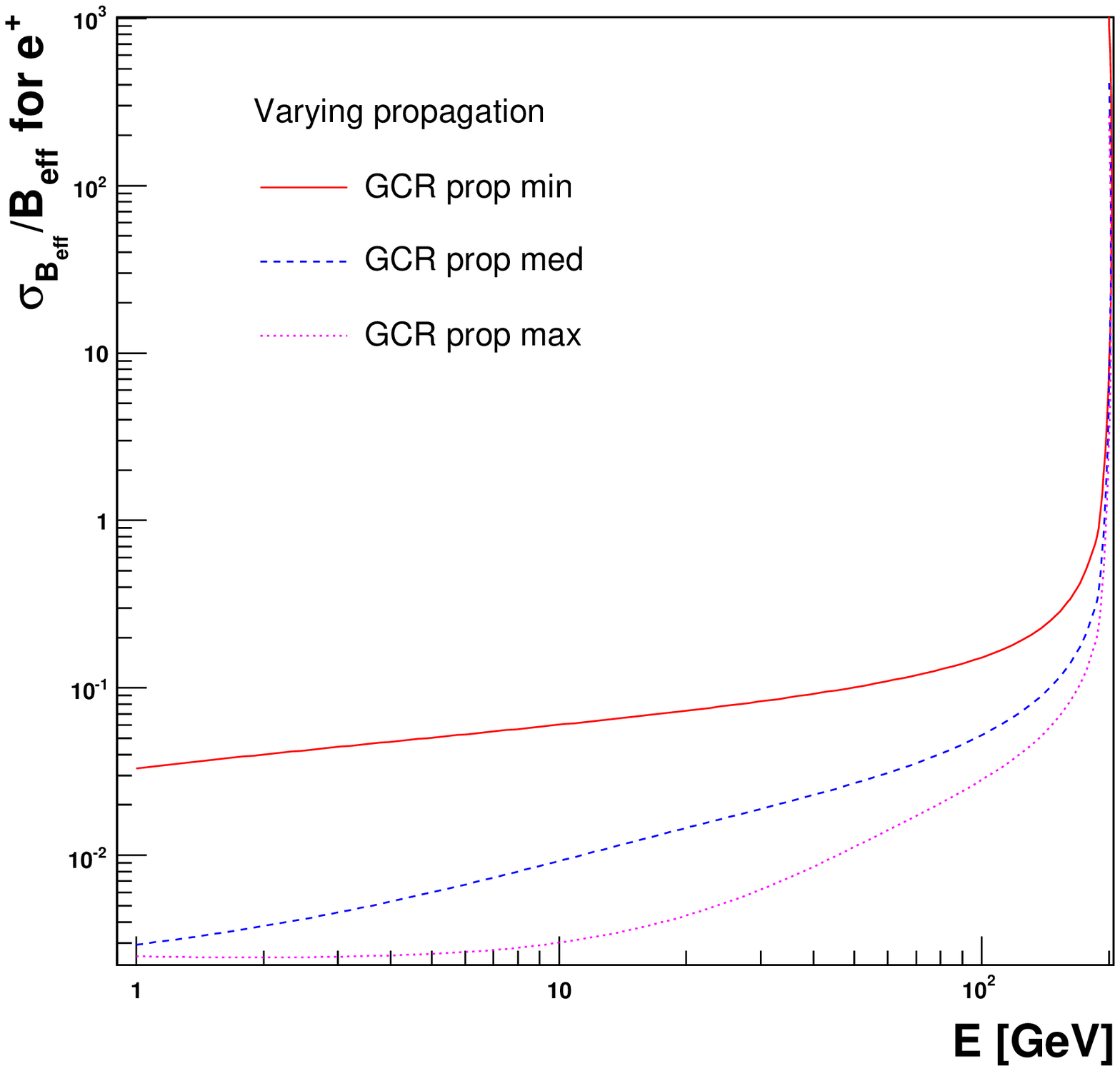}
\includegraphics[width=0.5\columnwidth, clip]
		{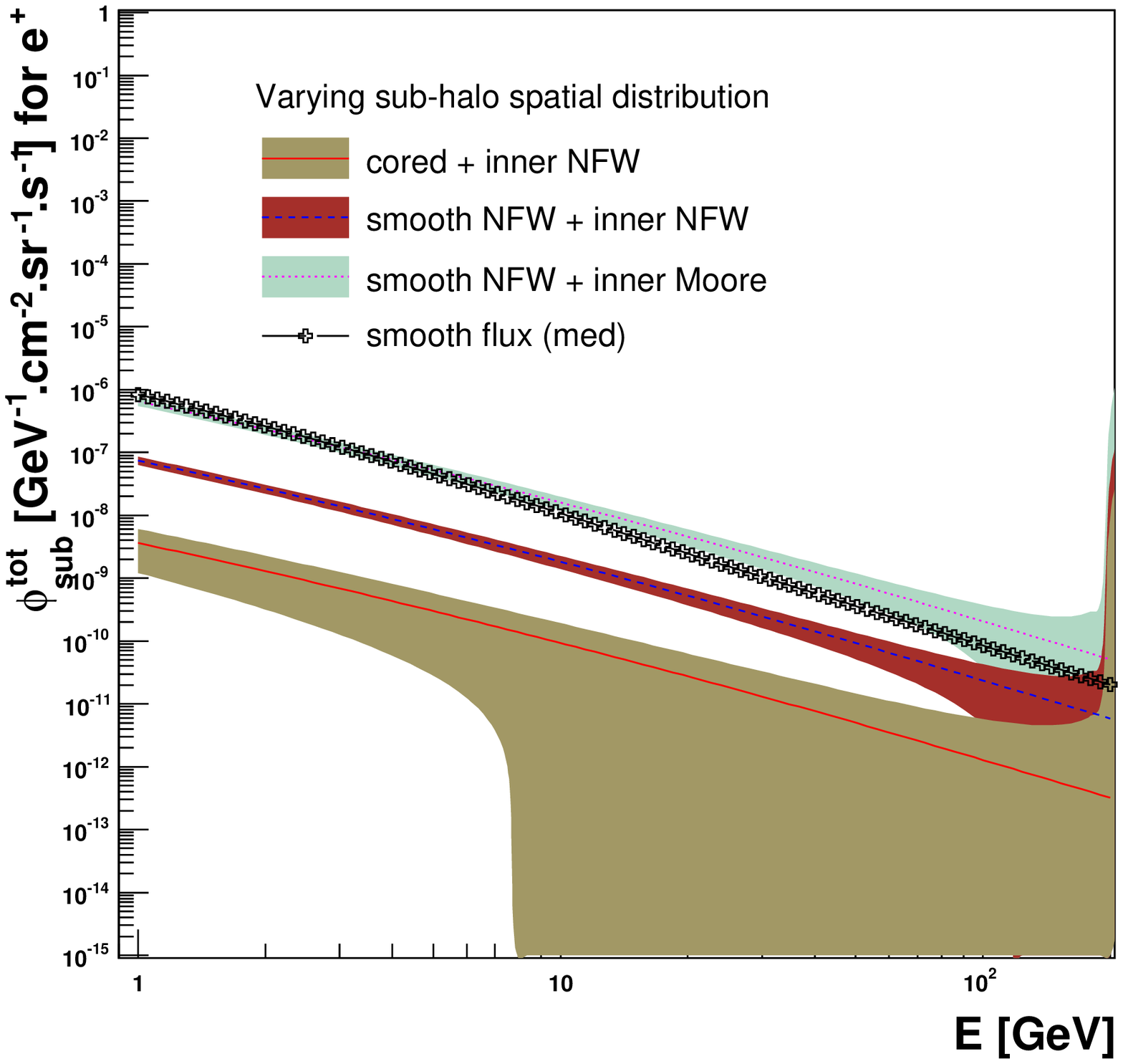}
\includegraphics[width=0.5\columnwidth, clip]
		{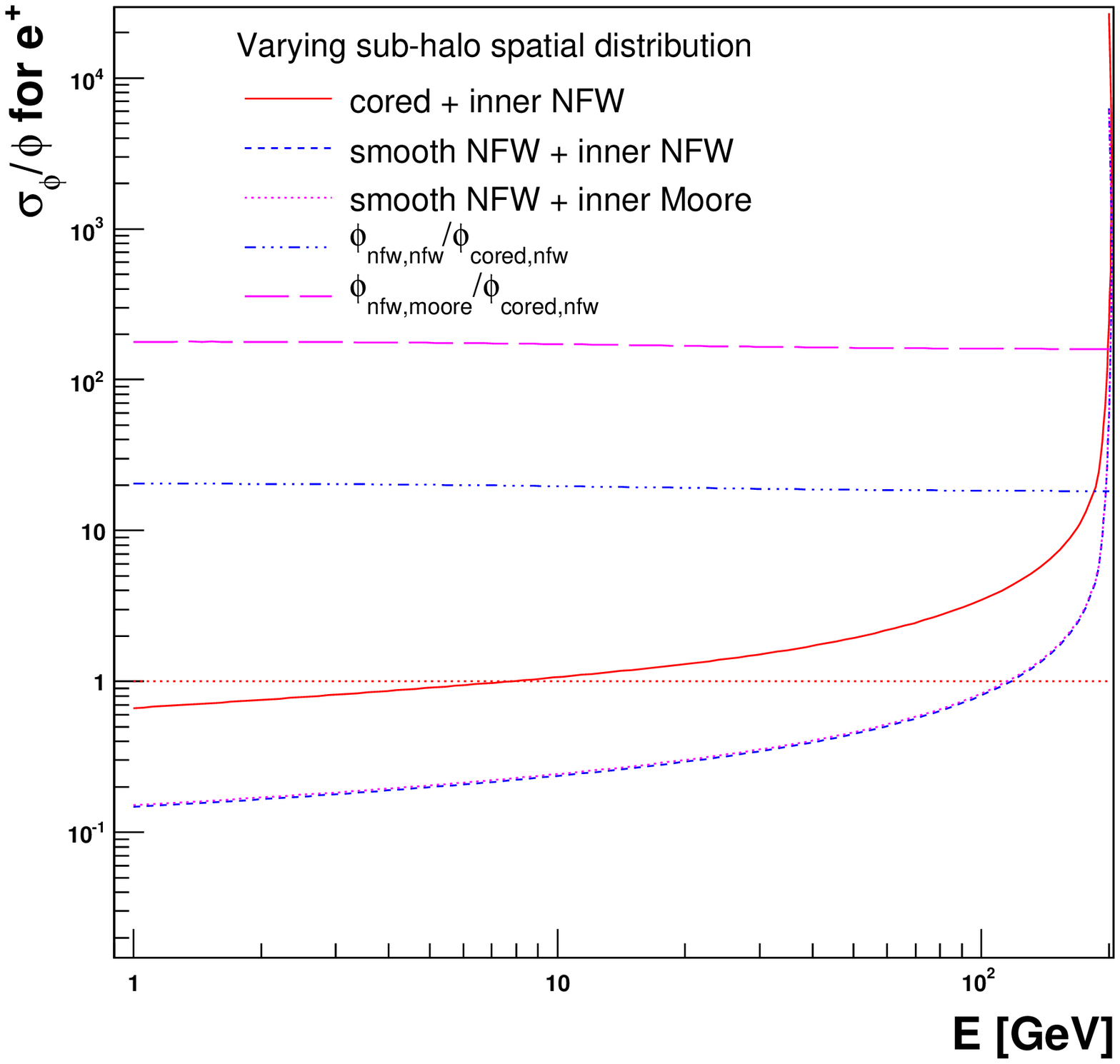}
\includegraphics[width=0.5\columnwidth, clip]
		{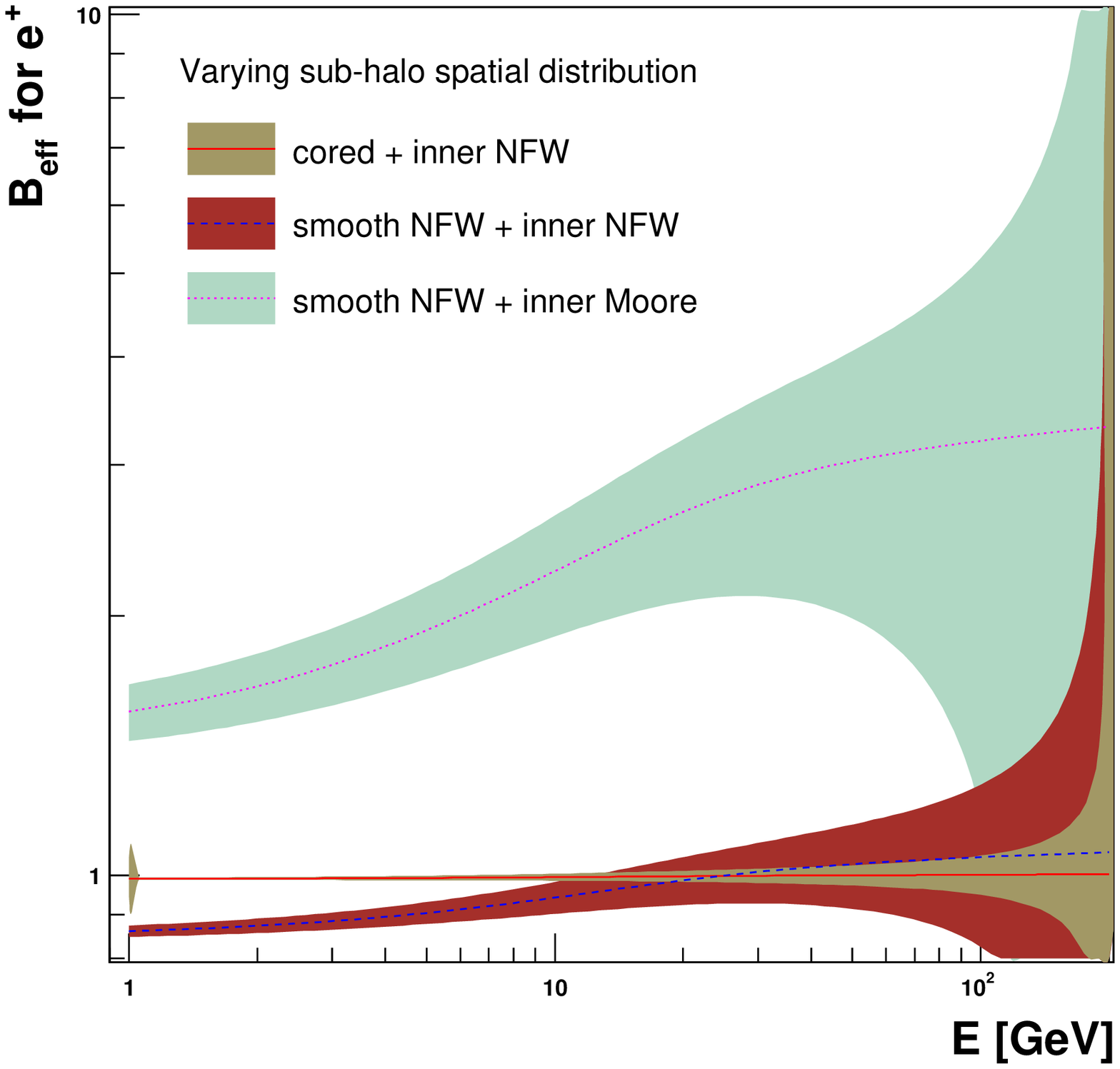}
\includegraphics[width=0.5\columnwidth, clip]
		{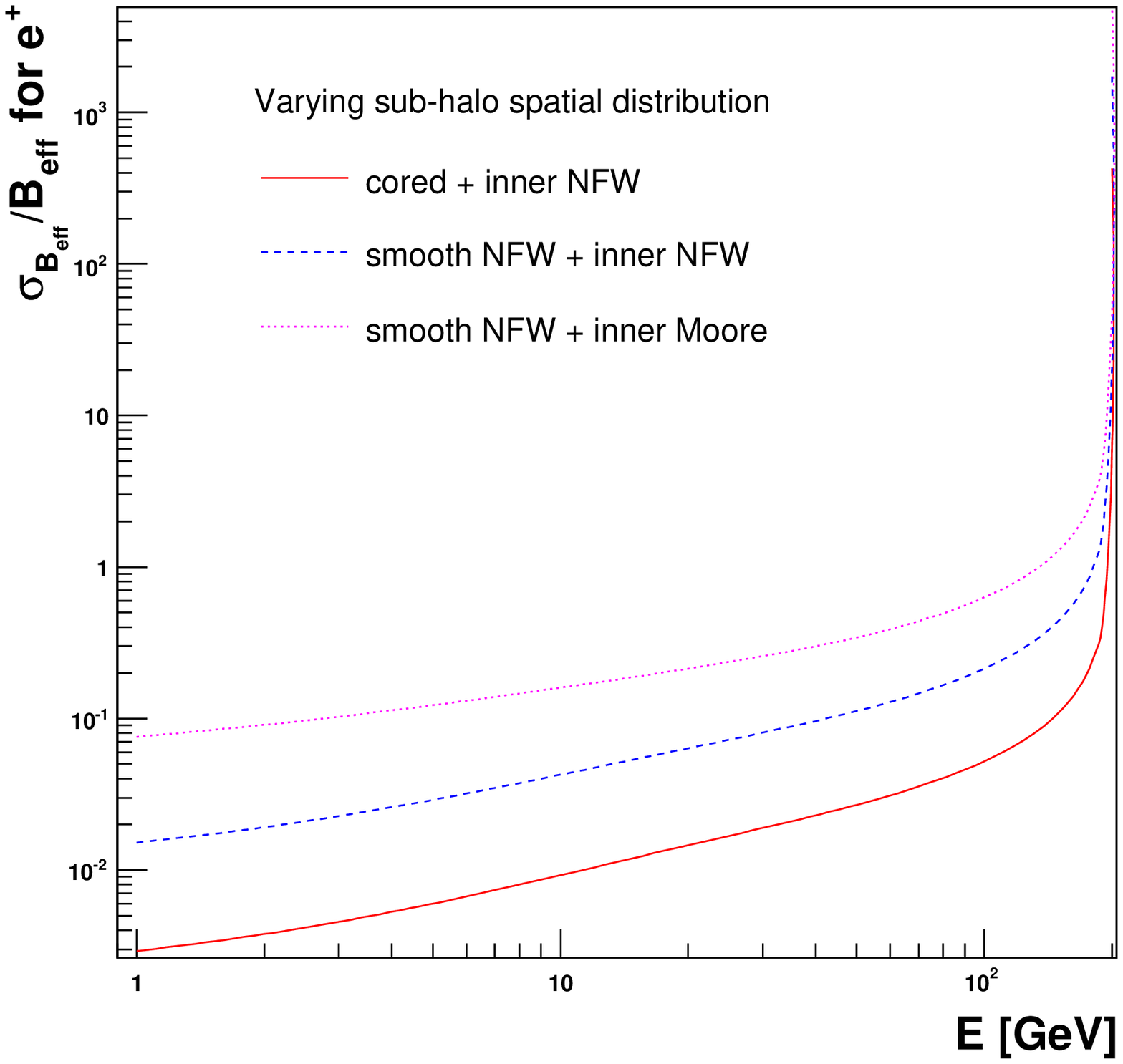}
\caption{\small \emph{Space-related effects on positrons, with the same panel 
    organisation as in Fig.~\ref{fig:mass-like_effects_pos}. First 
    row: Effects of varying the GCR propagation modelling, by using 
    the three propagation sets of parameters of Table~\ref{table:prop}. 
    Second row: Effect of varying the space distribution of sub-halos, going 
    from the cored isothermal space distribution to a case in which sub-halos 
    track the smooth NFW component (for completeness, we also take an example 
    where we also modify the inner sub-halo profile by taking a Moore instead 
    of a NFW).}}
\label{fig:space-like_effects_pos}
\end{center}
\end{figure*}

Figure~\ref{fig:mass-like_effects_pos} illustrates the mass-like effects, 
whereas Fig.\ref{fig:space-like_effects_pos} show the space-like ones.

\begin{itemize}
\item Mass-related effects
\end{itemize}
In the first line of Fig.~\ref{fig:mass-like_effects_pos}, we vary the minimal 
mass of the sub-halos, in other words the cut-off of the mass distribution. We 
actually compare three configurations by taking $\Mmin = 10^{-6}$, 1 and 
$10^{6}$ \Msol, the remaining parameters being those of the reference 
configuration given in Sect.~\ref{subsec:ref-conf}. The top left panel 
shows the whole contribution of sub-halos to the positron flux with 
the associated $1-\sigma$ contour, as well as the smooth contribution. We 
see that varying the minimal mass mainly influences the variance, while the 
mean values predicted for the flux remain of the same order of magnitude 
(only a factor of $\sim 5$ in flux between $\Mmin = 10^{-6}$ and $10^6\Msol$). 
The flux ratios of the three configurations are plotted in the second panel 
of the first line, taking the reference $\phi_{-6} = 
\phi(\Mmin = 10^{-6}\Msol)$. 
This is due to the fact that the product $\Ncl\times \langle \xi\rangle $ is 
almost independent of the minimal mass in this case. Should $\alpham$ have 
taken a value different than 1.9, the mean contribution of clumps would have 
a much stronger dependence on \Mmin. Nevertheless, the relative flux variance 
is different between the three configurations, as also shown in the bottom 
left panel. Actually, this comes from the total number of sub-halos, which is 
strongly depleted when \Mmin\ is increased ($\propto \Mmin^{1-\alpham}$). 
Therefore, the statistical flux variance, which scales like $1/\sqrt{\Ncl}$, 
is increased accordingly. Another effect comes from the energy loss of 
positrons. While the detected energy gets closer to the injected energy, the 
diffusion volume decreases, as explained in 
Sect.~\ref{subsubsec:space-like_pos}, and the number of clumps effectively 
contributing damps in the same way. Hence, the relative variance is enhanced 
when getting closer to the injected energy. However, the whole sub-halo 
contribution is finally far below the smooth flux, by about two orders of 
magnitude. This translates to an effective boost of $B_{eff}(E_d) \simeq 
(1-\fsol)^2 \sim 1$, with a very small variance, because it is also diluted by 
the smooth component.

In the second line panels of Fig.~\ref{fig:mass-like_effects_pos}, 
\alpham\ is varied, giving three different mass configurations: 1.8, 1.9 and 
2.0. As expected, the flux due to sub-halos is affected, and predictions 
slightly spread within one order of magnitude (a factor of $\sim 30$ between 
1.8 and 2.0). The relative flux variance is lower for large values of 
\alpham, as expected, because this 
increases the total number of sub-halos, more precisely the lighter ones. 
Nevertheless, varying \alpham\ within the reference configuration is not 
enough for sub-halos to strongly dominate over the smooth contribution: the 
averaged boost factors associated with the three examples lie around unity, 
even when getting closer to the injected energy (the maximum value is 1.08 
for $\alpham = 2.0$), with small statistical uncertainties.

The third line of Fig.~\ref{fig:mass-like_effects_pos} shows the consequences 
of varying the mass-concentration relation and the inner sub-halo profile. The 
reference model, which is inner NFW + B01, is compared with NFW+ENS01 (less 
concentrated sub-halos) and with Moore+B01 (more cuspy sub-halos). As 
expected, the flux obtained with the ENS01 concentration model is far below 
the reference one, by a factor of $\sim 20$, whereas the Moore sub-halos gives 
ten times more 
signal (this can also be seen from Fig.~\ref{fig:clump_parameters2}). These 
ratios are constant with the detected energy, as they are characterised by the 
ratios of $\langle \xi\rangle $'s. Again, we see that the expected boost 
factor is again negligible in all cases, around unity ($\sim 1.13$ for the 
best case, \ie~Moore inner profile + B01). Nevertheless, the increase of the 
variance associated with the latter happens at lower energies than previously, 
because the probability for a single clump to contribute more than the smooth 
component becomes sizable at farther distances.

\begin{itemize}
\item Space-related effects
\end{itemize}
The first line panels of Fig.~\ref{fig:space-like_effects_pos} illustrate 
how GCR propagation can strongly influence the predictions by using the three 
models of Table~\ref{table:prop} with the reference DM configuration. 
Differences grow when the detected energy is far below the injected one. This 
is mainly due to the change in the thickness of the diffusive slab. At 
energies close to the injected energy, the volume probed is very small, so 
propagation is not sensitive to the slab boundaries anymore. We see that for 
positrons, there is no huge differences between the maximal and medium 
propagation models, while the minimal one strongly depletes the positron flux 
at low energies. Indeed, the characteristic propagation length for positrons 
(a few kpc) is almost always contained in the maximal and medium slabs, 
whereas it is not the case for the minimal one. Anyhow, even the maximal 
propagation set is not enough to boost the sub-halo positron flux above the 
smooth contribution. Indeed, the values obtained (see the left panels) stack 
to unity, again with a small statistical uncertainty.

Finally, the second line panels of Fig.~\ref{fig:space-like_effects_pos} 
show the effect of changing the spatial distribution of clumps, from the 
reference cored isothermal to a situation in which they track the smooth NFW 
profile. For completeness, we do the exercise for both inner NFW and Moore 
profiles. In the left panels, we see that the sub-halo flux is enhanced when 
they track the smooth profile, of about one order of magnitude in this case. 
The effect is obviously stronger when an inner Moore density is taken, for 
which another order of magnitude arises. Nevertheless, the boost factors do 
not obey the same hierarchy. This is due to the way the smooth component is 
normalised when clumps are added. Indeed, we chose to readjust the smooth 
density by a factor $(1-\fsol)$, where the fraction density \fsol\ is defined 
in Eq.~(\ref{eq:fsol_def}), in order to get a constant local density 
\rhosol. When clumps track the smooth component, their number density is 
enhanced in the local neighbourhood, so that \fsol\ is enhanced accordingly. 
As the boost factor remains around $(1-\fsol)^2$ (the sub-halo contribution 
is negligible), except when considering inner Moore profiles, the case for 
which clumps track the smooth component is worse. However, taking an inner 
Moore profile for clumps gives a higher flux than the smooth alone, and 
the mean boost factor can reach an asymptotic value of $\sim 2-3$.

In summary, we have shown, by extensively playing with the maximum number of 
available parameters, that sizable boost factors to the positron flux are 
unlikely to arise from clumpiness. There could be situations in which a 
single sub-halo would be close enough to the Earth to dominate over the smooth 
component, but within the most reasonable modellings, the probability for 
this to happen is vanishingly small. Nevertheless, in order to provide more 
optimistic scenarios, even if less realistic according to the standard 
values of the parameters, a model characterised by $10^{-6}\Msol$ sub-halos 
with inner Moore profiles, with concentrations described by the B01 relation, 
spatially tracking the smooth DM density, would yield a mean boost factor 
whose asymptotic value would be around 3 for a logarithmic slope \alpham = 1.9 
(see lower panels of Fig.~\ref{fig:mass-like_effects_pos}). Taking \alpham = 
2 leads to a boost of $\sim 20$ (see Fig.~\ref{fig:extreme_boosts}). This is 
the most optimistic estimate that we can provide so far, but also the most 
unrealistic. Note finally that although the primary fluxes may vary by 2 
orders of magnitude due to uncertainties in the propagation parameters 
(see Table~\ref{table:prop} and upper panels 
of Fig.~\ref{fig:space-like_effects_pos}), 
the resulting average boost factors are unaffected because they are defined 
as flux ratios; the variance is nevertheless larger when the GCR horizon is 
reduced (\emph{min} configuration).

%######%
\subsubsection{Comments on the positron excess}
The HEAT experiment results for the 1994 flight hinted at the possibility of 
an excess of positrons near 8 GeV~\citep{1997ApJ...482L.191B}, which could not
be explained by a purely secondary production 
mechanism~\citep{1999APh....11..429C}. \citet{1999PhRvD..59b3511B} then found 
that neutralino annihilation could account for the missing flux providing that 
boost factors are larger than six; at that time, these authors estimate 
realistic boost factors to fall in the range $B_{\rm eff}\lesssim 100-1000$.
Note that such high boost factors would be ruled out in the present study.
However, later on, combining both 1994 and 1995 HEAT balloon flights,
 \citet{2001ApJ...559..296D} concluded that the positrons flux
was consistent with a secondary origin.  Results
from the MASS91 balloon-borne magnetic spectrometer above 7 GeV
\citep{2002A&A...392..287G} do not provide a definitive answer either.
As emphasised by these authors, very high energy $\sim 100$~GeV 
measurements are probably necessary to positively conclude for
standard or exotic mechanisms. Finally, from the most recent data
coming from the HEAT 2000 flight, \citet{2004PhRvL..93x1102B} cautiously
conclude that a primary contribution above a few GeV can still not
be ruled out.

Given these observations, several subsequent studies
have focused on finding a good DM candidate to
explain this possible excess. We do not wish to comment here on the
best candidate, but rather survey the boost factors used in the studies.
For example, for SUSY candidates, boost factors of 2.7 and 3.9 were used in 
\citet{2002PhRvD..65e7701K}, values in the range 30-100 in 
\citet{2002PhRvD..65f3511B}, from {\em small} to {\em large} boost factors in 
\citet{2002PhLB..536..263K}, in the range 1-5 in \citet{2006PhRvD..73e5004H}, 
and around 100 in \citet{2007MNRAS.374..455C}. \citet{2006JCAP...12..003M} 
favoured boosts of 5-10 to accommodate the expected measurements of PAMELA, 
for SUSY models with non-universal scalar and gaugino masses. For KK DM, 
\citet{2004PhRvD..70k5004H} found a range of 10-30. The boost factor used to 
fit the data depends of course on the WIMP candidate considered and its mass.

It appears that most of the models found so far to match the positron data 
require mild to significant boost factors. Such boost factors are disfavoured 
by our results if the clump parameters fall in the large ranges taken in this 
study. A high energy feature in the positrons data could still be an important 
clue to DM indirect detection, and it would be interesting in forthcoming 
studies to scan, e.g. the SUSY parameter space looking for models matching 
the data without boost factors. According to \citet{2007PhRvD..75f3506A},
the little Higgs model provides good options for detectability by the AMS-02
experiment, but could be short for PAMELA. In addition, it is worth noting 
that if some $\chi^2$-like searches for clump signatures are performed in the 
coming positron data, it will be very important to take the energy dependence 
of any boost factor into account, as soon as it is invoked.

To conclude, although having no boost factors may be less interesting
for SUSY theories to explain the data, any result that will be obtained 
when comparing to forthcoming data, if an excess is confirmed, will be more 
robust if no boost factor (an additional unknown parameter till now) is 
invoked. We recall that the two main uncertainties for WIMP annihilation 
induced antimatter signals are the propagation parameters in the Galaxy 
(a factor $\lesssim 100$) and the local DM density 
(a factor $\lesssim 2$, that shifts to 4 in terms of annihilation rate).

%---------------%
\subsection{Anti-protons}
\label{subsec:pbar}

First of all, it is worth quoting that contrary to positrons, for 
which an excess is still not understood, anti-proton present measurements are 
now well accounted for by purely \emph{standard} secondary production (e.g. 
\citealt{2001ApJ...563..172D}). This means that there is no need of DM, and 
obviously of any clump to fit the data: the present data has to be consider 
as an upper limit for the DM contribution. Things could change 
with the future results of PAMELA and AMS-02 at higher energies.

The flux enhancement for anti-protons has features different from positrons, 
as already stressed by \citet{2007PhRvD..75h3006B} and 
\citet{2007PhRvD..76h3506B}. 
This is mainly due to propagation, which is quite different from the positron 
case. Indeed, anti-protons do not lose energy, and can experience spallation 
processes and wind convection along their travel to the Earth, which 
occurs to be dominant at low energy. Nevertheless, as for positrons, the 
same classes of physical effects can be discussed. 

Regarding the space-related effects, the comments are the same as for 
positrons but with a reversal energy point of view (see 
App.~\ref{app:space-like_effects_pbar}). The relative variance is then maximal 
at low anti-proton energy. Note, however, that the three sets of 
propagation parameters give separate absolute fluxes (decreasingly according 
to \emph{max}, \emph{med} and \emph{min}), whereas \emph{max} and \emph{med} 
configurations give about the same fluxes for positrons.

%######%
\subsubsection{Overall effect on the anti-proton flux: boost factor estimate}

\begin{figure*}[t!]
\begin{center}
\includegraphics[width=0.5\columnwidth, clip]
		{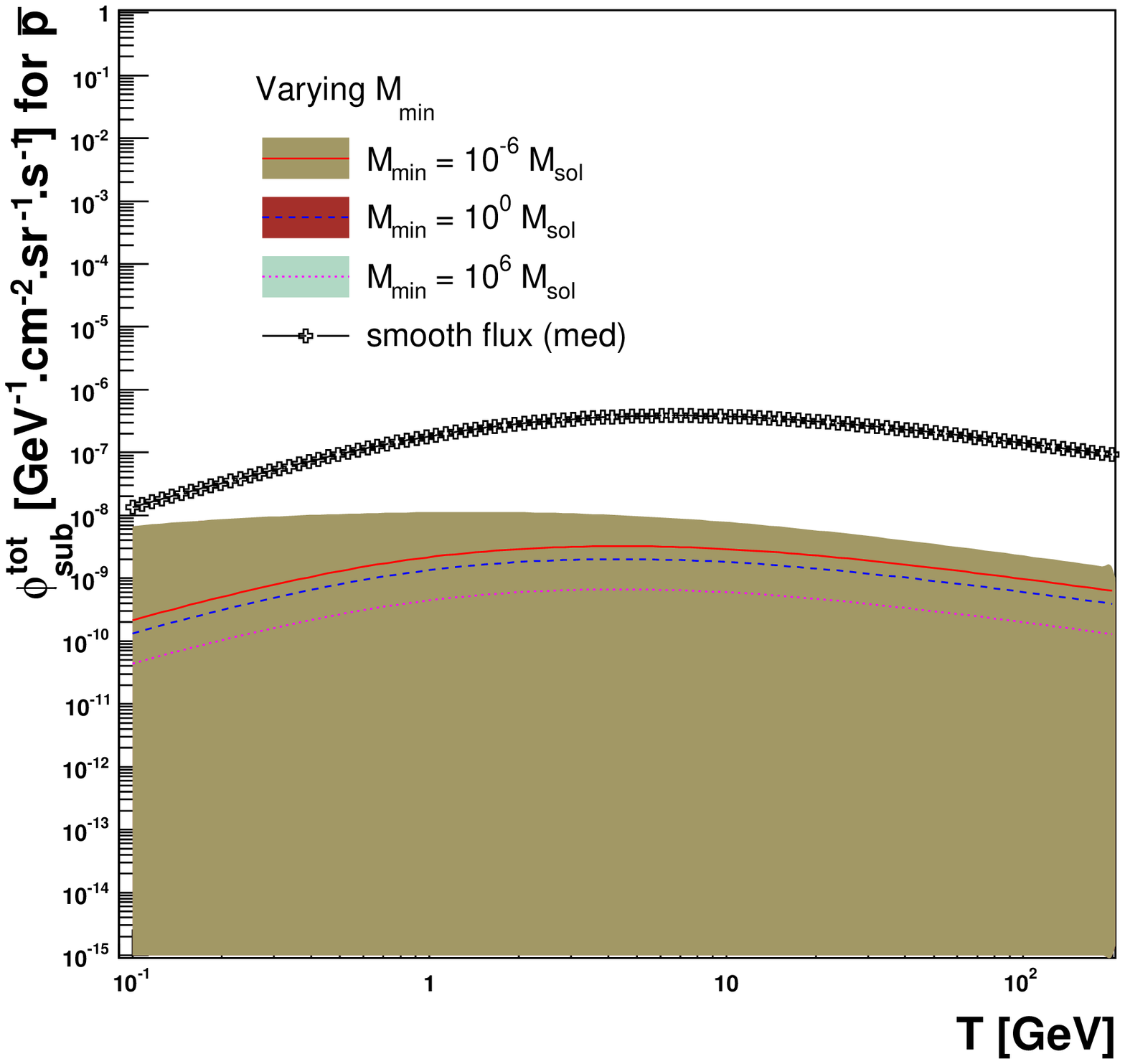}
\includegraphics[width=0.5\columnwidth, clip]
		{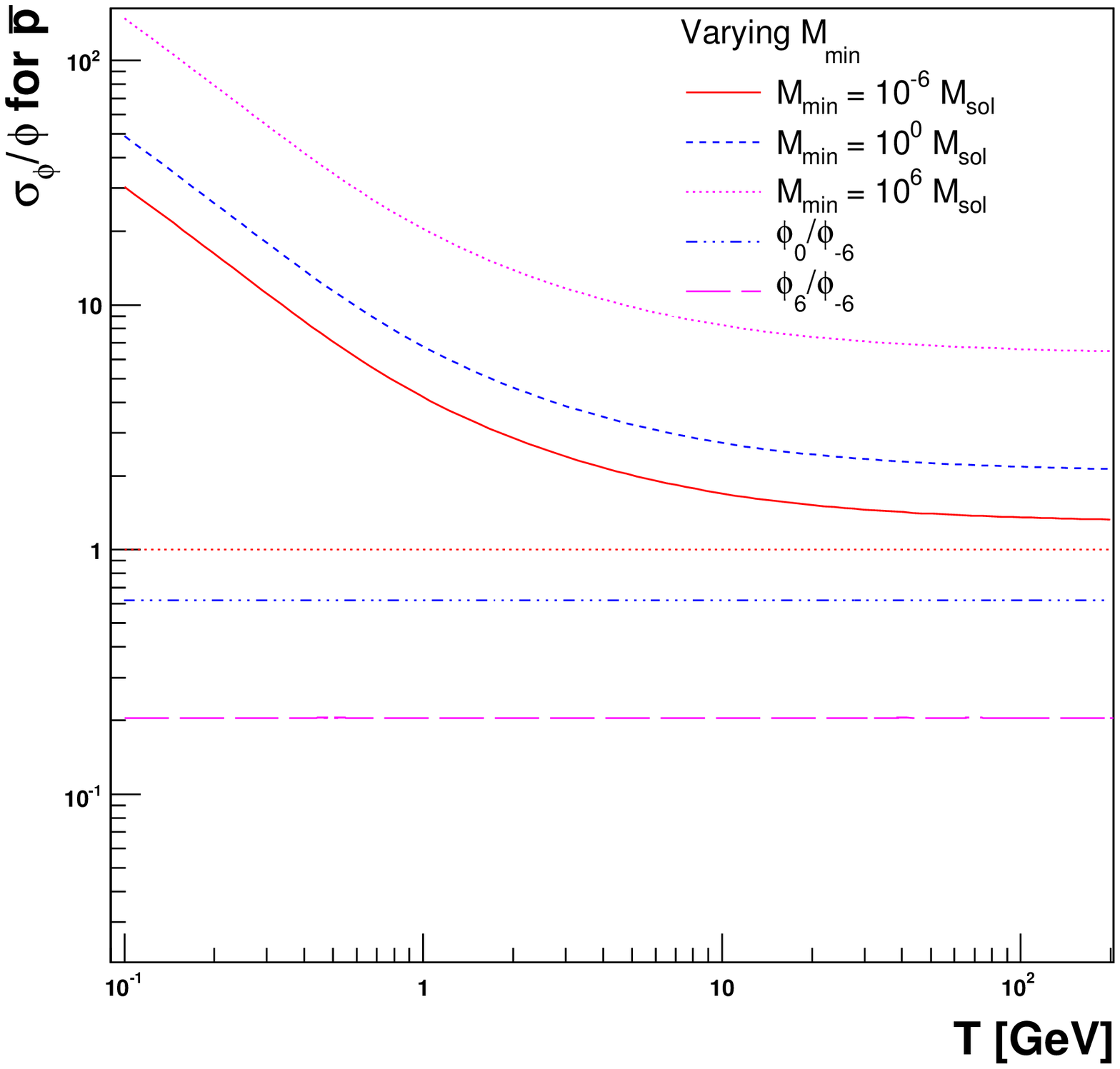}
\includegraphics[width=0.5\columnwidth, clip]
		{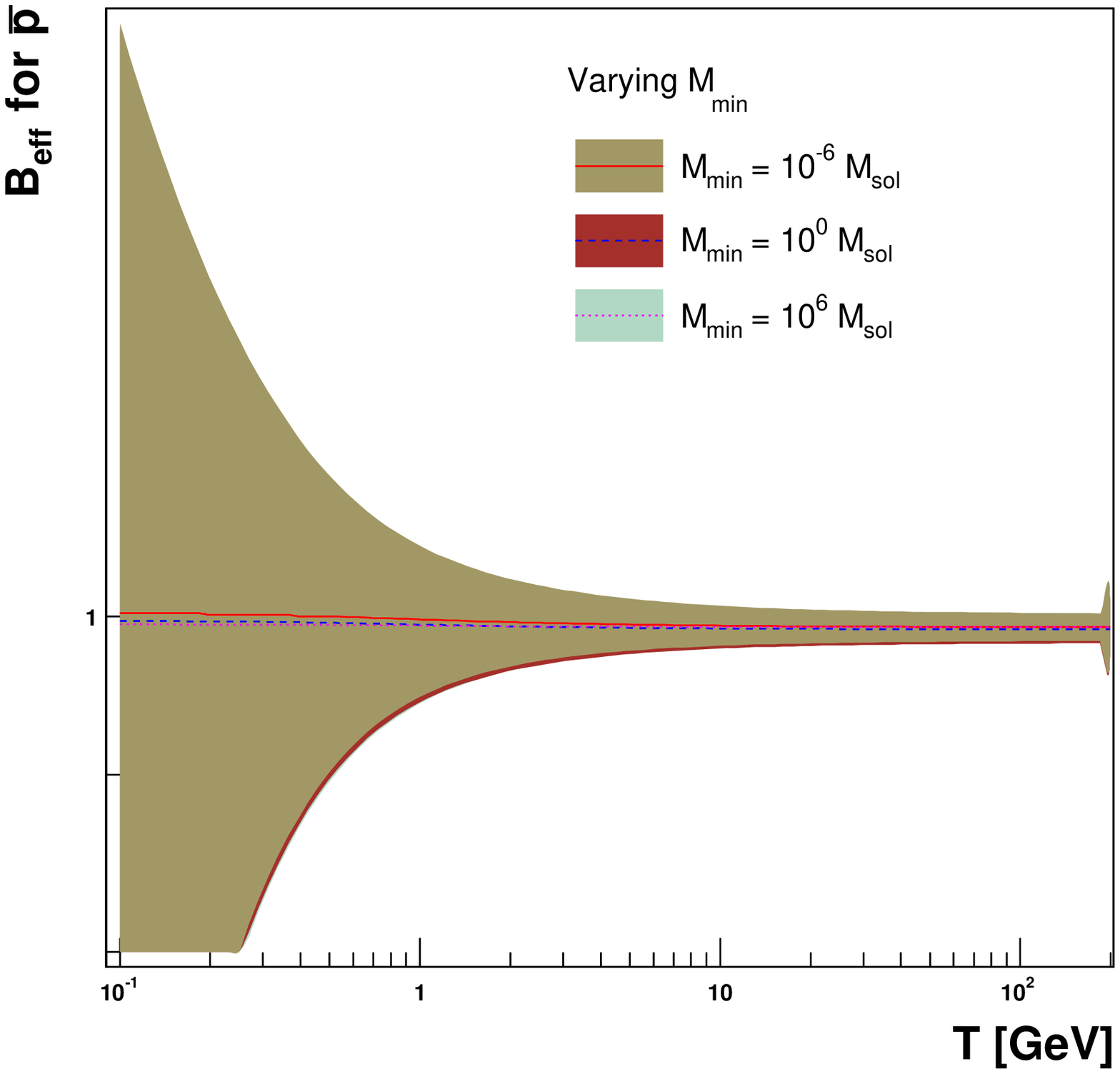}
\includegraphics[width=0.5\columnwidth, clip]
		{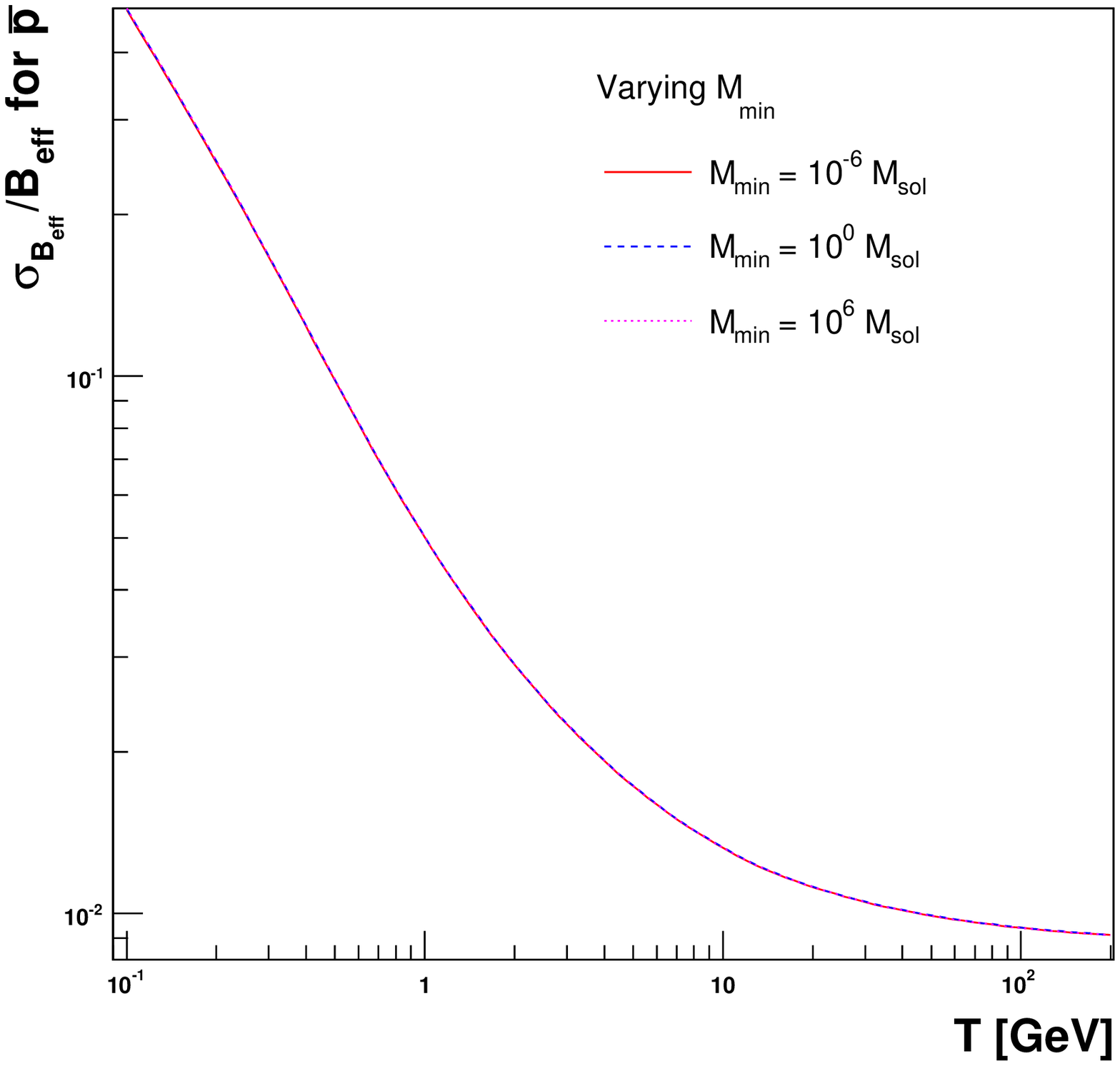}
\includegraphics[width=0.5\columnwidth, clip]
		{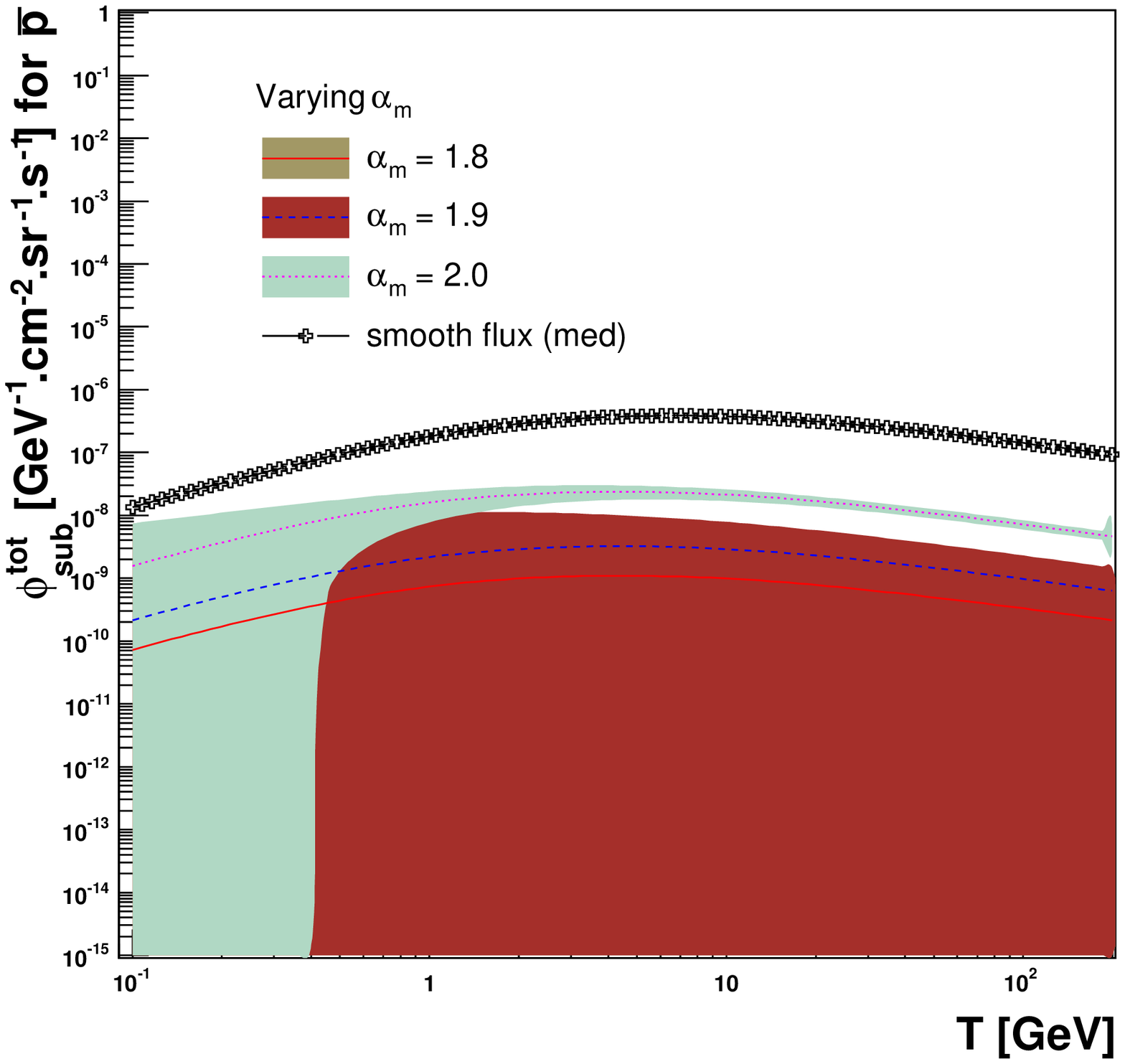}
\includegraphics[width=0.5\columnwidth, clip]
		{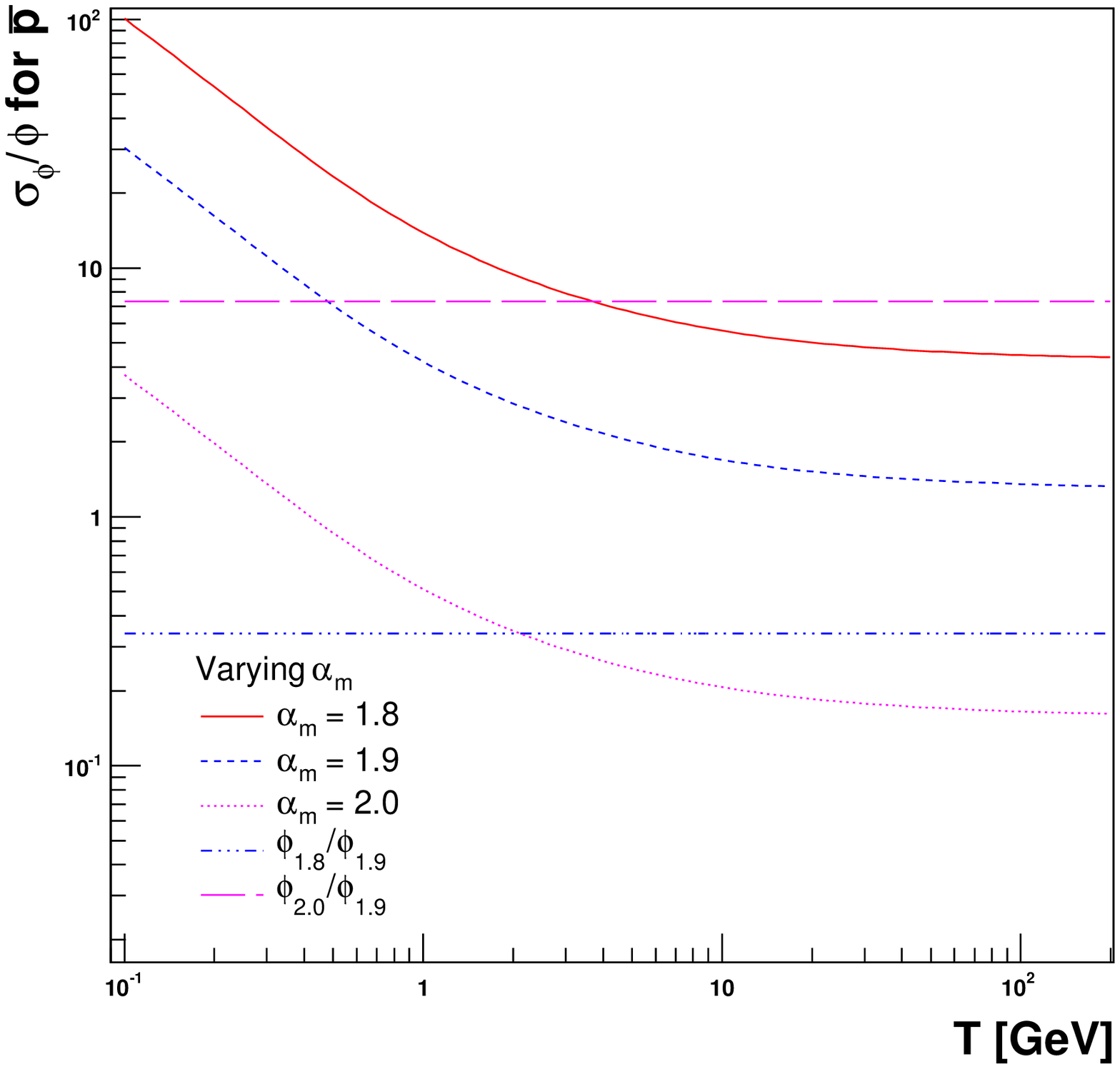}
\includegraphics[width=0.5\columnwidth, clip]
		{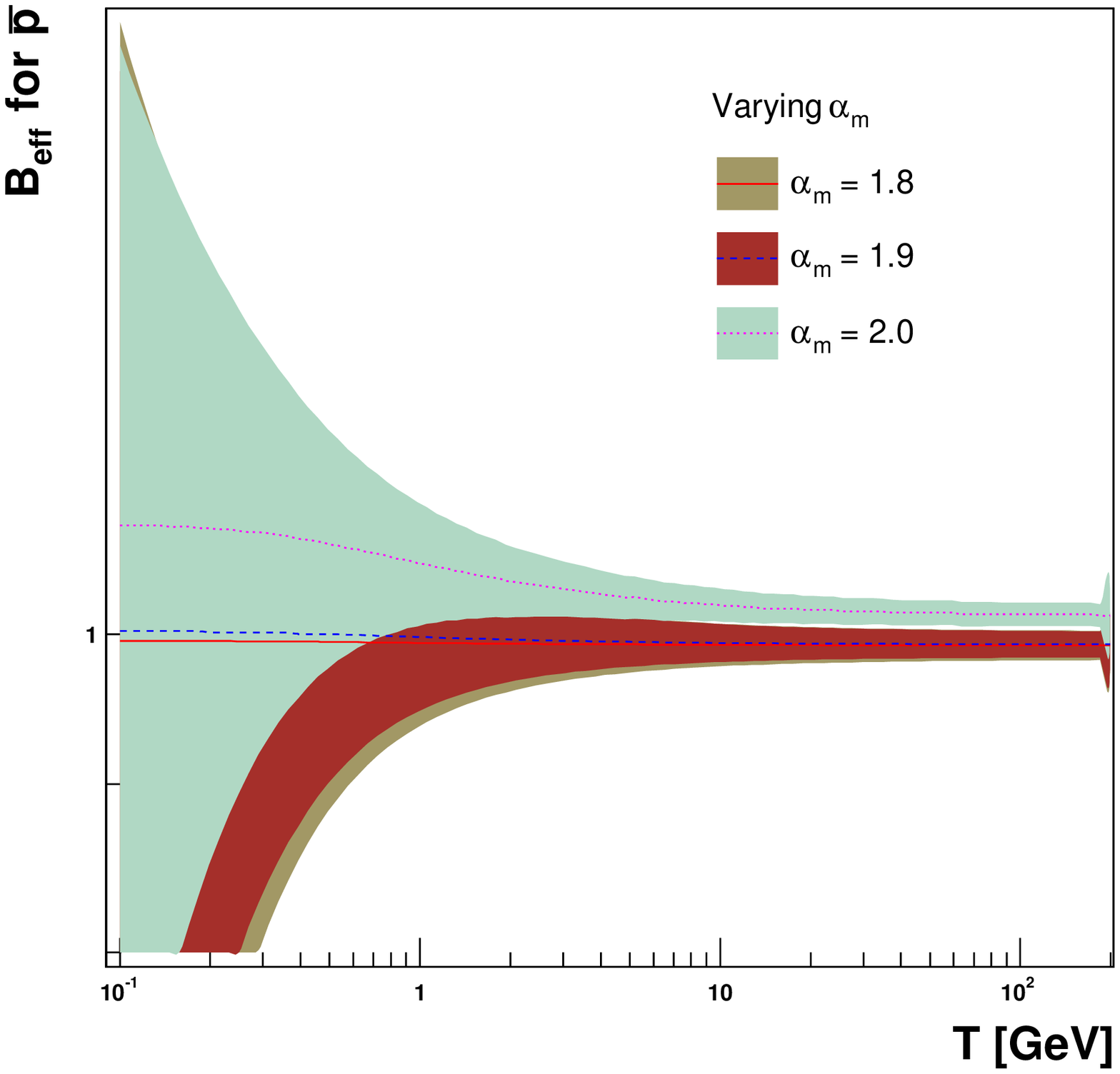}
\includegraphics[width=0.5\columnwidth, clip]
		{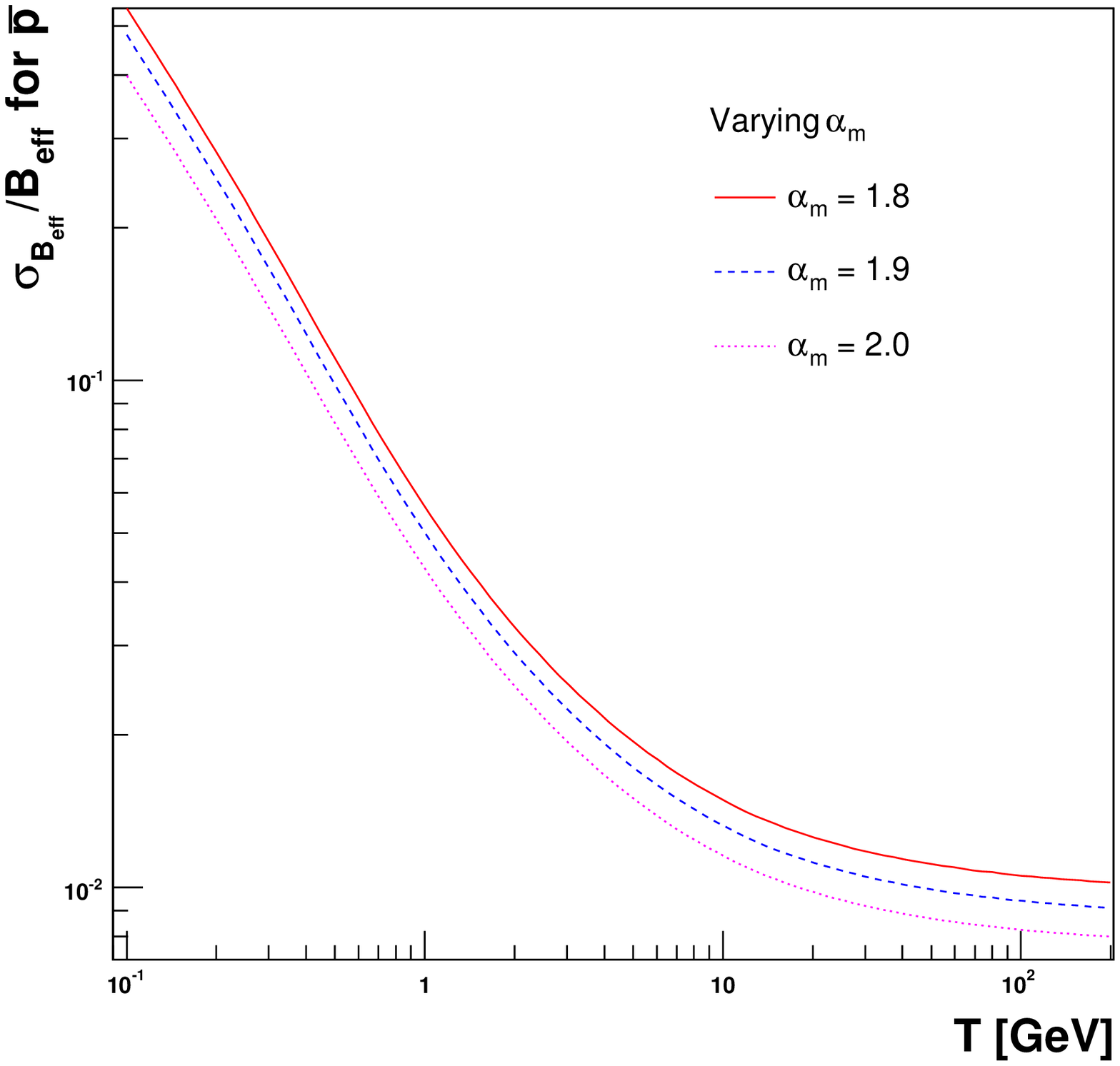}
\includegraphics[width=0.5\columnwidth, clip]
		{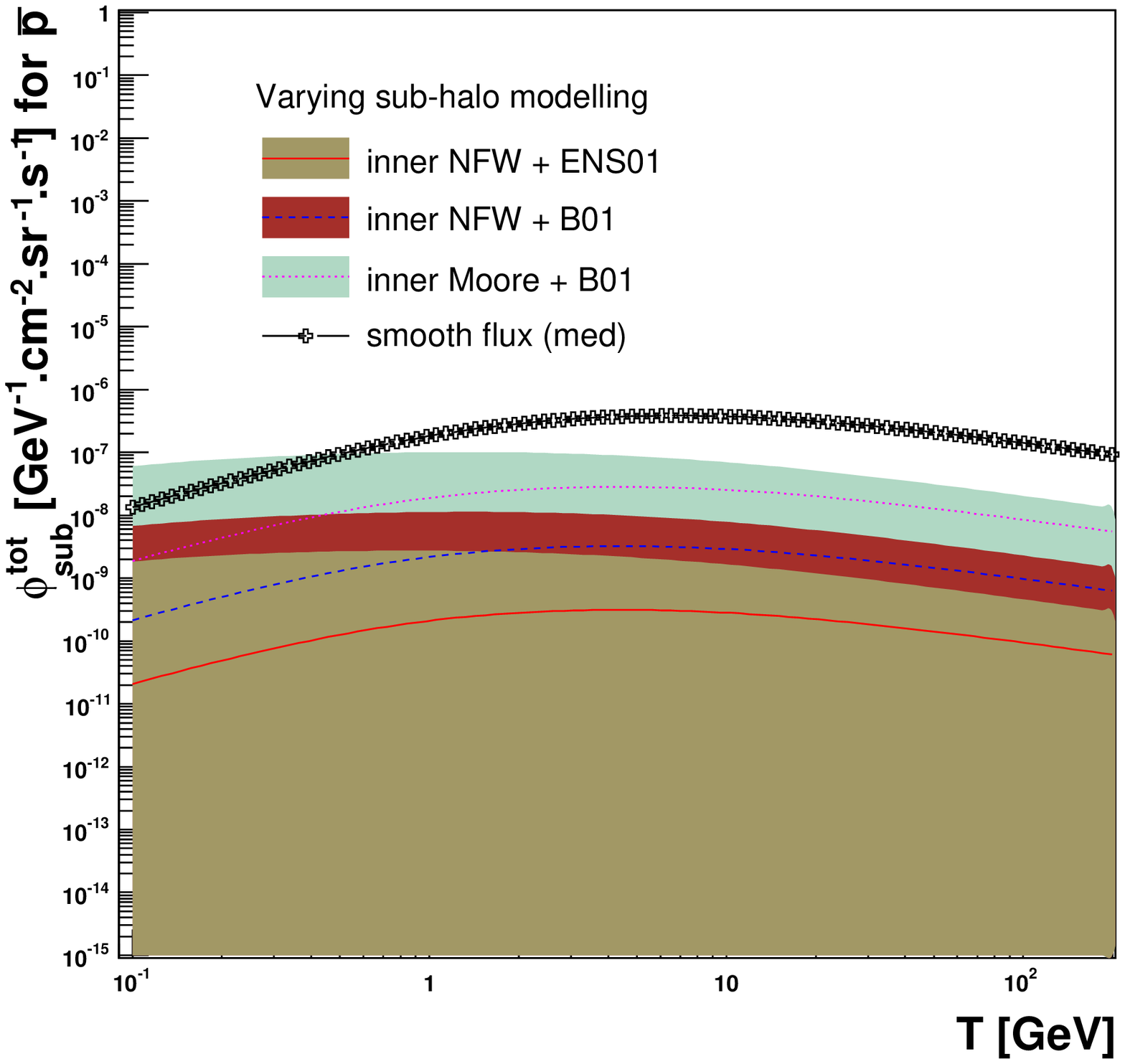}
\includegraphics[width=0.5\columnwidth, clip]
		{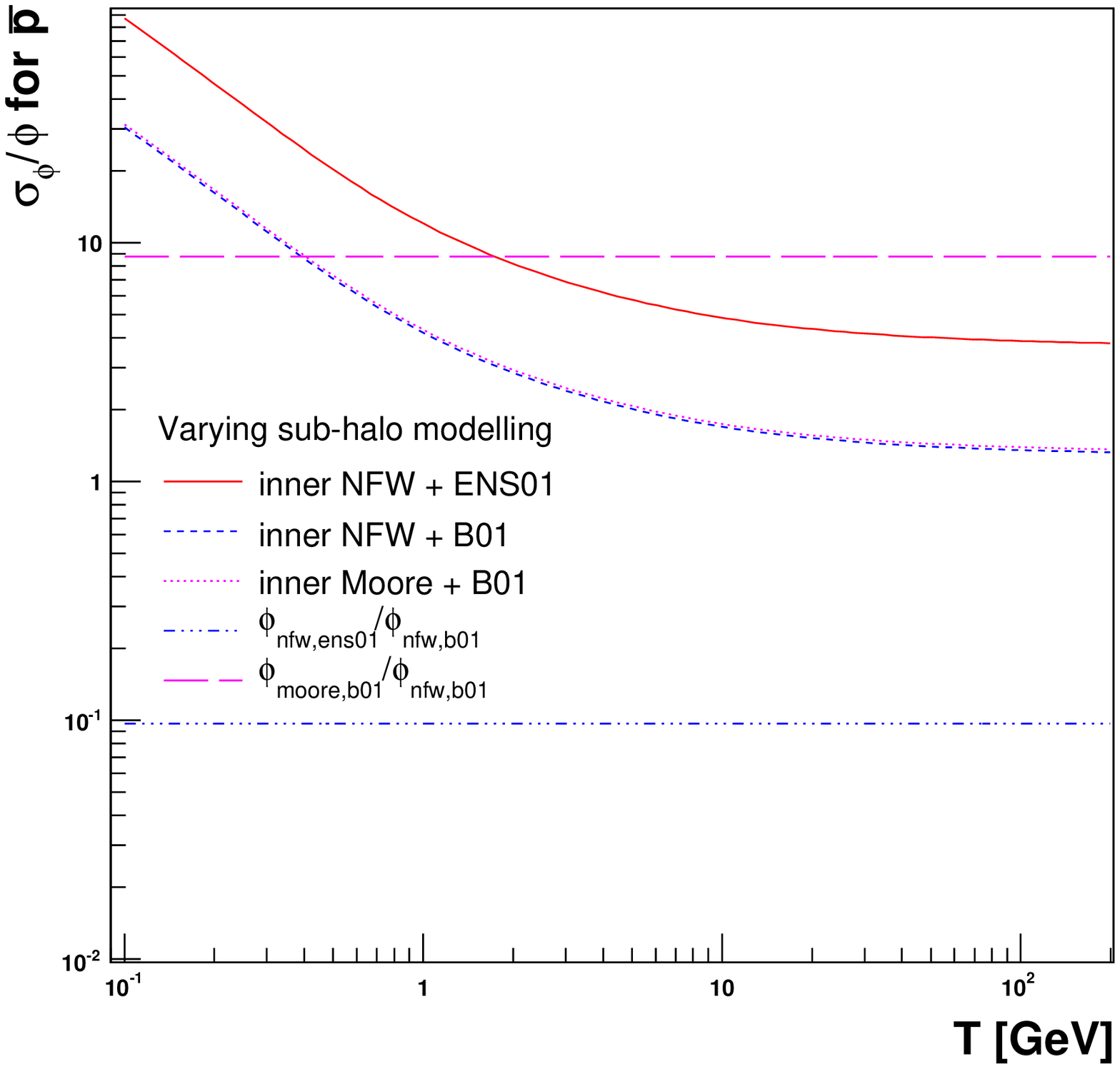}
\includegraphics[width=0.5\columnwidth, clip]
		{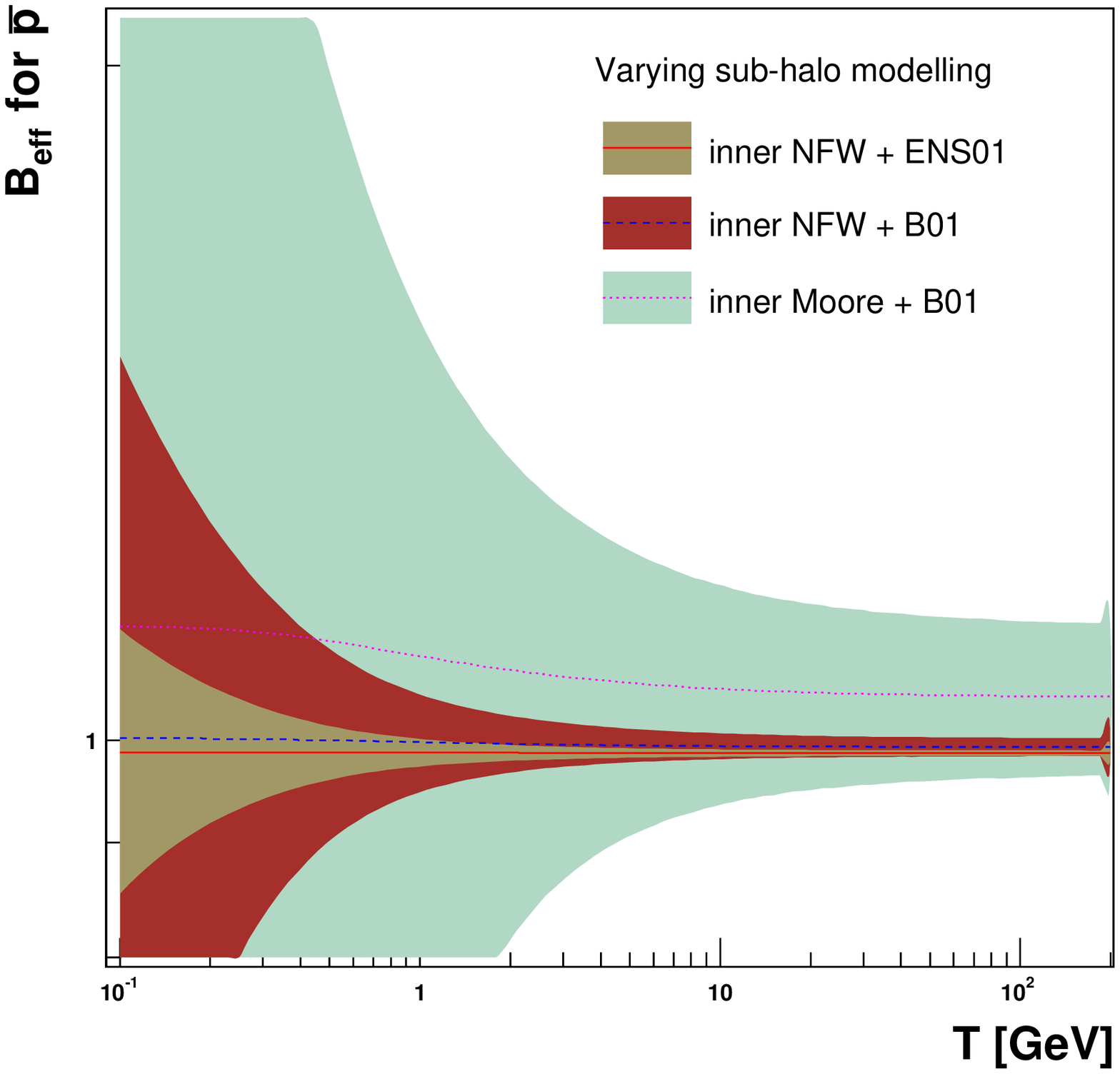}
\includegraphics[width=0.5\columnwidth, clip]
		{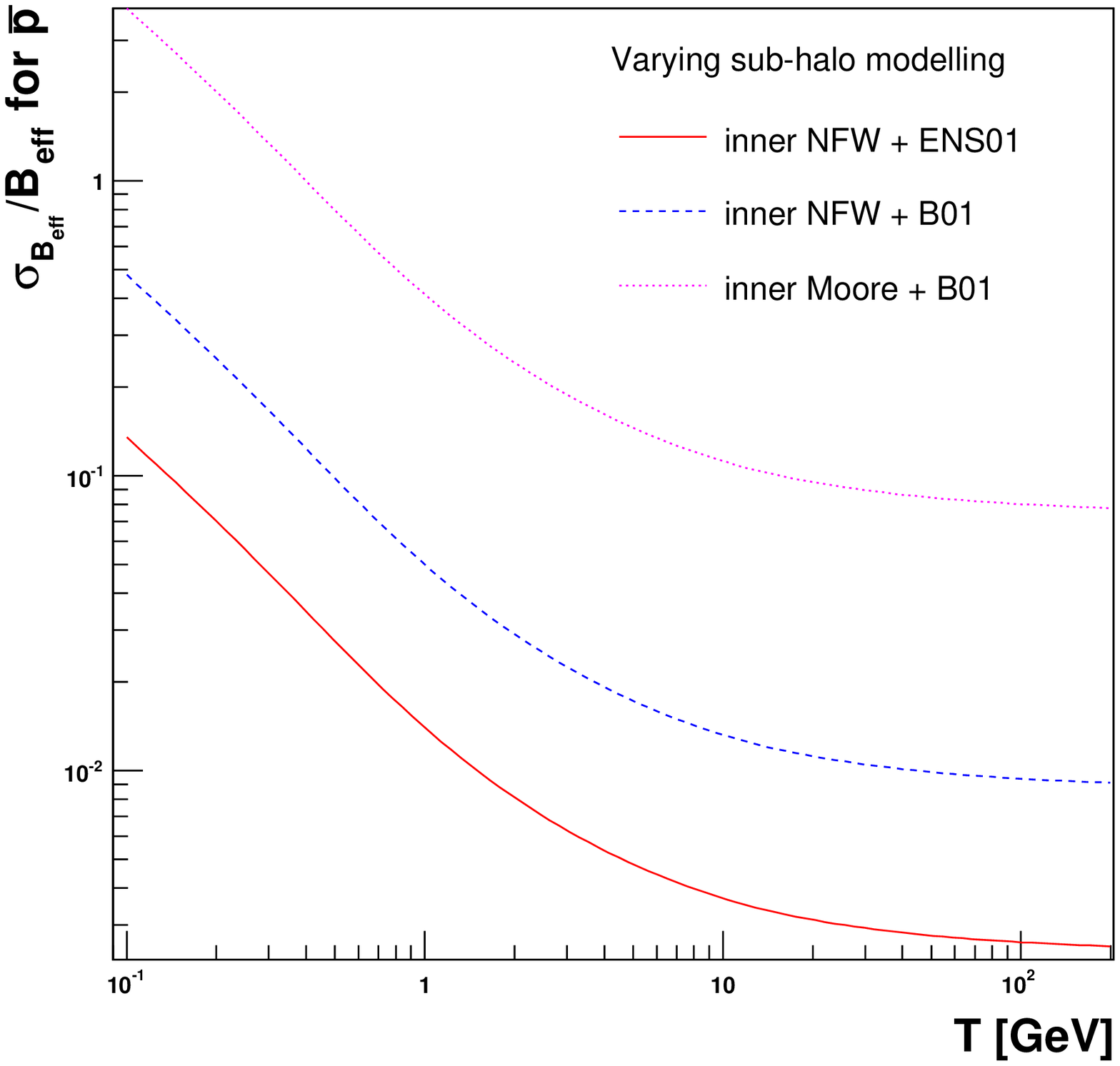}
\caption{\small \emph{Mass-related effects for anti-protons ; a toy flat 
    spectrum is injected at the sources, at a rate corresponding to 
    standard values of WIMP annihilation | same panel organisation 
    as for positrons (see Fig.~\ref{fig:mass-like_effects_pos}). First 
    row: Varying \Mmin. Second row: Varying \alpham. Third row: Varying 
    the sub-halo inner properties.}}
\label{fig:mass-like_effects_pbar}
\end{center}
\end{figure*}

\begin{figure*}[t!]
\begin{center}
\includegraphics[width=0.5\columnwidth, clip]
		{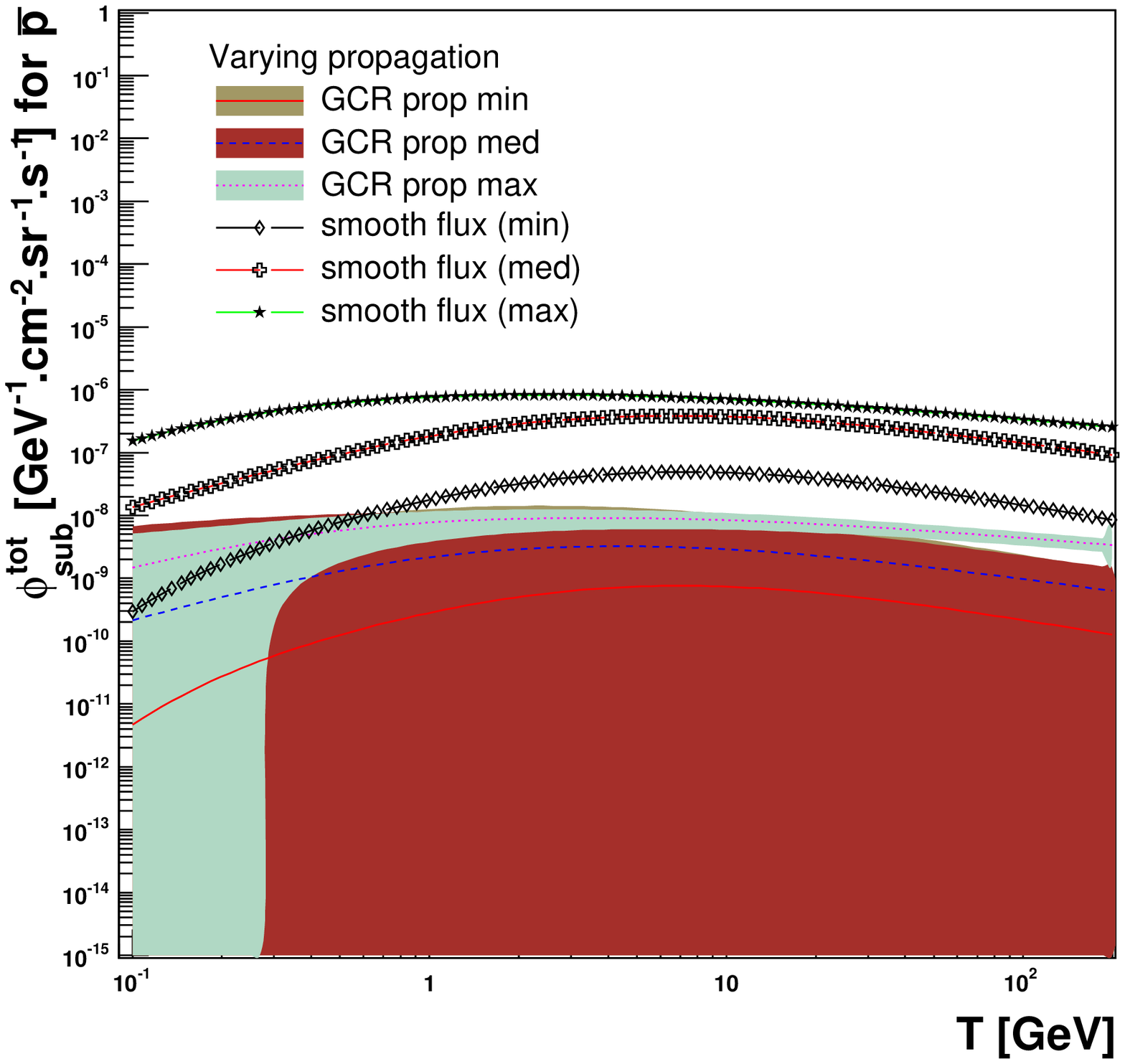}
\includegraphics[width=0.5\columnwidth, clip]
		{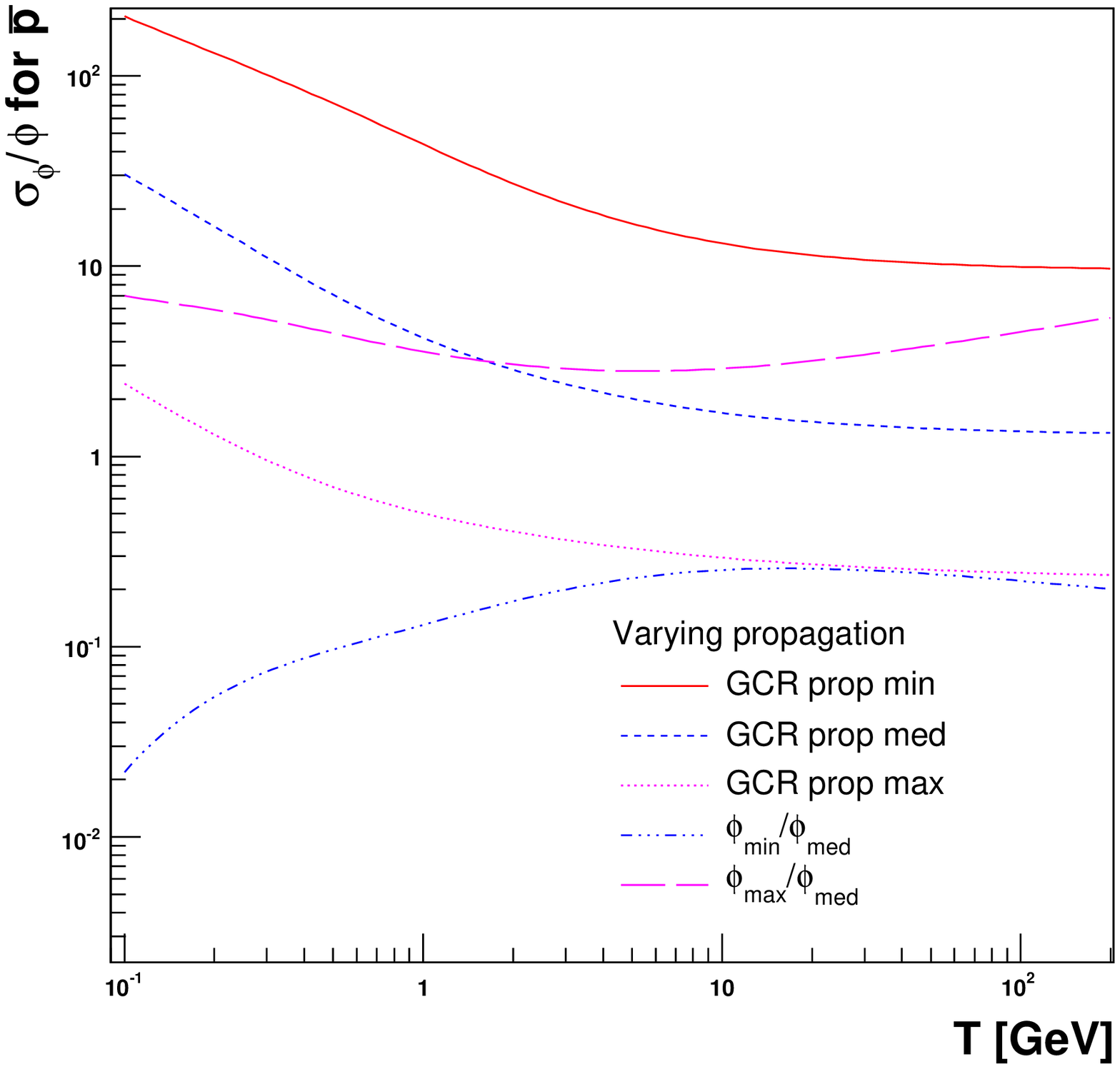}
\includegraphics[width=0.5\columnwidth, clip]
		{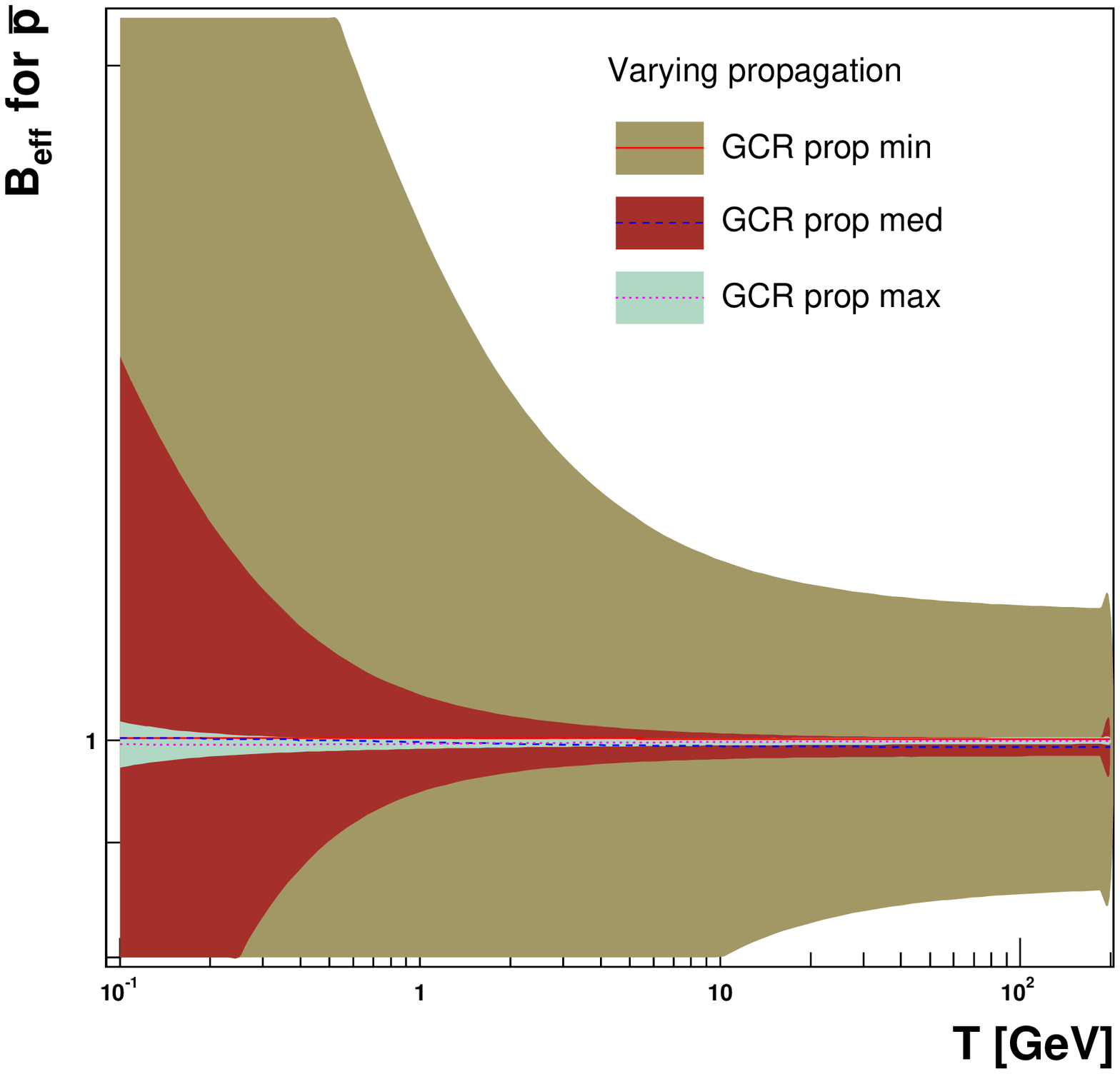}
\includegraphics[width=0.5\columnwidth, clip]
		{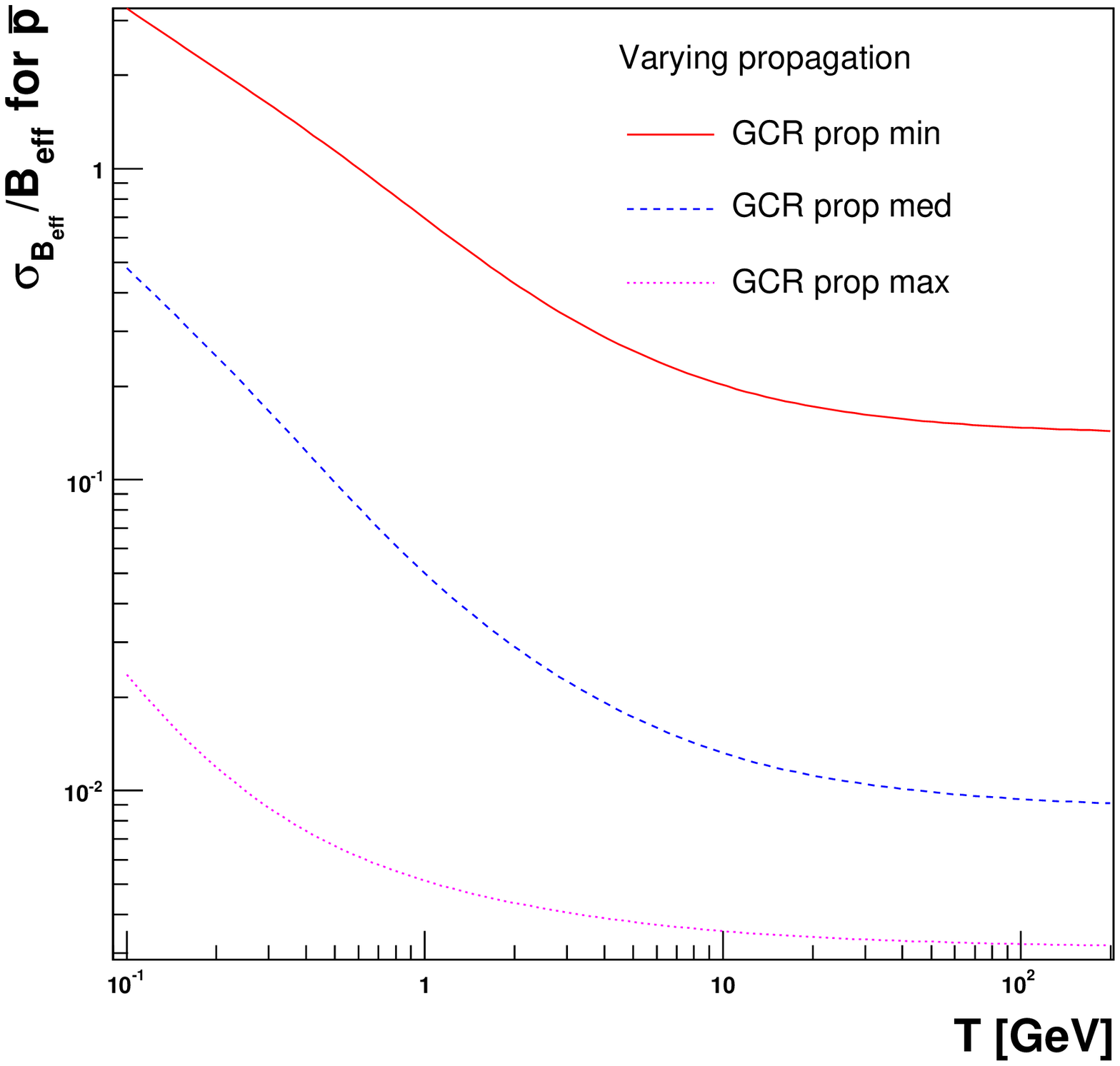}
\includegraphics[width=0.5\columnwidth, clip]
		{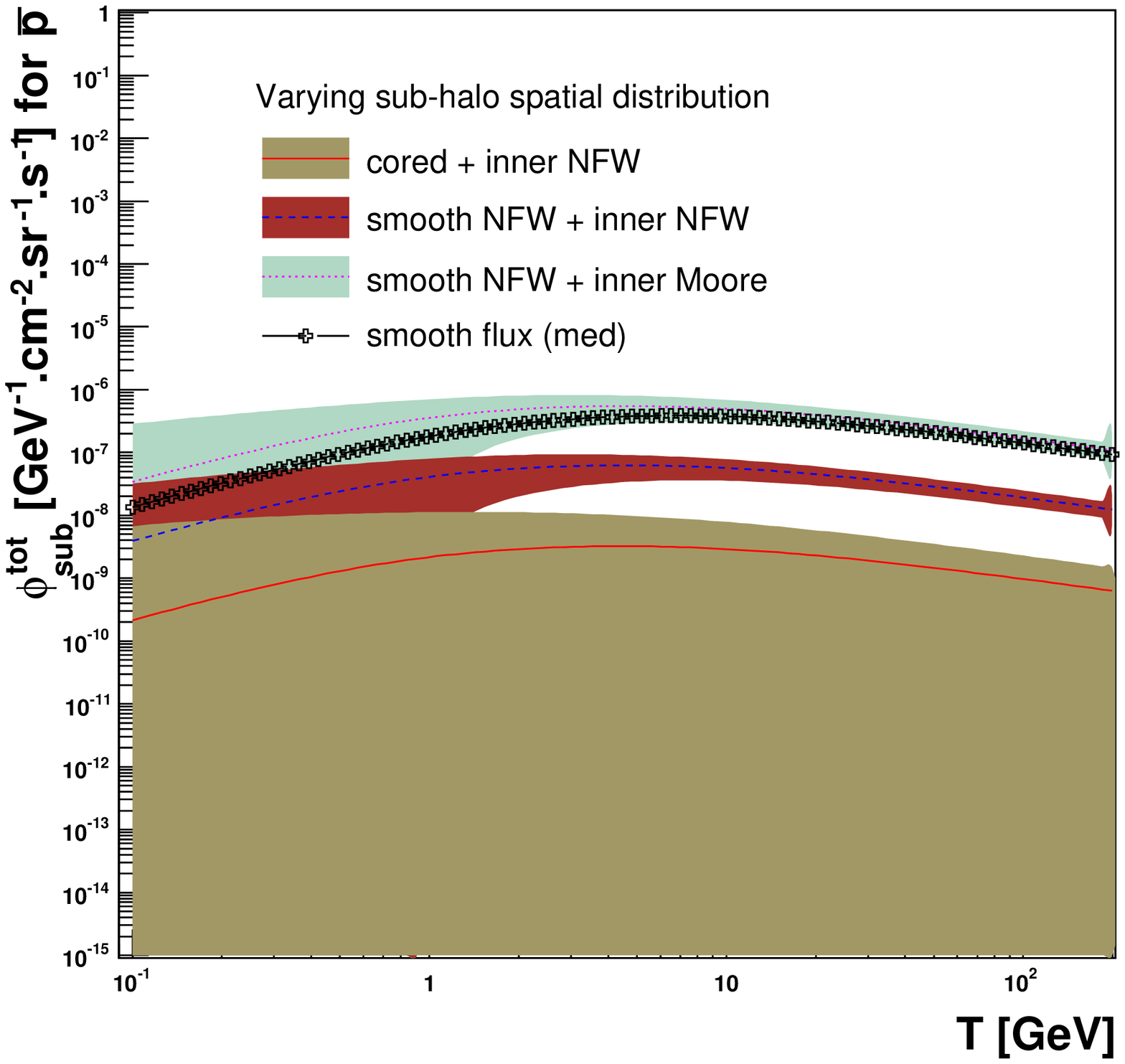}
\includegraphics[width=0.5\columnwidth, clip]
		{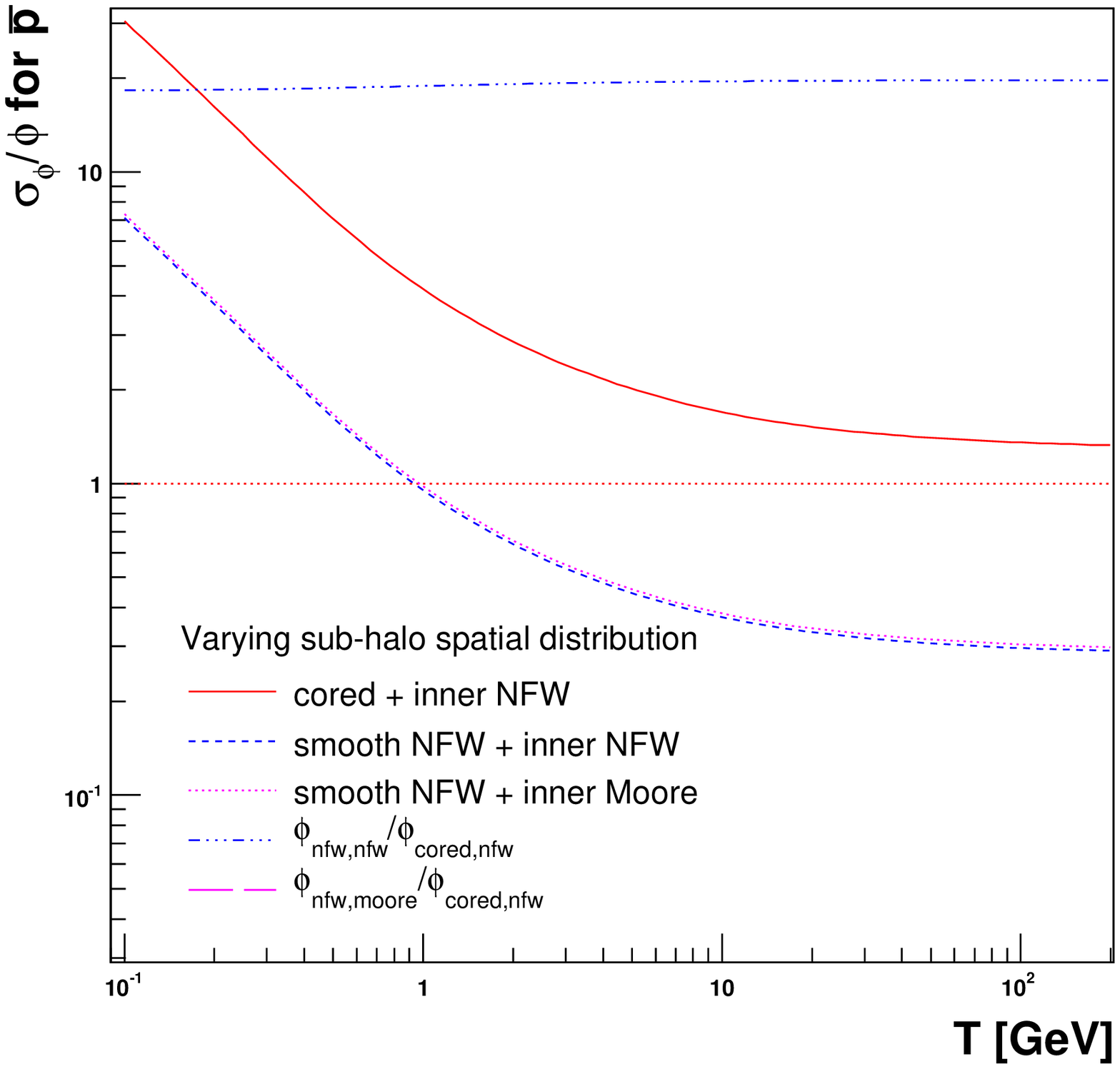}
\includegraphics[width=0.5\columnwidth, clip]
		{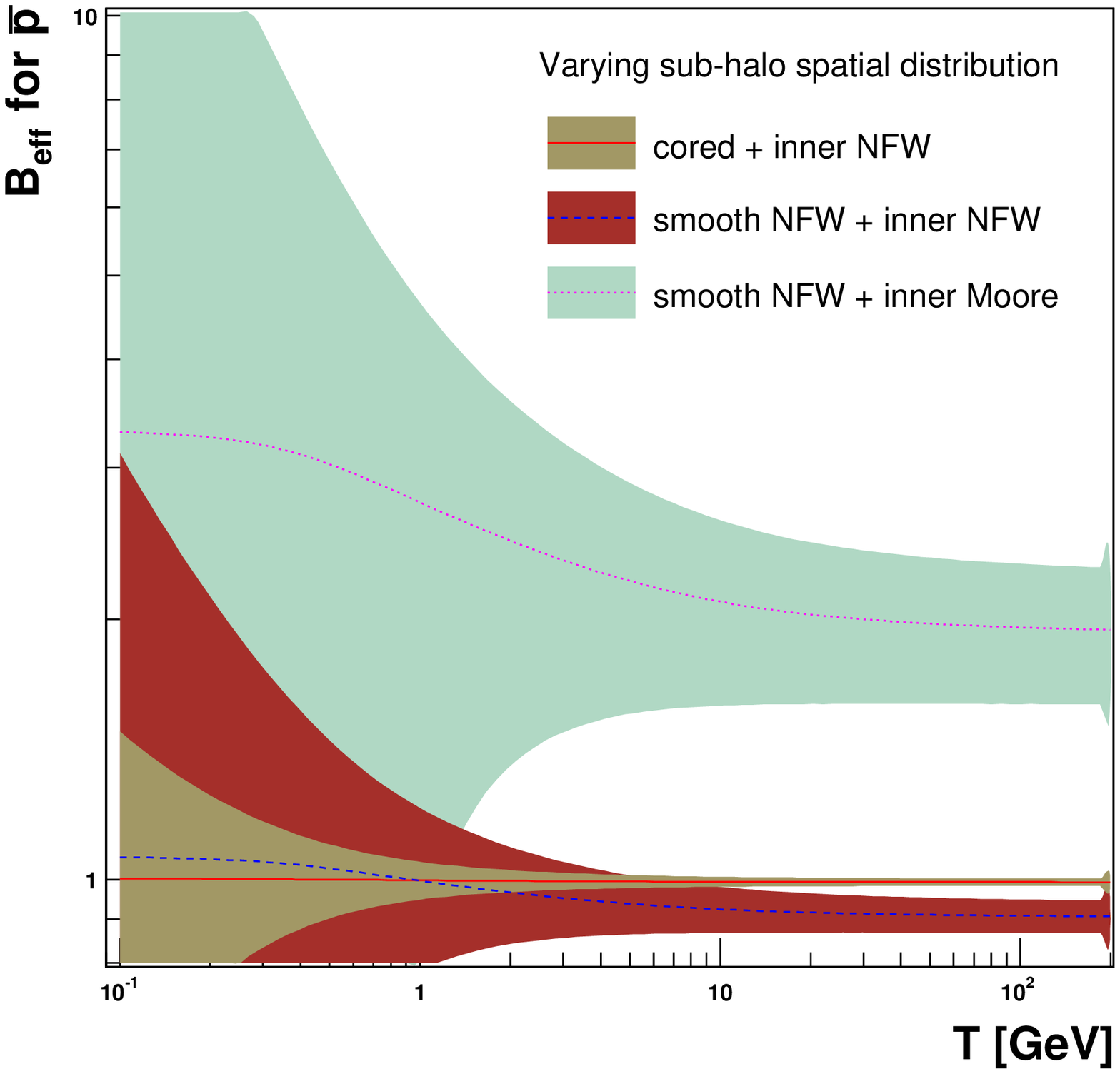}
\includegraphics[width=0.5\columnwidth, clip]
		{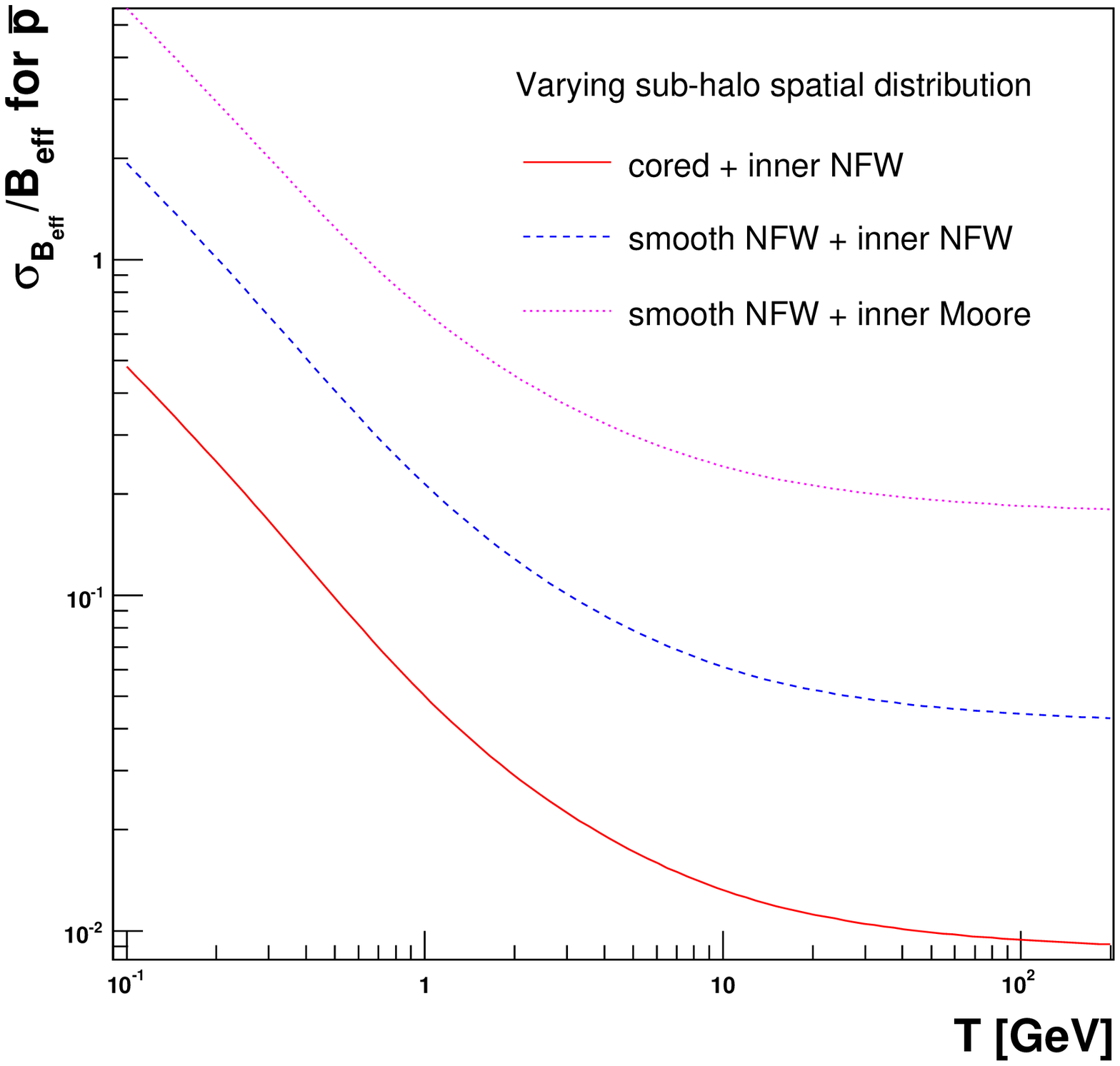}
\caption{\small \emph{Same than Fig.~\ref{fig:space-like_effects_pos}, but 
    focusing on space-related effects for anti-protons. First row: 
    Varying the propagation modelling. Second row: Varying the space 
    distribution (and the sub-halo inner profile).}}
\label{fig:space-like_effects_pbar}
\end{center}
\end{figure*}

We now discuss the origin of systematic differences when varying the 
DM configuration as well as the propagation modelling. 
Figure~\ref{fig:mass-like_effects_pbar} illustrates effects that are of 
mass type, while Fig.~\ref{fig:space-like_effects_pbar} shows the 
space-related ones. They are presented the same way as for positrons. 

For the mass-type category, the comments are exactly the same as those for 
positrons, and are already discussed in Sect.~\ref{subsubsec:pos_tot}.

Regarding space-related consequences, the picture is the reversal from that 
of  positrons, and the conclusion are the same as for positron, given the 
energy axis is read inversely.

In summary, the maximum boost factor occurs at low anti-proton energies 
when clumps are spatially distributed according to the smooth profile, and 
when they have an inner Moore profile. But even in this (disfavoured) 
configuration, the asymptotic mean value of the enhancement factor is 
$\lesssim 3$.

%/////////////////////////////////////////////////////////////////
\section{Summary and conclusions}
\label{sec:conclusions}

Clumpiness is a robust prediction of hierarchical structure formation, where
the bottom-up growth of structures is a consequence of the \lcdm\ cosmology.
Many issues remain about clumps, beside their survival, such as their number 
density, their mass and spatial distribution, as well as their intrinsic 
properties. In recent years, high resolution N-body simulations succeeded 
in tracing their gross features, although we are still far from a definite 
answer for several key parameters. Indeed, whereas numerical experiments 
now converge to a level better than 10\% over wide dynamic ranges 
\citep{2007arXiv0706.1270H}, the inclusion of baryons in hydrodynamical 
simulations, which could strongly modify the inner properties of sub-halos, 
remains a very difficult task.

The impact of the DM inhomogeneities on DM annihilation 
in the Galaxy was first underlined in~\citet{1993ApJ...411..439S}. 
With the achievement of cosmological N-body simulations in the last decade, 
and due to their positive results on clumpiness, many papers have 
subsequently focused on the consequences for the phenomenology of SUSY 
indirect detection in several channels, involving $\gamma$-rays and antimatter 
GCRs \citep[e.g.][]{1999PhRvD..59d3506B}. The effects for $\gamma$-rays 
have been extensively studied, whereas diffusion processes make the study 
much more difficult for antimatter GCRs. \citet{2003PhRvD..68j3003B} carried 
out a detailed study of boost factors for $\gamma$-ray signal (so the 
conclusions also hold for neutrinos), and concluded to a factor of 2 to 5 for 
different density profiles in a clump. However, whether or not primary 
antimatter fluxes may be boosted by clumpiness is an important issue for the 
interpretation of forthcoming data, especially for positrons (see the 
discussion below), and for putting more robust constraints on new physics.

In this paper, we have inspected the boost factors for antimatter GCRs 
as deeply as possible and the results are mostly independent from any 
WIMP model, in the context of cosmological substructures. To this aim, we 
have followed the semi-analytical method proposed in 
\citet{2007A&A...462..827L}, already used by \citet{2007PhRvD..75h3006B} 
and \citet{2007PhRvD..76h3506B} to compute the boost factors for 
positrons and anti-protons, but in a scenario in which DM 
inhomogeneities are due to the presence of intermediate mass black holes. 
We have also cross-checked our results with a more time-consuming MC 
simulation.

A full DM model was defined with (i) the host halo smooth DM profile, 
(ii) inner properties, minimal mass, space and mass distributions of 
sub-halos, and (iii) the mass-concentration relation. The reference set was 
chosen accordingly, with (i) NFW (kept fixed throughout the paper, see 
Table~\ref{table:fixed_values}), (ii) inner-NFW + $10^{-6}$ \Msol + cored 
isothermal + $\alpham = 1.9$, and (iii) B01 concentration. We then
extensively spanned over some extreme values of each parameter. We normalised 
all configurations so that the average local DM density was kept constant 
$\rho_{\rm tot}(\Rsol) = \rhosol=0.3$~GeV~cm$^{-3}$. For completeness, we 
also considered three sets of GCR propagation modelling, mainly characterised 
by the size of the diffusive halo, and by the amplitudes of diffusion and 
convection processes; all of them decrease from 
\emph{max} to \emph{min} settings (\emph{med} being the reference).

We found our results to depend on two main classes of effects, namely 
mass- and space-related. The mass-type effects characterise the full 
amount of antimatter produced in clumps through WIMP annihilations, given by 
the product $\Ncl \langle \xi\rangle _M$ (see Sect.~\ref{subsec:mass-like}), 
and its variance; they depend on the number of clumps (fixed by \Mmin\ at a 
given \alpham), their inner profiles (NFW or Moore) and mass distribution 
(\alpham). The space-related effects describe the average probability for the 
produced antimatter to reach the Earth, and is determined by $\langle \gtilde
\rangle $, and the associated variance; they therefore encode the whole 
spatial information, i.e. the propagation averaged on the clump space 
distribution, as well as the energy dependency. Our results are the following, 
some trends being consistent with some found in previous studies:
\bi
  \item larger \alpham\ values lead to larger luminosity of clumps 
    (hereafter ${\cal L}_{\rm cl} \propto \Ncl \langle \xi\rangle _M$), 
    because \Ncl\ increases faster than $\langle \xi\rangle _M$ decreases;
  \item diminishing \Mmin\ values lead to larger ${\cal L}_{\rm cl}$ if 
    $\alpham\ge 1.9$, otherwise this has no effect on the average luminosity 
    (massive clumps dominate the signal in average);
  \item the luminosity of a clump of a given mass with an inner Moore profile 
    is 10 times that obtained for the same clump with a NFW profile (which can 
    be analytically handled); hence, the whole clumpy flux can be evaluated 
    from a NFW profile and rescaled to any profile very easily, being for 
    instance 10 times larger for a Moore profile; this is independent of the 
    concentration relation assumed and the species considered;
  \item the maximum flux is obvioulsy obtained for maximal concentrations, 
    here the B01 $\cvir-\Mvir$ relation, and
    is lower for the ENS01 case (respectively a factor of $\sim 3$ and
    $\sim 30$ lower | mass dependent | for $\Mcl=10^{10}~\Msol$
    and $\Mcl=10^{-6}~\Msol$);
  \item spanning extreme ranges of parameters leads to the same 
    conclusion that no sizable enhancement is expected from clumpiness in 
    average, asymptotic values of \Beff varying from $(1-\fsol)^2 \sim 1$ 
    to $\sim 2.5$;
  \item boost factors for GCRs depend on energy, which characterises the size 
    of the effective volume of sensitivity that can be probed around the Earth;
  \item the relative statistical uncertainties on the whole sub-halo fluxes 
    and on the corresponding boost factors are dominated by mass-related 
    effects | linked to the internal properties of clumps and to their mass 
    distribution | as soon as small clump masses are considered 
    ($\lesssim 1\Msol$), and are hugely enhanced due to the mixing with the 
    (weaker) space-related ones | linked to the spatial distribution of clumps 
    and to propagation | especially when the GCR propagation scale gets 
    slender and slender, i.e. for low energy anti-protons, and detected (high) 
    energies close to the injected one for positrons (while the absolute 
    fluxes do depend strongly on propagation);
  \item statistical uncertainties are small at low energies for positrons 
    and at high energies for anti-protons (when the propagation scale is 
    larger).
\ei

As a first and mere conclusion concerning the whole sub-halo flux, the 
combination of smooth-like space distribution, small $\Mmin$, large $\alpham$, 
large concentrations, and very cuspy inner profiles will obviously lead to 
the largest mean flux (in the statistical sense of averaging over many DM 
outcomes for a given configuration). The maximum set of parameters that we 
considered for the sub-halos was defined by a spatial smooth-like NFW 
distribution, $\Mmin = 10^{-6}\Msol$, \alpham = 2, a Moore inner profile 
($\propto r^{-1.5}$), and the B01 concentration model. This led to asymptotic 
values of boost factors around 20 with small statistical 
errors (see Fig.~\ref{fig:extreme_boosts}). Conversely, a cored isothermal 
spatial distribution, large $\Mmin$, small $\alpham$, small concentrations 
(here ENS01) and mildly cuspy (e.g. NFW) inner profiles will lead to the 
smallest mean flux. We remind the reader that a very simple and 
straightforward way to estimate whether or not sub-halos may enhance the DM 
contribution to the antimatter fluxes is to verify whether the condition 
given by Eq.~(\ref{eq:boost_criterium}) is fulfilled.

Furthermore, note that the only relevant parameters for estimating the 
variance on the sub-halo flux are the minimal mass of clumps, their mass 
distribution and their local number density, beside the propagation length of 
the GCR species that defines an effective detection volume.

These results are in agreement with those of \citet{2004PhRvD..69j3509H},
who concluded, though mostly qualitatively, that it is very unlikely that 
significant boost factors occur for positrons. Our results, however, are more 
quantitative and detailed, apply to both anti-protons and positrons for any 
set of propagation parameters, and encompass the single configuration 
(NFW-ENS01) used in the above paper for clumps.  This is also consistent with 
the results of \citet{2003PhRvD..68j3003B} for $\gamma$-rays, where slightly 
larger boost factors were found: such a difference is expected because, as 
already underlined, $\gamma$-rays are integrated along the line of sight (we 
remind that the luminosity of clumps dominates over the smooth distribution
one beyond a few tens of kpc from the Galactic centre), instead of integrated 
inside a more {\em local} volume like for GCRs. 

It is interesting to ask the dependence of our calculation on the 
mass resolution achieved so far in N-body simulations, which we have 
referred to in defining our parameter sets. Indeed, the N-body numerical 
results are only valid at the spatial scale associated with the test particle 
mass, so that the Vlasov limit may not be reached at the smallest scales 
considered here (see discussion in Sect.~\ref{sec:DM}). Therefore, 
extrapolations of the physical properties of sub-halos down to $10^{-6}\Msol$ 
should always be taken cautiously, even if some numerical studies were able to 
survey such small systems at high redshifts \citep{2005Natur.433..389D}. 
Besides, even if DM sub-halos of $10^{-6}\Msol$ wander in the Galaxy, we can 
actually not know anything about their characteristics. Nevertheless, 
theoretical arguments based on the (inflation-motivated) scale invariance of 
the DM power spectrum down to the free streaming scale set by particle 
physics, the theoretical understanding of the DM mass function, the current 
knowledge, would it be far from complete, of hierarchical structure formation, 
and some numerical studies on the survival of very small sub-halos, somehow 
guarantee that our choice of parameter ranges is rather reasonable and 
sufficiently large to encompass a wide field of possibilities. Therefore, 
while this strongly asks for more detailed studies of the smallest DM 
structures, our results should also be taken as general statements that 
describe the effect of each considered parameter on boost factor predictions.

One could recover a sizable (energy dependent) boost factor by considering 
either a sub-halo which would be very massive ($\gtrsim 10^7 \Msol$) as well 
as very close to the Earth ($\lesssim$ 1 kpc, see e.g. 
Sect.~\ref{subsubsec:space-like_pos}), or very cuspy inner profiles combined 
with a significant local abundance of sub-halos. The latter case would 
correspond to a clump configuration given by a smooth-tracking spatial 
distribution, $\Mmin\lesssim 10^{-6}\Msol$, $\alpham\gtrsim 2$, concentration 
$\gtrsim$ B01, and $r^{-\beta}$ inner profiles with $\beta \gtrsim 1.5$. Such 
a situation is very improbable given the current theoretical results of 
gravitational collapse or mass function studies, and also considering the most 
likely configurations of clumpiness found among N-body results. Thus, it may 
not be taken as a natural prediction of structure formation. Regarding 
the former case, we stress that the statistical probability to find such a 
massive object in the solar neighbourhood is vanishingly small (such masses 
are now well resolved in numerical simulations, and are not expected to be 
numerous). Moreover, some observational constraints might exist on the 
presence of such a massive and close object. Anyway, the calculation of 
fluxes originating from a single nearby source is straightforward, and one can 
very easily model its required features. The price to pay would be to invoke 
some kind of Galactic \emph{lottery} in order to explain \emph{why} a single 
clump would wander \emph{here} and \emph{now}.

%%%%%%%%%%%%%%%%%%%%%%%%%%%%%%%%%%%%%%%%%%%%%%%%%%%%%%%%%%%%%%%%%%%%%%
\begin{acknowledgements}
We warmly thank C.~Tao and all the organisers of the {\em 2$^{\rm nd}$
Sino-French Workshop on the Dark Universe}, during which this work has been 
initiated. We are grateful to J.~Diemand, P.~Salati and R.~Taillet for 
enlightening exchanges and comments, and to V.~Eke for providing us with his 
concentration code. J.L. and D.M. acknowledge the support from the French 
GDR-SUSY, as well as the incentive of its current headmaster J.~Orloff. 
D.M. thanks all people at IHEP for warm hospitality during his stay. X.-J.B. 
is supported by the NSF of China under the grant Nos. 10575111, 10773011 and 
also in part by the Chinese Academy of Sciences under the grant No. 
KJCX3-SYW-N2.
\end{acknowledgements}

%%%%%%%%%%%%%%%%%%%%%%%%%%%%%%%%%%%%%%%%%%%%%%%%%%%%%%%%%%%%%%%%%%%%%%
\bibliographystyle{aa}
\bibliography{lymb}

%%%%%%%%%%%%%%%%%%%%%%%%%%%%%%%%%%%%%%%%%%%%%%%%%%%%%%%%%%%%%%%%%%%%%%
%%%%%%%%%%%%%%%%%%%%%%%%%%%%%%%%%%%%%%%%%%%%%%%%%%%%%%%%%%%%%%%%%%%%%%
%\appendix
\begin{appendix}
%%%%%%%%%%%%%%%%%%%%%%%%%%%%%%%%%%%%%%%%%%%%%%%%%%%%%%%%%%%%%%%%%%%%%%
%%%%%%%%%%%%%%%%%%%%%%%%%%%%%%%%%%%%%%%%%%%%%%%%%%%%%%%%%%%%%%%%%%%%%%
\section{Effective propagation volumes and the threshold mass 
$\mathbf{M_{\rm th}}$}
\label{app:volumes}

On the one hand, the spatial distribution and the number of clumps
in each mass decade is known. On the other hand, the propagation
properties define {\em effective} volumes $V_{\rm eff}$
\citep{2003A&A...402..971T,2003A&A...404..949M},
which enclose and pre-select the sources contributing to the flux.
This is all we need for a quick estimate of the variance on the fluxes 
(for the clumps in that given mass range), hence
the estimate of the threshold mass $M_{\rm th}$. Effective volumes 
$V_{\rm eff}$ are reminded in App.~\ref{app:Veff} and $M_{\rm th}$
is given in App.~\ref{app:thresh_absolute}. In App.~\ref{app:m_th},
we show that $M_{\rm th}$ can be set to a higher value than that derived
from the effective volumes.

The method is general, and is discussed below for $\alpham=1.9$ 
(Sect.~\ref{subsubsec:mass}, Eq.~\ref{eq:mass-distrib})
and the cored distribution of clumps (Sect.~\ref{subsbubsec:spat}, 
Eq.~\ref{eq:dPdV}).

%---------------%
\subsection{Effective volumes $V_{\rm eff}$}
\label{app:Veff}
All DM sources beyond the boundary $z=L$ (size of the diffusive halo of the 
Galaxy) can be safely discarded~\citep{2002A&A...388..676B}. Furthermore,
in a diffusive process, a source located at a radial distance $r$ gives
a negligible contribution if $\eta\equiv r/L$ is larger than a few 
\citep{2003A&A...402..971T,2003A&A...404..949M}. These two boundaries
generate a cylinder $V_{\rm eff}=\pi \eta^2L^2 \times 2L$; sources out of it 
may be considered to add negligible contribution to the total 
flux\footnote{An analysis of these effective volumes has been presented and 
discussed in great details in \citet{2003A&A...404..949M}, for both \pbar\ 
and $e^+$.}.

In addition to the parameter $L$, effective volumes may be further decreased 
depending on the value of the galactic convecting wind $V_c$: the effective 
halo size $L^* = 2K(E)/V_c$ plays a similar role as $L$ (exponential cut-off 
of the contributions, \citealt{2003A&A...402..971T}). At low energy, 
$L^\lesssim L$, decreasing $V_{\rm eff}$. This set the effective volume for 
\pbar,
\begin{equation}
V_{\rm eff}^{\bar{p}} = 2\pi \eta^2 \; \{\min(L,L^*)\}^3.
\label{eq:Mth_pbar}
\end{equation}

Because of energy losses, the typical distance travelled by positrons
is $r_* = \sqrt{4K(E) \tau_{\rm loss}}$
(e.g. \citealt{2003A&A...404..949M}; see \citealt{2007PhRvD..76h3506B}
and App.~\ref{app:space-like_effects_pos} for a more precise description).
If $r_*\gg L$, we recover $V_{\rm eff}^{\bar{p}}$, but if
$r_*\lesssim L$, the effective volume for positrons is
\begin{equation}
V_{\rm eff}^{e^+} = \frac{4}{3} \pi \eta^3  r_*^3.
\label{eq:Mth_e+}
\end{equation}

Note that the above volumes are in practise distorted in various
directions (also because of the intrinsic spatial distribution of DM
sources), but they suffice for a qualitative estimate.

%---------------%
\subsection{Threshold mass $M_{\rm th}$ for the three propagation sets}
\label{app:thresh_absolute}
In Eqs.~(\ref{eq:Mth_pbar}) and (\ref{eq:Mth_e+}), we set $\eta=10$.
We remind that this parameter sets the distance beyond which sources can
be discarded. Taking a high value for $\eta$ gives a conservative estimate of 
$M_{\rm th}$. The effective volumes, which do depend on energy, completely 
determine the number of clumps $N_{\rm eff}$ contributing to the flux. As 
fluctuations in $N_{\rm eff}$ generate fluctuations in the signal, the mass 
threshold $M_{\rm th}$ is obtained demanding that $N_{\rm eff}[M_{\rm th}-
  10M_{\rm th}]\gtrsim 10$. We recall that for antimatter primaries, the most 
relevant contribution comes from the local neighbourhood, so that the local 
number density of sub-halos $d{\cal P}(\Rsol)/dV$ can be used in the next 
approximations.

\paragraph{Anti-protons:} the effective number of clumps is given by
\ben
N^{\bar{p}}_{\rm eff}[M-10M] &\approx& N_{\rm tot}[M-10M]\times
\frac{d{\cal P}}{dV}(\Rsol) \times V_{\rm eff}\nonumber\\
&\approx &10^5\left (\frac{M}{M_{\odot}}\right 
)^{-0.9}\times \left\{\frac{\min(L,L^*)}{1~{\rm kpc}}\right\}^3\nonumber,
\een
leading to
\begin{equation}
M^{\bar{p}}_{\rm th} \approx 10^4 M_\odot \times 
\left\{\frac{\min(L,L^*)}{1~{\rm kpc}}\right\}^3.
\label{eq:mth_pbar}
\end{equation}
For example, for the reference halo size $L=4$~kpc and no galactic
wind, we get the energy independent result $M^{\bar{p}}_{\rm th}\sim 
6.4\times10^5 M_\odot$.

\paragraph{Positrons:} using $\tau_{\rm loss} = 300{\rm~Myr}\times 
1~{\rm GeV}/E$, and plugging the diffusion coefficient $K(E)\approx 
K_0E^\delta$ in $r_*$, we get, at high energy,
\[
N^{e^+}_{\rm eff}[M-10M]\approx 10^9\left (\frac{M}{M_{\odot}}\right )^{-0.9}
	\left(\frac{K_0\times E^{\delta-1}}
	{1~{\rm kpc}^2~{\rm Myr}^{-1}}\right)^{3/2}
\]
and
\begin{equation}
M^{e^+}_{\rm th} \approx 10^8 M_\odot \times 
\left(\! \frac{K_0\times E^{\delta-1}}{1~{\rm kpc}^2~{\rm Myr}^{-1}}\!\!
\right)^{3/2}.
\label{eq:mth_e+}
\end{equation}
For example, for the best propagation parameters set (Table~\ref{table:prop}), 
i.e. $\delta=0.7$ and $K_0=0.0112$, we get at $E=10$~GeV, 
$M^{e^+}_{\rm th}\sim 5 \times 10^4 M_\odot$.

\paragraph{Comparison:} for the three sets of propagation parameters (as 
reminded in Table~\ref{table:prop} of Sect.~\ref{subsec:propag_param}) and for 
various energies, Table~\ref{tab:Mth} gathers the threshold mass for \pbar\ 
and $e^+$, calculated from Eqs.~(\ref{eq:mth_pbar}) and (\ref{eq:mth_e+}).
\begin{table}
\centering
\begin{tabular}{c c c c c}
\hline\hline
 $E_k$&    0.5~GeV    & 1~GeV  & 10 GeV &  100 GeV\\
  \!\!\!$M_{\rm th} (\times \!10^5M_\odot)$\!\!\!\!\!\!\!\! &   
  $\bar{p}\,|e^+$ & $\bar{p}\,|e^+$  & $\bar{p}\,|e^+$ &  $\bar{p}\,|e^+$\\ 
  \hline
{\em max}& 340./37.         & 340./21.         &  340./3.3    & 340./0.51\\
{\em med}& 0.32/1.6         & 1.2/1.2          & 6.4/0.42     & 6.4/0.15\\ 
{\em min}& 0.0007/0.08  & 0.003/0.06  & 0.1/0.04     & 0.1/0.02\\ 
\hline
\end{tabular}
\caption{\label{tab:Mth} Threshold mass (as defined in the text)
in units of $10^5M_\odot$ for \pbar\ and $e^+$ at various energies for
the three representative sets of propagation parameters 
(see Table~\ref{table:prop} in Sect.~\ref{subsec:propag_param}).}
\end{table}
For anti-protons, we repeat that the leading parameter is $L$, except at low 
energy when the Galactic wind blows particles efficiently out of the diffusive 
volume. If the halo size $L$ is large ($max$ set, i.e. $L=15$~kpc), the 
variance associated with the flux of a clump mass range is sizable only for 
masses above $\gtrsim 3.10^7M_\odot$, independent of the energy. A small halo 
size ($min$ set, i.e. $L=1$~kpc) is associated with a strong wind, for which 
the threshold mass decreases down to $\sim 10^2M_\odot$ at 0.5 GeV 
(unmodulated). For positrons, the leading parameter is the value of the 
diffusion coefficient, hence $K_0$ and $\delta$. As for anti-protons, this is 
the $min$ propagation set that leads to the smaller threshold mass. The range 
spanned by the various configurations is, however, tighter than that for 
anti-protons. The smaller values are observed at high energy (strong energy 
losses). At low energy and for the $min$ set of parameters, as emphasised 
above, taking into account the wind would give extremely small $M_{\rm th}$ as 
for anti-protons.

From these numbers, we may already predict that propagation parameters
corresponding to small $V_{\rm eff}$ will lead to smaller fluxes 
\citep{2002A&A...388..676B,2005PhRvD..72f3507B,2004PhRvD..69f3501D,
2006astro.ph.12714M}, but also a larger associated variance than the 
configurations with large $V_{\rm eff}$.

%---------------%
\subsection{Choosing a higher $M_{\rm th}$ for the MC simulation}
\label{app:m_th}

In principle, in the MC realisations (Sect.~\ref{subsec:methodMC}),
all sub-halos with a mass larger than $M_{\rm th}$ need to be generated.
We just provided an {\em absolute} criterion in App.~\ref{app:thresh_absolute}.
Using a {\em relative} criterion, a higher $M_{\rm th}$ (useful
for reducing the computational time) can be found.

The mean flux from clumps in the mass range $[M, M+dM]$ is given by the number 
of clumps $N(M)$ in the diffusive volume, and the variance of the signal is 
just proportional to $\sqrt{N}$, so that the variance-flux-ratio is 
(see also \citealt{2007A&A...462..827L})
\begin{equation}
\frac{\sigma_\phi}{\langle\phi\rangle}(M)\propto \frac{1}{\sqrt{N(M)}}\;.
\end{equation}
Using the relation $N(M)\propto M^{1-\alpham}$ of Eq.~(\ref{eq:mass-distrib}),
we get
\begin{equation}
\frac{\sigma_\phi}{\langle\phi\rangle}(M) \propto M^{(\alpham-1)/2}\;,
\label{appeq:sig}
\end{equation}
thus recovering the results of~\citet{2007A&A...462..827L}, who used 
sub-halos of equal masses and properties. As discussed in 
Sect.~\ref{subsec:lum-profile}, for $\alpham=1.9$, the luminosity is roughly 
constant per logarithmic mass bin, so that considering the sub-halo mass range 
$M_i\equiv [10^i-10^{i+1}]M_{\odot}$, we get
\begin{equation}
\sigma_\phi (M_i) \propto M_i^{0.45}\;.
\end{equation}
This equation states that the clumps maximally contributing to the variance 
are the heaviest ones, and only these need to be taken into account. If we 
adopt $M_{\rm th}=10^7M_{\odot}$, the neglected part will contribute only 
$0.1\%$ to the total variance.

\begin{figure}[t!]
\begin{center}
\includegraphics[width=\columnwidth, clip]
{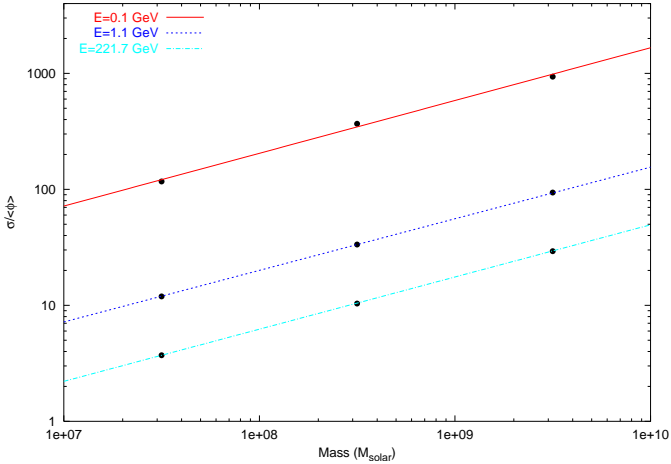}
\caption{Relative error $\sigma/\langle\phi\rangle$ as a function of mass
for several energies. The three lines are fitting results using function 
$A M^{\beta}$, with $\beta=0.455,\,0.445$ and $0.45$ from the top to bottom, 
respectively. The quantity $10^{(2i+1)/2}$ is used to represent the mass bin 
$M_i$.}
\label{fig:ratio-mass}
\end{center}
\end{figure}

As an illustration, Eq.~(\ref{appeq:sig}) is shown to be in full
agreement with the result of the MC simulation
in Fig.~\ref{fig:ratio-mass} (for anti-protons and $\alpham=1.9$).
The latter graphs are independent of the intrinsic profile of the clumps
and of the propagation parameters. In Sect.~\ref{subsec:methodMC}, we take 
advantage  of this higher value to speed up the MC calculations.

\end{appendix}

\begin{appendix}

%///////////////////////////////////////////////////////////////
\section{More details on mass- and space-related effects on fluxes and boosts}
\label{app:mass_space_complement}

We recall that the total contribution of sub-halos to the antimatter flux has 
a mean value given by:
\beq
\phicltot = \frac{v}{4\pi} S \times \Ncl \times \langle \xi\rangle _M \times \langle \gtilde\rangle _V\;,
\eeq
which can be expressed as the product of two main terms. The first one is the 
integrated luminosity of clumps:
\beq
L_{\rm cl} = \Ncl \times \langle \xi\rangle _M,
\eeq
which is purely mass-related in the sense that it depends only on the 
mass distribution and internal features of sub-halos. The relevant 
parameters are \Mmin, \alpham, the mass-concentration relation and 
the inner sub-halo profile. The second is merely the propagation term 
$\langle \gtilde\rangle _V$ averaged over the spatial distribution of 
sub-halos, which is then of purely space type.

%--------------------%
\subsection{Mass-related effects}
\label{app:mass-like_effects}
A simple analysis of the behaviour $L_{\rm cl}$ is helpful, and gives already 
interesting insights on the final results. In the limit for which 
$\Mmin\ll\Mmax$ and $\alpham \sim 2$, we find that the total number of clumps 
$\Ncl \propto M_{\rm ref}^{\alpham-1} \Mmin^{1-\alpham}$ 
(see Eq.~\ref{eq:Ncl_approx}), and that the mean value $\langle \xi\rangle_M 
\propto \Mmin^{\alpham-1}\times\ln(\Mmax/\Mmin)\sim \Mmin^{\alpham-1}$ 
(assuming $\xi \propto \Mcl$), at variance of small factors. This means that 
$L_{\rm cl} \propto M_{\rm ref}^{\alpham-1} \times \ln(\Mmax/\Mmin)$ increases 
with \alpham, and slightly (logarithmically) depends on \Mmin. Actually, we 
find that spanning (\alpham,\Mmin) respectively from ($2,10^{-6}\Msol$) to 
($1.8,10^{6}\Msol$) makes a decrease of only $\sim 40$ in the integrated 
luminosity (the above approximation gives $\sim 150$, but we recall that 
the actual dependence of $\xi$ in \Mcl~is not merely linear). Some numerical 
values are given in Table~\ref{tab:mass-like_effects}.

The pure mass-induced relative fluctuations of the sub-halo flux are given by 
$\sigma_\xi /(\sqrt{\Ncl}\langle \xi\rangle _M)$ (see Eq.~\ref{eq:sig_ju}). As 
$\sigma_\xi\propto \Mmin^{(\alpham-1)/2}$ from the same arguments as above, 
those relative uncertainties approximately scale like 
$(\ln (\Mmax/\Mmin))^{-1}$, and are thus expected to only slightly 
(logarithmically) decrease when \Mmin~increases. They are actually found to 
lie in the range 10-0.1\% for (\alpham,\Mmin) going respectively from 
($2,10^{-6}\Msol$) to ($1.8,10^{6}\Msol$) (see 
Table~\ref{tab:mass-like_effects}). This may appear surprising because we 
would naively expect the relative variance to scale like 
$\Ncl^{-1/2}\propto\Mmin^{(\alpham-1)/2}$, and then to depend much more 
strongly on \Mmin. To summarise, the dropping of the total number of 
clumps, which reduces the statistical sample, is compensated by smaller 
fluctuations around the mean luminosity $\langle \xi\rangle $ from clump to 
clump (the range \Mmin-\Mmax~gets thinner), so that the mass-induced relative 
uncertainties remain roughly constant.

%double xi[3][3] = { {8.49e-10, 1.e-10, 2.96e-11},
%                    {4.88e-5, 1.57e-5, 7.36e-6},
%                    {1.83, 1.3, 9.8e-1}};
%double sigma_xi[3][3] = { {1.77e-4, 3.17e-5, 5.68e-6},
%                          {4.46e-2, 1.59e-2, 5.68e-3},
%                          {1.1e1, 7.86, 5.59}};
%double Ncl[3][3] = { {1.63e13, 4.05e14, 1.01e16},
%                     {2.58e8, 1.61e9, 1.01e10},
%                     {4.08e3, 6.41e3, 1.01e4}};
%double Nxi[3][3] = { {1.38e+04,4.05e+04,2.99e+05},
%                     {1.26e+04,2.53e+04,7.43e+04},
%                     {7.47e+03,8.33e+03,9.90e+03}};
%
\begin{table*}
\centering
\begin{tabular}{c c c c c}
\hline
\hline
 $M_{\rm min}$ &  $\Ncl$  & $\langle \xi\rangle _M$  & 
$\sigma_\xi / \langle \xi\rangle _M$  & $\Ncl \times\langle \xi\rangle _M$  \\
$(\Msol)$ & & (kpc$^{3}$) &   & (kpc$^{3}$)\\
\hline
$10^{-6} $  &  $( 1.6 | 4.1 | 1.0 )\times 10^{ 13 | 14 | 16 }$  & 
$( 85 | 10 | 3 )\times 10^{-11}$ & $( 2.1 | 3.2 | 1.9 )\times 10^{5}$ & 
$( 1.4 | 4.1 | 30.0 )\times 10^{4}$    \\
 $1$  &  $( 2.6 | 1.6 | 1.0 )\times 10^{ 8 | 9 | 10 }$  &
$( 49 | 16 | 7 )\times 10^{-6}$  &$( 9.1 | 10.1 | 7.7)\times 10^{2}$  &  
$( 1.3 | 2.5 | 7.4 )\times 10^{4}$  \\
$10^{6}$  & $( 4.1 | 6.4 | 10.1 )\times 10^{3}$   & 
$( 1.8 | 1.3 | 1.0 )$ &  $ 6.0 | 6.0 | 5.7 $ &  
$( 7.5 | 8.3 | 9.9 )\times 10^{3}$ \\
\hline
\end{tabular}
\caption{Total number of clumps $\Ncl$, mean effective annihilation volume 
  $\langle \xi\rangle _M$ and its relative variance $\sigma_\xi/\langle \xi
  \rangle _M$, and product $\Ncl \times f\langle \xi\rangle _M$ (proportional 
  to the total number of primary GCRs produced in clumps) for different mass 
  models. We choose three different mass ranges varying the minimal mass from 
  $10^{-6}$ to $10^6 \Msol$, and for different logarithmic slopes 
  $\alpham = \{1.8|1.9|2.0\}$ of the mass distribution.}
\label{tab:mass-like_effects}
\end{table*}

%--------------------%
\subsection{Space-related effects for positrons}
\label{app:space-like_effects_pos}
The space-related effects for positrons are characterised by the mean value 
and the variance of \gtilde~over the spatial sub-halo distribution. For 
positrons, the relevant scale is the energy loss scale, which sets the 
characteristic propagation length, as stressed in \citet{2007A&A...462..827L}. 
This propagation length is given by the following equation:
\beq
\lambda_D \equiv \sqrt{4 K_0 \tau_E
  \left(\frac{\epsilon^{\delta - 1}
  -\epsilon_{S}^{\delta-1}}{1-\delta} \right)  } \; \;,
\label{eq:propscale_pos}
\eeq
where $K_0$ and $\delta$ are the normalisation and the logarithmic slope of 
the diffusion coefficient, respectively, and $\epsilon \equiv (E/\{E_0 = 1\;
\rm{GeV}\})$. If we take the medium propagation parameters of 
Table~\ref{table:prop}, and a typical timescale for energy loss of $\tau_E 
\approx 10^{16}\,\text{s}$, then for a 200 GeV injected energy, we find a 
propagation length $\lambda_D \simeq 6.9 \; \text{kpc} \times 
\sqrt{\epsilon^{-0.3}-0.2}$, which ranges from 0.4 kpc at a detected energy 
$E_d$ of 190 GeV to 5.7 kpc at 1 GeV. This shows that for positrons, the main 
contributions to the flux are likely to come from regions close to the solar 
neighbourhood. From Eq.~(\ref{eq:propscale_pos}), we also see that a larger 
diffusion coefficient or/and a lower detected energy will allow the 
integration of contributions over a larger volume (the former case is, 
however, generally associated with a smaller diffusion slab model, which 
erases those extra contributions).

Assuming now that all relevant contributions are those inside a volume $V_D$ 
around the Earth bounded by $\lambda_D \lesssim L$ ($L$ is half the vertical 
extension of the diffusive halo), and that the propagation is roughly constant 
over this small volume, we can simplify the propagator \gtildepos\ in the 
limit of infinite 3D diffusion:
\beq
\gtildepos \simeq \frac{\tau_E}{E_0 \epsilon^2}\times
\frac{\theta(\lambda_D - |{\bf r}-{\bf r}_{\odot}|)}
     {(\pi \lambda_D^2)^{3/2}}\;.
\eeq
Assuming also that the spatial distribution of clumps does not vary that 
much within $V_D$, thus given by $d{\cal P}(\Rsol)/dV$, we get:
\beq
\langle \gtildepos\rangle _{V} = \int_V d^3{\bf x}\; \gtildepos 
\frac{d{\cal P}}{dV}({\bf x}) 
\simeq \frac{\tau_E}{E_0 \epsilon^2} \times \frac{d{\cal P}}{dV}(\Rsol)\;.
\eeq
The interpretation is trivial, as well as the consequences for the boost 
estimate. Besides, the relative variance straightforwardly reads:
\beq
\frac{\sigma_\gtildepos}{\langle \gtildepos\rangle } \propto \left( V_D^2 
\times \frac{d{\cal P}}{dV}(\Rsol)\right)^{-1/2}\;.
\eeq
This means that the relative variance decreases when the spatial probability 
function of clumps raises and when the effective detection volume increases, 
which is physically obvious but better quantified with the previous equations.
If we argue in terms of local number density of clumps, this only says 
that the global space-related variance scales like $N_{\rm obs}^{-1/2}$, where 
$N_{\rm obs}$ is the number of clumps inside $V_D$.

A more quantitative information is given in Table~\ref{tab:propag_effects}. In 
this table, we calculate the mean value and the variance of \gtildepos\, 
as defined by Eqs.(\ref{eq:def_gmean}) and (\ref{eq:sig_g}), respectively, for 
the three propagation models detailed in Table~\ref{table:prop}. We consider 
the injection of 200 GeV positrons in sources, ($Q(E_S) = \delta(E_S - 200 
\;{\rm GeV})$), and compute the propagator averaged on the spatial 
distribution of sub-halos together with the associated statistical variance. 
We show the results obtained for detected energies of 150 and 10 GeV, which 
correspond to diffusion lengths $\lambda_D$ of $\sim$0.9 and 3.8 kpc, 
respectively. Such quantities are parts of the flux probability function 
related to a single clump, but encoding only the spatial and propagation 
information.

\begin{table*}
\centering
\begin{tabular}{c | c c | c c }
\hline
Propagation & $\langle \gtildepos\rangle $ & $\sigma_{\gtildepos}/\langle 
\gtildepos\rangle $ & $\langle \gtildepbar\rangle $ & 
$\sigma_{\gtildepbar}/\langle \gtildepbar\rangle $ \\
& (s.kpc$^{-3}$.GeV$^{-1}$) &  & 
(s.kpc$^{-3}$.GeV$^{-1}$) &  \\
\hline
\emph{Min} &  $5.462\times 10^{6} | 1.723\times 10^{5}$ & $287.1 | 1089$ & 
$2.335\times 10^{8} | 4.850\times 10^{7}$ &  $829.2 | 622.2$ \\
\hline
\emph{Med} &  $2.840\times 10^{7} | 1.744\times 10^{5}$ & $67.95 | 446.6$ & 
$9.220\times 10^{8} | 2.320\times 10^{8}$ & $106.2 | 84.6$ \\
\hline
\emph{Max} &  $3.666\times 10^{7} | 1.742\times 10^{5}$ & $25.84 | 267.3$ & 
$2.652\times 10^{9} | 1.157\times 10^{9}$ & $18.42 | 15.22$\\
\hline
\textcolor{blue}{\emph{Med$^{\star}$}} & \textcolor{blue}{$5.589\times 10^{8} 
| 3.191\times 10^{6}$} & \textcolor{blue}{$15.02 | 104.3$} & 
\textcolor{blue}{$1.801\times 10^{10} | 4.561\times 10^{9}$} & 
\textcolor{blue}{$23.34 | 18.53$} \\
\hline
\end{tabular}
\caption{\emph{Mean value and variance of \gtilde\ (single clump) for 
positrons (left-hand side) and anti-protons (right hand side), each line 
accounting for each set of propagation parameters of 
Table~\ref{table:prop}. The results for positrons correspond to particles 
injected at 200 GeV in sources, and detected at $10|150$ GeV. For 
anti-protons, a flat spectrum $dN/dT = 1 \;{\rm GeV}^{-1}$ is injected at 
sources and detected kinetic energies of $10|150$ GeV are also considered. For 
both positrons and anti-protons, the DM configuration is the reference one, 
except for \textcolor{blue}{Med$^{\star}$}, for which we take a space 
distribution of sub-halos that tracks the smooth NFW component.}}
\label{tab:propag_effects}
\end{table*}

We see from this table the expected behaviour when varying the propagation 
model: at a given detected energy, $\langle \gtildepos\rangle $ increases from 
the minimal to the maximal propagation configuration, and also increases when 
the positron is detected at a lower energy (its mean free path is somehow 
longer). There is a factor of $\sim 20$ between the minimum (\emph{min} model, 
high $E_d$) and maximum values (\emph{max} model, low $E_d$). The 
space-associated contribution to the relative variance affecting the single 
clump flux is also given in Table~\ref{tab:propag_effects}, and is in the 
range 10-1000.  It has to be compared with the relative mass-induced variance 
of Table~\ref{tab:mass-like_effects}, i.e. that on $\xi$. We see that while 
the relative variance on \gtildepos\ is large, the one affecting $\xi$ almost 
always dominates, unless the minimal mass of clumps is $\gtrsim 10^6\Msol$. 
Thus, though propagation uncertainties are important, the mass-induced effects 
are likely to outclass the systematic uncertainties over a large energy range. 
Nevertheless, they re-enter the game as soon as the propagation scale gets 
very short (detected energies very close to the injected energy). As the 
crossing space-mass term dominates the global relative variance, we can 
determine the systematic errors affecting the global flux predictions, taking 
the previous ranges obtained for mass-like contributions. For the average 
clumpy contribution to the flux, we get $\sim 40\times 20$, which are three 
orders of magnitude. For the associated relative variance, we find ranges 
$0.1-10\%$ (clump mass and number) and 10-1000 (space-induced), which yield 
a total of four orders of magnitude. This provides the systematic 
uncertainties on the flux and its variance. Nevertheless, such uncertainties 
are diluted for the boost factor estimate, as only a small part of the 
parameter space gives a sub-halo contribution greater than that of the smooth.

%--------------------%
\subsection{Space-related effects for anti-protons}
\label{app:space-like_effects_pbar}

The same reasoning used for positrons can apply to anti-protons, that 
is the use of the propagation effective volume. Therefore, complementary 
to the following discussion, we refer the reader to the arguments and 
conclusion of the previous paragraph.

The typical diffusion length for anti-protons depends mainly on the convective 
wind, and can be expressed as:
\beq
\Lambda_D \equiv \frac{K(E)}{V_{\rm conv}},
\eeq
where $K(E)$ is the diffusion coefficient at energy $E$ and $V_{\rm conv}$ 
is the velocity of convection. This is quite different to that used for 
positrons in the sense that this length is much lower at low energy. With the 
medium set of propagation parameters, we get $\Lambda_D \simeq 1.4 | 86 $ kpc 
for kinetic energies of $0.1|10$ GeV, respectively. However, the vertical 
boundary $L$ of the diffusive halo limits that range to a few times $L$, so 
that the actual characteristic propagation length for anti-protons is usually 
comparable with the size of the slab. Anyhow, this means that above a few GeV, 
anti-protons can almost probe the entire diffusive slab, as they can originate 
from far away regions. Hence, the picture for anti-protons is the reversal 
of that for positrons. Besides, the characteristic diffusion length is larger 
for anti-protons than for positrons, so that the arguments using local 
quantities are less relevant here, unless for asymptotic values of the 
boost factor (which occur at low energy for anti-protons, see 
Eq.~\ref{eq:bsol}).

Numerical values for the mean value and variance of \gtildepbar\ are given in 
Table~\ref{tab:propag_effects}, where we have taken a flat injection 
spectrum for anti-protons, $dN/dT = 1\;{\rm GeV}^{-1}$, and we have considered 
two detected kinetic energies of 10 and 150 GeV (no energy losses for 
anti-protons). We recover the same range for systematic uncertainties as for 
positrons (see App.~\ref{app:space-like_effects_pos}).

\end{appendix}

%%%%%%%%%%%%%%%%%%%%%%%%%%%%%%%%%%%%%%%%%%%%%%%%%%%%%%%%%%%%%%%%%%%%%%

\end{document}